\documentclass[manuscript,screen]{acmart}

\usepackage{algorithmic}
\usepackage{graphicx,color}
\usepackage{textcomp}
\usepackage{xcolor}
\usepackage{hyperref}
\usepackage{tabularx}
\usepackage{booktabs}

\usepackage{float}
\usepackage{multirow}

\usepackage{graphicx}
\usepackage{listings}
\usepackage{fancyvrb}

\usepackage{tikz}
\usetikzlibrary{positioning,fit,backgrounds}

\usepackage{array}
\usepackage[caption=false,font=normalsize,labelfont=sf,textfont=sf]{subfig}
\usepackage{stfloats}
\usepackage{url}
\usepackage{verbatim}

\AtBeginDocument{%
  }

\setcopyright{none}
\copyrightyear{This is a preprint under review}
\acmYear{2026}
\acmDOI{}
\acmISBN{978-1-4503-XXXX-X/2018/06}

\begin{document}

\title{Privacy Engineering: A Systematic Literature Review}

\author{Nemania Borovits}
\email{n.borovits@tilburguniversity.edu}
\affiliation{%
  \institution{Tilburg University}
  \country{the Netherlands}
}

\author{Damian Andrew Tamburri}
\email{datamburri@unisannio.it}
\affiliation{%
  \institution{University of Sannio - JADS/NXP Semiconductors}
  \country{Italy}
}

\author{Willem-Jan Van Den Heuvel}
\email{w.j.a.m.v.d.heuvel@tue.nl}
\affiliation{%
  \institution{Eindhoven University of Technology}
  \country{the Netherlands}
}

\renewcommand{\shortauthors}{Borovits et al.}

\begin{abstract}
Privacy obligations under GDPR increasingly shape software engineering. We synthesize 90 studies from 2018 to 2025 using a systematic review with thematic synthesis to chart privacy engineering. Thirteen dimensions form two recurrent cores: Privacy Enhancing Technologies (PETs) with Privacy Metrics (PM) and Verification and Testing (VT) and Governance and Accountability (GA) with Transparency and Communication (TC) and Organizational Measures (OM). Modeling and Specification (MS) mediates between the cores. Lifecycle mapping shows concentrations at requirements and design (MS, GA), at implementation and verification (PETs, VT, PM, TC) and at operation and decommissioning (GA, OM, Data Subject Rights Management (DSRM), Incident Response and Management (IRM), Lifelong Management (LM)). Handoffs link models to rules and tests, mechanisms to metrics and deployments such as enclaves and ledgers to governance records. Domains reweight but do not alter structure: healthcare weights GA with VT and PETs, IoT and edge weight PETs with VT and PM at device and edge, web measurement weights TC with VT, AI and ML weight PETs with PM. IRM, LM and Data Minimization and Purpose Limitation (DMPL) are less often primary foci, signaling priorities for future work. The results provide a practical map and a replication-ready scaffold for assessment and updates.
\end{abstract}

\ccsdesc[500]{Security and privacy~Privacy protections}
\ccsdesc[500]{Social and professional topics~Privacy policies}
\ccsdesc[500]{Security and privacy~Privacy-preserving protocols}

\keywords{Privacy Engineering, Systematic Literature Review}


\maketitle

\section{Introduction}\label{intro}

Privacy engineering translates data-protection principles into requirements, architectures, operational processes and evidence for software systems. It aims to make privacy a verifiable property by specifying who may do what with which data under which conditions and by implementing technical and organizational controls that enforce those decisions and leave auditable traces~\cite{gurses2016privacy,stallings2019information}. In contemporary practice, this scope spans requirements and threat modeling, user-facing communication artifacts (for example, consent and policy representations), Privacy-Enhancing Technologies (PETs) in data pipelines and platforms, verification and measurement practice, governance and accountability processes and lifecycle operations such as retention, erasure and redress~\cite{de_chaves_privacy_2023,saltarella2024translating,kosenkov2025systematic}. Regulatory frameworks, notably the General Data Protection Regulation (GDPR) and the California Consumer Privacy Act (CCPA), increased demand for demonstrable controls and operational evidence in daily systems, rather than policy text alone and encouraged platform designs that link organizational procedures to technical artifacts and provenance~\cite{jones_profile_2020,zhang_privacy-by-design_2022}. Supervisory authorities can impose sanctions when systems fail to protect personal data in practice; since May 2018, authorities across the European Union have imposed fines totaling almost € 7 billion for GDPR violations\footnote{\url{https://www.enforcementtracker.com/?insights}}, which shows that inadequate controls translate into direct financial loss and disruption to business. This compliance-driven view presents privacy mainly as a constraint that prevents sanctions, thereby emphasizing loss avoidance. Industry reports also describe privacy engineering as part of a broader strategy in which organizations pursue compliance alongside gains in reputation and the perceived reliability of their products and services; these accounts link privacy engineering to repeatable workflows from design to deployment, clearer roles across legal, engineering and operations teams and automation for activities such as access review, configuration management and incident handling and they describe reuse of the resulting models, controls and evidence in other governance and risk-management efforts that extend beyond privacy~\cite{porsche2021data,forbes2022privacy,bloomberg2023data,linkedin2023leveraging,ibm2023data}. In this perspective, privacy engineering serves both compliance and competitive advantage, because it reduces exposure to enforcement while also contributing to reputation and to the optimization of data-handling processes.

Evidence from deployed ecosystems shows that the controls and logs that are executed depend on the controls and logs that are executed. Web-scale measurements reported that consent banners and post-consent configurations materially change tracking behavior and page composition, suggesting that measurements that ignore banners understate the extent of tracking~\cite{woods2022commodification}. Longitudinal studies of encrypted DNS deployments documented certificate hygiene issues and centralization risks between resolvers and browsers~\cite{sassala2025path}. Large-scale analysis of mobile ecosystems uncovered widespread inconsistencies between code-level practices and stated policies, demonstrating that policy text is not a reliable proxy for run-time behavior~\cite{fan2020empirical}. In parallel, stewardship platforms for research and health data, Trusted Research Environments (TRE), Personal Health Trains (PHT) and Personal Data Stores (PDS), formalized approvals, secure settings and disclosure control, tying organizational workflows to provenance and release gates so that access and reuse remain accountable~\cite{oskoui2025developing,pammi2025digital,bindawas2025distribution}. Mechanism families matured at the same time: differential privacy extended from structured analytics to unstructured media~\cite{demelius2025recent}, homomorphic encryption and related cryptographic computation advanced implementation practice~\cite{diaz2023connecting} and federated and distributed learning brought locality and secure aggregation into mainstream pipelines~\cite{nguyen2021federated,issa_blockchain-based_2023}. Metric practice developed around calibration and robustness, supporting comparative evaluation of privacy–utility trade-offs and attack resistance~\cite{wagner_technical_2019}. Requirements and architectural methods contributed ontologies, goal/asset models, misuse and data-flow diagrams and tactic/pattern catalogs that make privacy decisions visible and traceable across design and operation~\cite{kunz2023privacy,hills2024holistic,sion2025robust}. Together, these strands establish a broad but heterogeneous landscape that spans mechanisms, interfaces, governance artifacts and lifecycle controls across domains and deployment settings.

A consolidated view is useful when it relates policy-facing artifacts and organizational processes to executable controls and verifiable evidence. Interfaces and stewardship components mediate consent, purpose and sharing and their effectiveness depends on coupling to controller logic and provenance so that user intent carries into execution and audit~\cite{kounoudes2020mapping,murmann2021design,theys2025understanding}. Mechanism-centric work refined evaluation conventions (for example, privacy–utility curves and robustness metrics), yet reporting remains uneven across stacks and domains, which complicates cross-paper comparison and reuse~\cite{wagner_technical_2019,lo_systematic_2022}. A field-level map that connects artifacts and evidentiary practices would support design planning - what artifacts should exist and when - and appraisal - what substantiates claims in operation - while improving comparability across domains~\cite{jones_profile_2020,zhang_privacy-by-design_2022}.

Several needs arise from previous work. First, incident handling and redress appear mainly within accountability narratives or metric catalogs; End-to-end exercises and longitudinal reporting are limited, which restricts the assessment of operational effectiveness~\cite{xia_towards_2024,salem_comprehensive_2023}. Second, lifecycle operations such as retention, erasure and ownership transfer are specified architecturally or procedurally, yet evidence on scalability, cross-party coordination and conflict resolution (for example, legal holds versus erasure requests) remains sparse~\cite{shishkov_privacy_2021,pallas2024privacy,carmichael_personal_2024}. Third, minimization outside locality-first designs is under-evaluated; IoT, edge and split-computing pipelines reduce exposure by design, but systematic minimization-first evaluations beyond those settings are less common~\cite{benhamida_pyff_2021,shafee2025privacy}. Fourth, multiple ecosystems show that transparency and user control do not automatically propagate to enforcement and provenance; policy–code inconsistencies and decoupled interfaces lead to drift between intent and execution~\cite{jha_internet_2022,wang_as_2024,javed2024systematic}. Fifth, governance accounts document roles and processes more often than they evaluate linkages from policies to verifiable execution paths such as contracts, attestation and temporal or audit tables, at scale and over time~\cite{jones_profile_2020,zhang_privacy-by-design_2022,mohanta2025protecting,yang_secudb_2024}. Finally, heterogeneity in the reporting of metric and the design of the benchmark hinders comparability, especially for architectural PET and cross-domain studies~\cite{wagner_technical_2019,lo_systematic_2022}.

To address the needs outlined above, the contributions of this study are as follows: (i) we derive thirteen dimensions and model their co-occurrence, linking mechanisms with measurement and testing and policies with interfaces, enforcement and logging; these relations form two interacting cores, with Modeling \& Specification acting as a mediator; (ii) we map every dimension to both the software development lifecycle and the personal data lifecycle and we identify hand-offs from specification artifacts to implementation decisions, from PET outputs to verification artifacts and from deployment choices to governance records; (iii) we characterize domain-specific emphases without changing the dimension set. In healthcare, Governance \& Accountability frequently pairs with Verification \& Testing and with Privacy-Enhancing Technologies. In IoT and edge settings, Privacy-Enhancing Technologies pair most often with Verification \& Testing and with Privacy Metrics. On the web, Transparency \& Communication pairs with Verification \& Testing. In AI/ML pipelines, Privacy-Enhancing Technologies pair with Privacy Metrics. Finally, within the included studies, Incident Response \& Management, Lifelong Management and Data Minimization \& Purpose Limitation appear only as secondary assignments. These findings articulate the state of the art in the privacy engineering literature and outline directions for future research.

The remainder of the paper is organized as follows. Section~\ref{sec:research_meth} presents the review protocol and synthesis procedure. Section~\ref{sec:reporting} reports the results: emergent dimensions and findings per dimension, with cross-cutting co-occurrence, lifecycle analyses and application domains. Section~\ref{sec:discussion} discusses the results and their implications. Section~\ref{sec:threads} presents the threads to validity and Section~\ref{sec:conclusion} concludes the study.

\section{Research Methodology}\label{sec:research_meth}

\begin{figure*}[!h]
    \centering
    \includegraphics[width=0.60\textwidth]{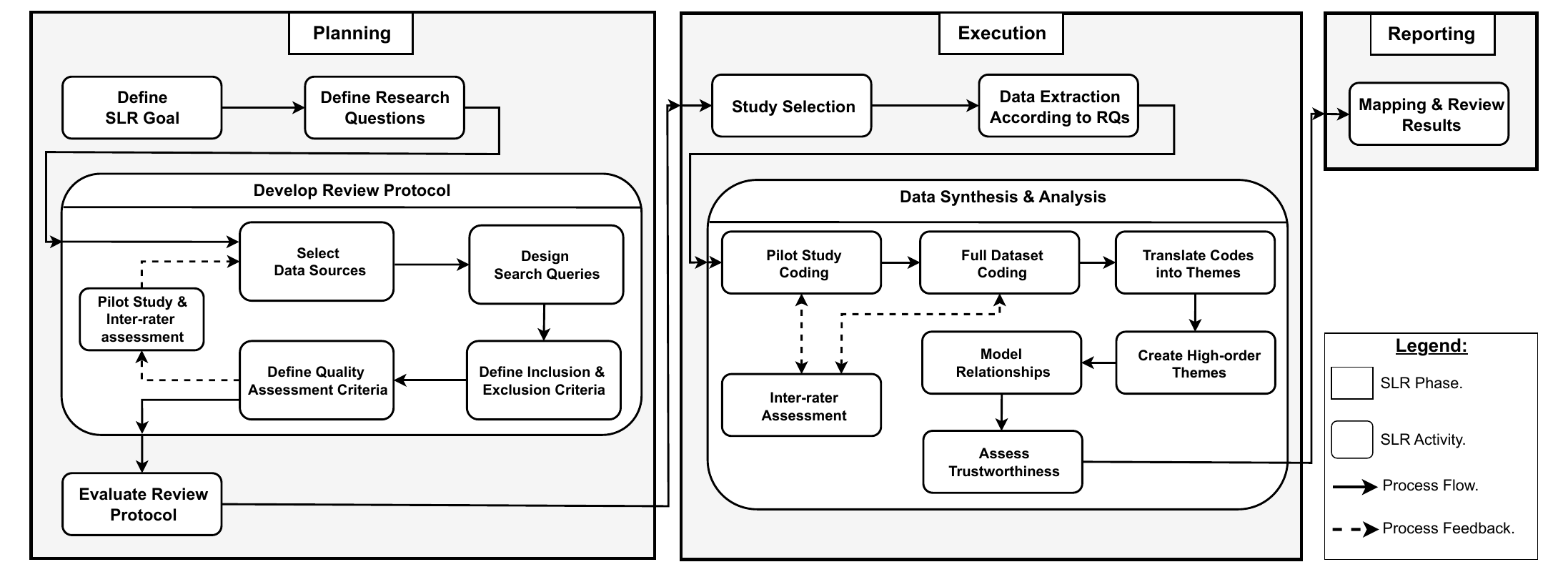}
    \caption{Overview of the SLR process following the guidelines by Kitchenham et al~\cite{kitchenham2004procedures,kitchenham2015evidence}.}
    \label{fig:slr_steps}
\end{figure*}

\subsection{Planning}\label{planning}
We planned the review in accordance with Kitchenham’s guidance for evidence-based software-engineering studies and prepared for thematic synthesis following the guidelines by Cruzes \& Dybå. Figure~\ref{fig:slr_steps} provided the scaffold for this phase (define goal and questions; develop the review protocol; evaluate the protocol) and we reported the steps in that order~\cite{kitchenham2004procedures,kitchenham2015evidence,cruzes2011recommended}. We aimed to characterize privacy engineering, based on published evidence, as an engineering activity. The focus covered artifacts and operations that teams implemented and monitored: requirements and architectural decisions; user-facing communication; mechanisms in data pipelines and platforms; measurement and verification; governance controls linked to provenance; and lifecycle actions such as retention, erasure and redress. Prior work showed that observable outcomes depended on executed controls and logs and that platform and process accounts bound approvals and assessments to enforcement points. We therefore framed questions that identified recurrent concerns across studies, captured how those concerns appeared together in deployed or evaluated systems and located them in software development and personal data lifecycles. Accordingly, we formulated the research questions as follows:

\begin{itemize}
\item RQ1. \textit{Dimensions.} Which privacy engineering dimensions emerged from the corpus and how did we define and delimit each in operational terms?
\item RQ2. \textit{Interactions.} How did these dimensions co-occur within the same studies and which pairs or triads recurred most strongly?
\item RQ3. \textit{Lifecycle.} Where did these dimensions appear along the software development lifecycle (requirements, design, implementation, verification, deployment, operation, decommissioning) and the personal-data lifecycle (collection, transfer, storage, processing/analytics, sharing/release, retention, deletion/erasure) and which hand-offs did the data support?
\item RQ4. \textit{Domains.} How did emphases and couplings vary across major domains (e.g., healthcare, IoT/edge, web/AI) without changing the overall structure?
\end{itemize}

We required each retrieved record to include (i) privacy engineering vocabulary, (ii) a software-engineering context and (iii) either evidence of implementation/evaluation or an operational governance component. This constraint maintained topical relevance and ensured that the corpus contained implemented or evaluated practice rather than policy discussion alone. We selected the following research databases, namely  IEEE Xplore (\href{https://ieeexplore.ieee.org}{ieeexplore.ieee.org}), Scopus (\href{https://www.scopus.com}{www.scopus.com}), ACM Digital Library (\href{https://dl.acm.org}{dl.acm.org}) and PubMed (\href{https://pubmed.ncbi.nlm.nih.gov/}{pubmed.ncbi.nlm.nih.gov}).

To avoid biasing the synthesis, we deliberately refrained from inserting taxonomy- or outcome-derived terms into the core query. Instead, we used a single library-neutral boolean operator and adjusted only the field scopes (Title/Abstract/Keywords) per engine. The Boolean was:
\begin{equation}
\label{eq:core-query}
\mathrm{Query} \;=\; (P) \;\wedge\; (SE) \;\wedge\; \big( (E) \;\vee\; (G) \big).
\end{equation}
In Eq~\eqref{eq:core-query}, \emph{P} denotes privacy engineering, \emph{SE} denotes software engineering, \emph{E} denotes empirical/evidence terms and \emph{G} denotes governance/operations terms. In Appendix~\ref{sec:protocol_tables}, Table~\ref{tab:facet-terms} lists the facets and the representative terms that we used in the libraries. We piloted these terms on a stratified slice to assess recall and precision, then froze the engine-specific strings~\cite{kitchenham2004procedures,kitchenham2015evidence}.

We selected these facets because \emph{P} anchored privacy-engineering and regulatory vocabulary, \emph{SE} constrained results to software artifacts/processes/evaluations, \emph{E} focused the corpus on implemented or evaluated practice and \emph{G} captured links from organizational artifacts to executable controls and logs. Together, \emph{P} and \emph{SE} ensured topical and disciplinary relevance, while \emph{E} or \emph{G} ensured operational substance, matching Eq~\eqref{eq:core-query}. We piloted the strings across venues and years, froze engine-specific variants and logged exact queries, dates and result counts in a search log, consistent with Kitchenham’s planning guidance~\cite{kitchenham2004procedures,kitchenham2015evidence}. We defined inclusion and exclusion prior to selection (Appendix~\ref{sec:protocol_tables}, Table~\ref{tab:incl-excl-privacy}): we included peer-reviewed primary or secondary studies that proposed, implemented, evaluated, audited, or operationalized privacy-engineering artifacts (methods, architectures, tools or measurements, or organizational processes tied to executable controls). We limited the search window to May 2018--March 2025 and excluded items without technical or procedural substance or outside software/system contexts. For each candidate, we recorded design type, context, data/tool availability, assumptions or threat model, links to standards or regulations (when applicable) and evaluation details (metrics, baselines/benchmarks, measurement protocols) to support transparent decision-making without assigning a numeric quality score~\cite{kitchenham2004procedures}. The first two authors piloted title/abstract and full-text screening on a balanced subset, reconciled disagreements, refined notes and then froze the protocol. We prepared a single extraction form that maintained traceability and enabled constant comparison~\cite{cruzes2011recommended}; for each included study, we captured publication metadata, context and verbatim findings with page anchors, together with analysis-ready fields linking to execution (codeable evidence for open coding, lifecycle placement in both frames and minimal tags for theme co-occurrence), preparing the dataset for synthesis and modeling in the execution phase.

\subsection{Execution}\label{execution}
\begin{figure*}[!t]
    \centering
    \includegraphics[width=0.50\textwidth]{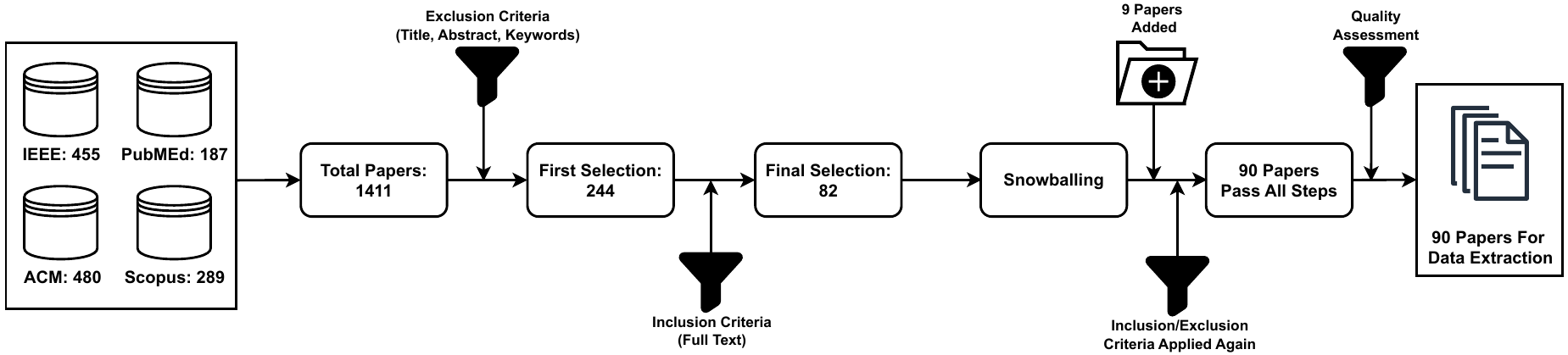}
    \caption{Overview of the papers selection process.}
    \label{fig:slr_rpcoess_exec}
\end{figure*}

We executed the protocol in Figure~\ref{fig:slr_steps} following Kitchenham's guidelines and the thematic synthesis procedure by Cruzes and Dybå and reported it for auditability~\cite{kitchenham2004procedures,cruzes2011recommended,kitchenham2015evidence}. Figure~\ref{fig:slr_rpcoess_exec} reports counts. We removed duplicates; screened titles/abstracts and full texts against inclusion/exclusion criteria, logged exclusions with reasons and recorded decisions. Two reviewers screened a pilot subset independently, reconciled disagreements, refined notes and applied the refined protocol to the set. The complete screening yielded 90 studies.

We used a standardized extraction form, defined extraction questions (EQs) and recorded items (Appendix~\ref{sec:protocol_tables}, Table~\ref{tab:extraction-questions}). We retained the form with verbatim anchors, page references and lifecycle notes (software development; personal data). We grouped studies by type (reviews/surveys, architectures/frameworks, deployed-system measurements/audits, algorithmic/systems PET proposals) and application area (healthcare, IoT/edge, Web/mobile, AI/ML incl.\ federated learning, blockchain/ledgers, data platforms, secure research environments) for coding. We recorded design, datasets/tools, threat model/assumptions, regulation/standards links and evaluation details; we also recorded search transparency/selection/coding method (qualitative/review) and privacy--utility or robustness metrics with baselines, formal analysis, or measurement protocols (technical/empirical). The second author cross-checked extractions; we reconciled discrepancies, memoed rationales and updated the versioned codebook~\cite{kitchenham2004procedures,kitchenham2015evidence}.

We conducted five synthesis rounds: pilot calibration; four corpus-wide passes (open coding, grouped codes, DT translation, dimension modeling). We calibrated the form and initial codebook; the first two authors coded a pilot subset independently, reconciled differences and versioned the codebook with memos; Cohen's kappa reached \(k = 0{,}835\)~\cite{landis1977measurement}. We coded the dataset with constant comparison and archived all versions of extraction forms, mapping sheets, memos and codebook versions. We created Descriptive Themes (DTs) from codes when at least two studies supported a pattern, assigned identifiers (area prefixes) and logged wording changes without changing identifiers. After DT stabilization, we derived dimensions from the DT structure rather than predefined categories or direct promotion of codes~\cite{cruzes2011recommended}. We built a paper-by-DT incidence matrix, computed DT--DT co-occurrence counts, inspected adjacency groups and applied axial coding; we kept the hierarchy codes/grouped codes, DTs and dimensions~\cite{cruzes2011recommended}. We marked saturation when two successive passes preserved the DT set.

We assigned DTs only when methods, analyses, or findings provided evidence beyond mere mention and we recorded rationales and disconfirming cases. We assigned dimensions via the stabilized DT-to-dimension map~\cite{cruzes2011recommended}: each paper received one primary dimension supported by its DTs and papers received secondary dimensions only when they substantiated DTs mapped elsewhere; mentions without elaboration yielded neither DTs nor secondary dimensions. We documented decision rules and justification notes, ran an inter-rater assessment on a sample, reconciled disagreements, rechecked stability and obtained \(\bar{k} = 0{,}842\); sensitivity check removed borderline cases and confirmed stability. We counted studies per DT/dimension, produced a DT identifier catalog for verbatim use, placed summary tables in results, kept the paper--DT--dimension matrix in an appendix and released replication artifacts (extraction form; versioned codebook with examples/change history; DT catalog with support counts; audit log; mapping matrix) in machine- and human-readable formats; the replication package provides the dataset and synthesis results\footnote{\url{https://github.com/nboro/slr_priv_eng}}.

\section{Reporting - Study Results}\label{sec:reporting}
\subsection{Dataset Meta-Analysis}
\begin{figure}[!t]
  \centering
  \includegraphics[width=0.65\linewidth]{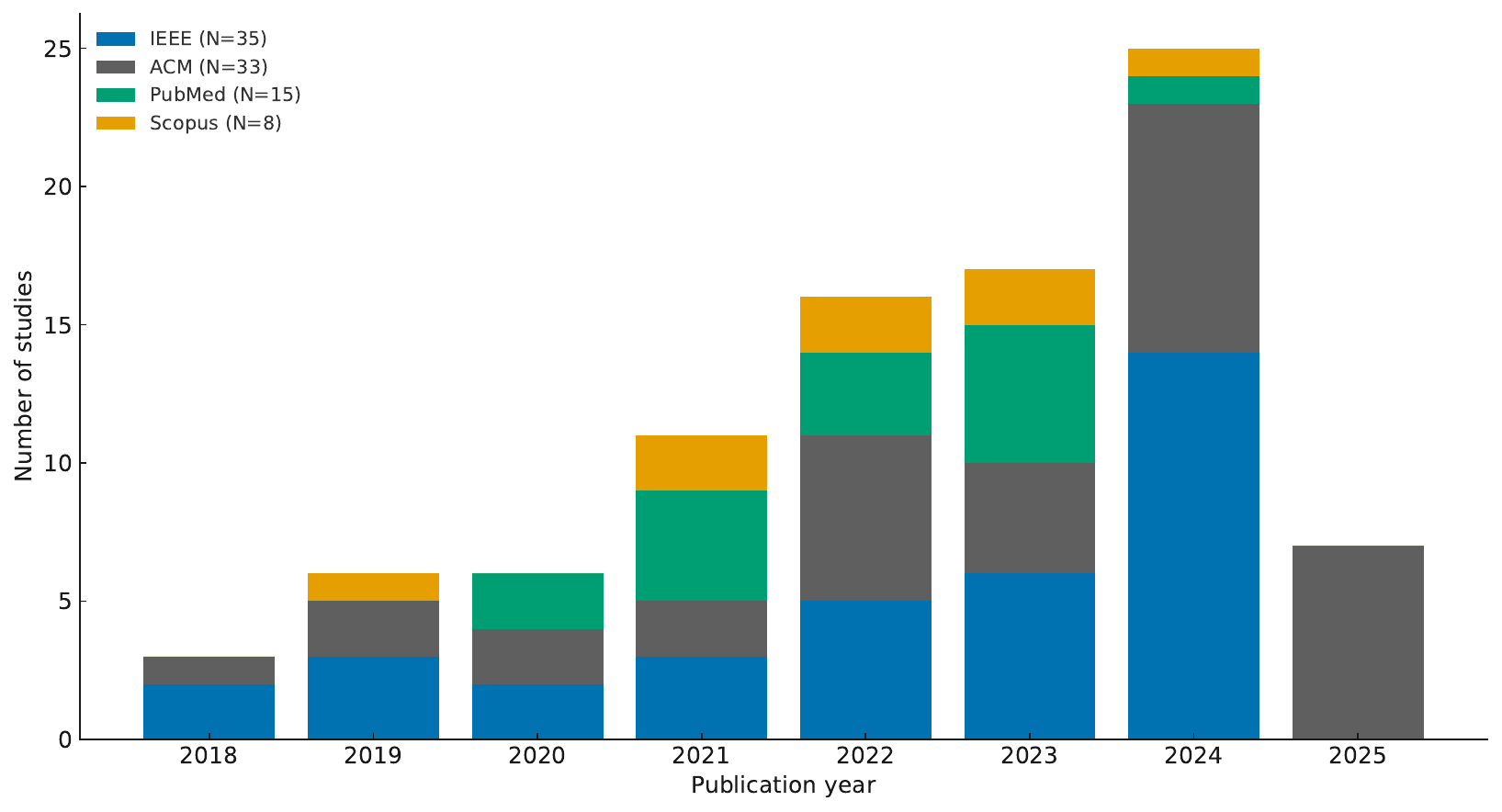}
  \caption{Publication trend of accepted studies for per year and by source.}
  \label{fig:slr_studies_year_source}
\end{figure}

Before proceeding to the deeper analysis of the results, it is worth reporting some meta-information on the prior studies included in this systematic literature review.
Figure~\ref{fig:slr_studies_year_source} reports source-by-year counts for privacy-engineering aspects across the 90 selected works.
The total publication volume increases consistently after 2018 and rises markedly in 2024; the lower 2025 bar reflects publications through March 9, 2025, which was the cutoff for inclusion.
IEEE and ACM provided most entries overall.
It is noticeable that between 2020 and 2023, PubMed contributed at levels similar to and in several years higher than, either IEEE or ACM individually.
Scopus contributed at a lower but steady level.
In the privacy engineering literature, these observations demonstrate rapidly increasing interest in engineering privacy for contemporary software systems and, given the 2024 rise, indicate that additional papers are likely in the near future.
They also indicate sustained attention in software engineering venues with substantial input from health and data-intensive domains.
The clear upward trend in publications, together with the diverse attention across application domains, further motivates our work, namely a systematic literature review that analyzes which aspects warrant consideration when engineering privacy in contemporary and data-intensive software systems and identifies key limitations for future research to address.

\subsection{Results Overview}

\begin{figure*}[!h]
    \centering
    \includegraphics[width=0.70\textwidth]{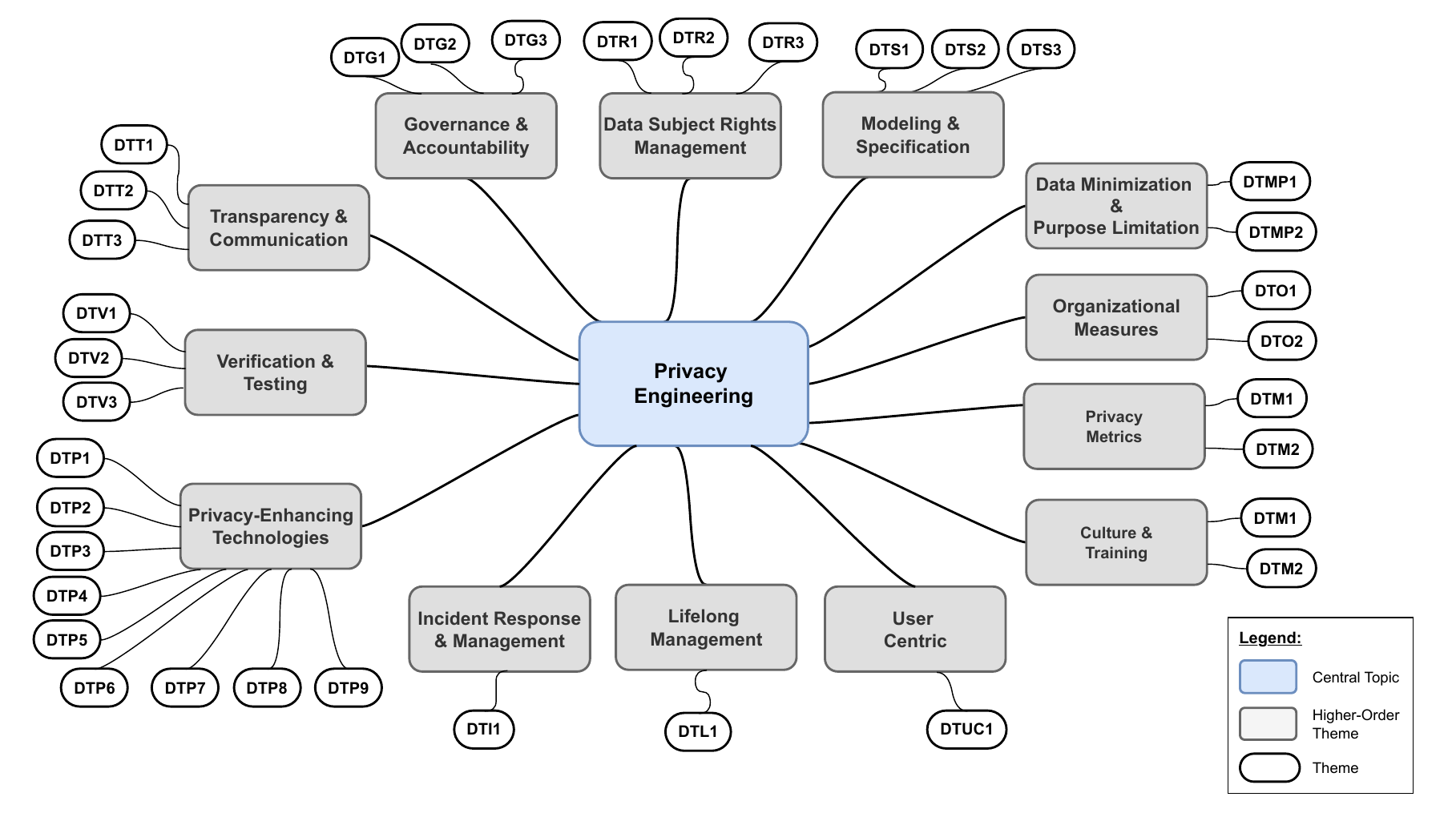}
   \caption{Thematic map instantiation of privacy engineering derived from the selected dataset as introduced by Cruzes \& Dybå (2011)~\cite{cruzes2011recommended}. Descriptive Themes (DTs) denote cross-paper themes that generalize into higher-order themes, reported in our study as the resulting privacy engineering dimensions.} 
    \label{fig:them_synth_res}
\end{figure*}

Our thematic synthesis yielded a catalog of 36 cross-paper DTs, which in turn produced thirteen high-order dimensions for privacy engineering. In Appendix~\ref{sec:studies_to_dimensions}, Table~\ref{tab:dts_dims2} lists the DT codes and titles mapped to their respective dimension and Figure~\ref{fig:them_synth_res} illustrates the resulting thematic map~\cite{cruzes2011recommended}. Privacy-Enhancing Technologies (PETs), expressed in DTP1--DTP9, covers mechanism families: differential privacy, cryptographic computation, Trusted Execution Environments (TEEs), obfuscation, federated and distributed learning and blockchain-backed enforcement. Governance \& Accountability (GA), represented by DTG1--DTG3, covers access control with provenance and audit, roles and standards and incentive or reputation mechanisms in distributed settings. Verification \& Testing (VT), captured by DTV1--DTV3, covers benchmarks and measurement studies, verification and reproducibility artifacts and robustness or certification testing. Transparency \& Communication (TC), via DTT1--DTT3, covers consent and explainability interfaces, machine-readable policies and operational transparency in systems. Modeling \& Specification (MS), expressed in DTS1--DTS3, covers ontologies and requirements models, architectural patterns and service components and privacy-by-design tactics. Data Subject Rights Management (DSRM), via DTR1-DTR3, addresses rights handling, rolling consent and user control modalities and data subject rights tooling.

Minimization \& Purpose Limitation (DMPL), represented by DTM1--DTM2, covers on-device Local Data Processing (LDP) and Federated Learning (FL) and purpose binding via policy-to-rule mapping. Organizational Measures (OM), captured by DTO1--DTO2, covers privacy-by-design backlogs and practices and training-enabled organizational processes. Privacy Metrics (PM), expressed in DPM1--DPM2, covers calibration of privacy–utility trade-offs and robustness or attack-surface metrics. Culture \& Training (CT), represented by DTC1--DTC2, covers privacy awareness programs and organizational privacy culture and climate. Lifelong Management (LM), expressed in DTL1, addresses retention, deletion and other lifecycle operations; Incident Response \& Management (IRM), captured by DTI1, addresses tamper-evidence, forensics and audit trails; User-Centric (UC), represented by DTUC1, addresses user-centered privacy in context.

\begin{figure}[h!]
  \centering
  \includegraphics[width=0.70\linewidth]{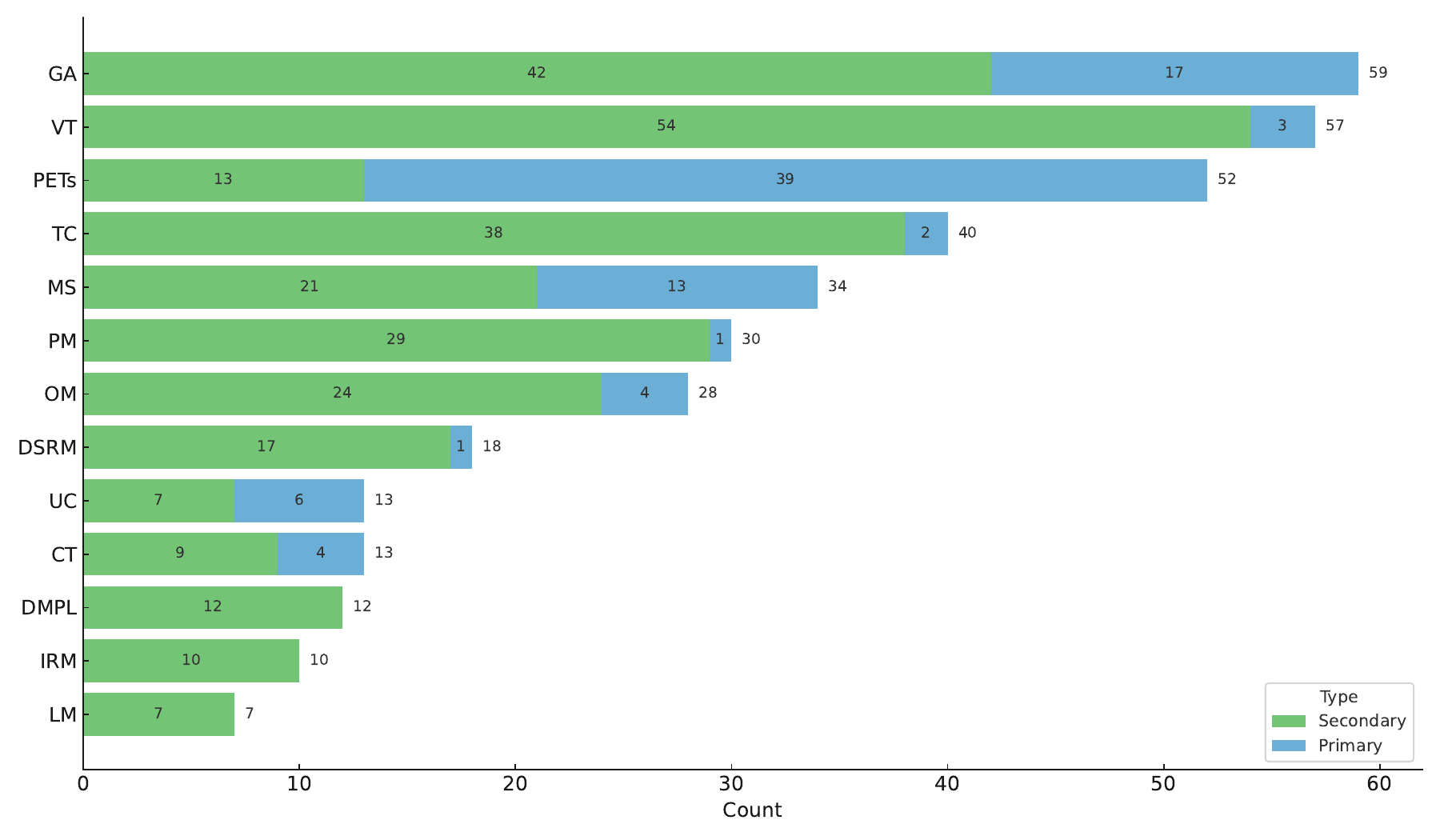}
  \caption{Counts per privacy engineering dimension across all studies, aggregating every occurrence of a dimension appearing as either the Primary or Secondary theme assignment for each study. Label abbreviations: GA = Governance \& Accountability; VT = Verification \& Testing; PETs = Privacy-Enhancing Technologies; TC = Transparency \& Communication; MS = Modeling \& Specification; PM = Privacy Metrics; OM = Organizational Measures; DSRM = Data Subject Rights Management; CT = Culture \& Training; UC = User-Centric; DMPL = Data Minimization \& Purpose Limitation; IRM = Incident Response \& Management; LM = Lifelong Management.}
  \label{fig:dimension-counts}
\end{figure}

Figure~\ref{fig:dimension-counts} indicates that, among the dimensions, PETs dimension is the only one with more – and specifically, significantly more –primary contribution assignments in our data set than any other dimension.
There are three additional dimensions that are approximately balanced in terms of the primary–secondary distribution, namely MS, CT and UC.
For the other dimensions, we observe that primary contributions are significantly fewer than secondary contributions.
For GA, although it has the second-highest count of primary assignments, the distribution remains clearly skewed toward secondary relative to primary.
For VT, TC, PM, OM and DSRM, we observe a significantly larger number of secondary contributions than primary contributions.
Finally, there are three dimensions that contain only secondary assignments, namely DMPL, IRM and LM.
Taken together, these observations have several implications for interpreting the corpus.
First, they indicate that the PETs literature is more mature than the literature associated with the remaining dimensions, since PETs topics constitute the primary focus of a larger share of the selected studies in our dataset.
Second, the pattern indicates rising attention to the other dimensions, as reflected by the absolute counts observable for GA, VT and TC.
However, because primary contributions are relatively few—and in some cases absent—this provides a clear indication of future research directions in the privacy engineering domain.
These findings can also be confirmed by Table~\ref{tab:dts_dims2} in Appendix~\ref{sec:studies_to_dimensions}, since the cross-paper themes, as presented by the respective DT codes for PETs, exhibit greater granularity with broader topics.
This indicates, both in magnitude and in topical breadth, the potential exploration space that the remaining dimensions could realize.
Finally, these observations further motivate our study, namely, a systematic review of the literature that elucidates the dimensions of privacy engineering and identifies specific directions for future studies.
In Appendix~\ref{sec:studies_to_dimensions}, Table~\ref{tab:pe-dim-primary2} presents the mapping of included studies to the dimensions of privacy engineering. For each dimension, it reports the assignment of each study, distinguishing primary from secondary.

\subsection{Findings by Emergent Dimensions}\label{sec:per_dimension}
This section reports results per dimension that emerged from clustering the DTs. For each dimension, we define the scope of each dimension as derived from the corpus, describe the synthesized representative evidence, report concise findings and noted systematic interactions with other dimensions. 

\subsubsection{Privacy-Enhancing Technologies.}
This dimension captured technical mechanisms that enforce privacy requirements through computational means, restricting information disclosure and minimizing data exposure at rest, in transit and in use, rather than relying primarily on organizational policy. Contributing DTs included DTP1 (differential-privacy operationalization)~\cite{zhao_survey_2022,zhao_scenario-based_2024,mazmudar_cache_2022,yu_dop-sql_2024}, DTP2 (cryptographic PETs)~\cite{marcolla_survey_2022,zhang_security_2020}, DTP4 (obfuscation and attribute-inference prevention across visual, biometric and tabular data)~\cite{melzi_overview_2024,zhao_survey_2022}, DTP5 (federated and distributed learning)~\cite{lo_systematic_2022,nguyen_federated_2023,chen_advancements_2025,weng_faster_2024,zhang_ppfed_2024}, DTP6 (blockchain-backed privacy and accountability)~\cite{zhang_security_2020,liang_identity_2024,alghuried_blockchain_2025,zukaib_blockchain_2023} and DTP9 (augmentation via synthetic data or digital twins)~\cite{wang_security_2025,jordan_selecting_2022}.

\begin{figure}[!h]
    \centering
    \includegraphics[width=0.55\linewidth]{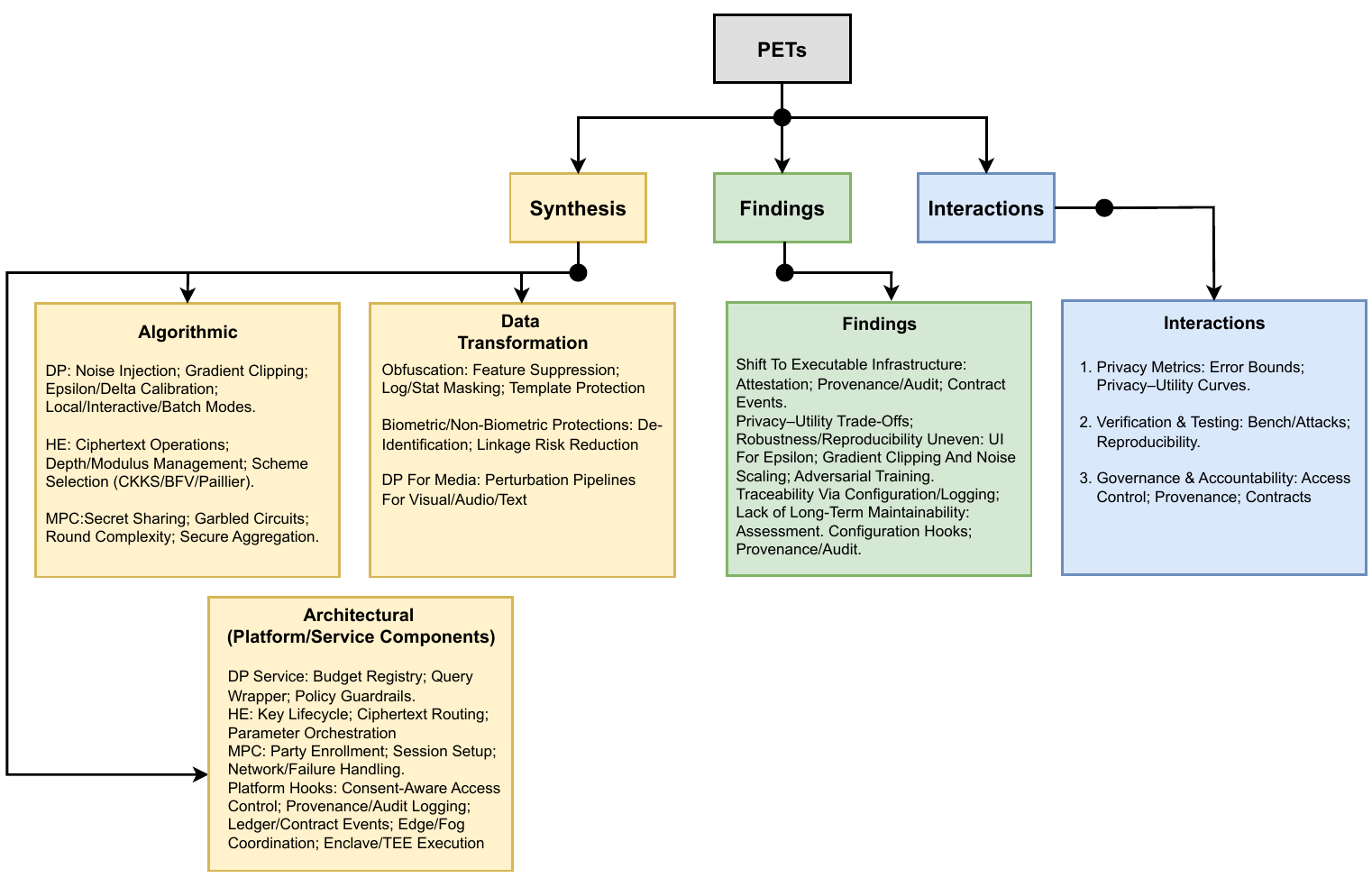}
    \caption{Overview of the insights for the PETs dimension.}
    \label{fig:pets_insights}
\end{figure}

This dimension captured mechanisms that implemented privacy by computation or by architecture rather than by organizational policy. Contributing DTs included
DTP1 (Differential Privacy operationalization (interactive, batch, local; utility-aware calibration))~\cite{mazmudar_cache_2022,yu_dop-sql_2024,wang_pp-csa_2024,zhao_survey_2022,zhao_scenario-based_2024,wang_security_2025,beg_data_2022,pramod_privacy-preserving_2023,jordan_selecting_2022,noauthor_guest_2020,weng_faster_2024,yu_insights_2024,liu_privacy_2021,pal_privacy_2019,wu_privacy-preserved_2022,liu_privacy-preserving_2024,odema_privynas_2024,sun_toward_2022},
DTP2 (Cryptographic PETs (HE/MPC/ABE/ZK/NIZK) and privacy-preserving computation)~\cite{alghuried_blockchain_2025,li_longitudinal_2023,mo_security_2024,zhang_security_2020,nguyen_federated_2023,chen_advancements_2025,melzi_overview_2024,liang_identity_2024,ali_privacy-preserved_2025,zhao_scenario-based_2024,almarshoud_security_2024,beg_data_2022,pramod_privacy-preserving_2023,sun_data_2020,daoudagh_data_2021,mastrolembo_ventura_enhancing_2023,khalid_enhancing_2023,zhang_privacy-by-design_2022,jordan_selecting_2022,benhamida_pyff_2021,saksena_rebooting_2021,hammoudeh_service-oriented_2021,witt_decentral_2023,zukaib_blockchain_2023,noauthor_guest_2020,yu_insights_2024,liu_privacy_2021,liu_privacy-preserving_2024,iwaya_security_2020,marcolla_survey_2022},
DTP3 (Enclave/TEE-based enforcement and verifiable computation)~\cite{alghuried_blockchain_2025,yang_secudb_2024,liang_identity_2024},
DTP4 (Obfuscation \& attribute-inference prevention (visual, biometric, or tabular).)~\cite{du_privategaze_2024,zhao_visual_2025,zhao_survey_2022,melzi_overview_2024,wang_security_2025,wilkowska_interdisciplinary_2023,yang_approaching_2024,zhang_ppfed_2024},
DTP5 (Federated/Distributed learning mechanisms (secure aggregation, personalization, communication efficiency))~\cite{lo_systematic_2022,issa_blockchain-based_2023,nguyen_federated_2023,chen_advancements_2025,ali_privacy-preserved_2025,pramod_privacy-preserving_2023,zhang_privacy-by-design_2022,jordan_selecting_2022,witt_decentral_2023,zukaib_blockchain_2023,solis_exploring_2024,zhang_ppfed_2024,weng_faster_2024,yu_insights_2024,liu_privacy-preserving_2024},
DTP6 (Blockchain-backed privacy and accountability (smart contracts, auditability, incentives))~\cite{issa_blockchain-based_2023,nguyen_federated_2023,chen_advancements_2025,liang_identity_2024,ali_privacy-preserved_2025,wang_security_2025,almarshoud_security_2024,pramod_privacy-preserving_2023,sun_data_2020,khalid_enhancing_2023,zhang_privacy-by-design_2022,salem_comprehensive_2023,semantha_conceptual_2021,witt_decentral_2023,zukaib_blockchain_2023,solis_exploring_2024,yu_insights_2024,wu_privacy-preserved_2022,liu_privacy-preserving_2024,hoel2019privacy},
DTP7 (Threats \& Attacks on ML (adversarial, poisoning, extraction/inversion) \& defenses)~\cite{mo_security_2024,liu_privacy_2021},
DTP8 (Robust aggregation and attack-resilient FL algorithms)~\cite{lo_systematic_2022,chen_advancements_2025,witt_decentral_2023}
and DTP9 (Augmentation PETs (synthetic data \& digital twins))~\cite{chen_advancements_2025,zhao_scenario-based_2024,wang_security_2025,jordan_selecting_2022}. The insights for PETs are summarized in Figure~\ref{fig:pets_insights}.

Across the corpus, studies proposed PETs at three layers. \emph{Algorithmic} PETs implemented DP mechanisms for training and querying and combined them with secure computation and aggregation when applicable~\cite{zhao_scenario-based_2024,mazmudar_cache_2022,yu_dop-sql_2024,marcolla_survey_2022}. \emph{Architectural} PETs embedded privacy into platforms and services through consent-aware access control, provenance on ledgers and edge/fog coordination~\cite{daoudagh_data_2021,wu_privacy-preserved_2022,sun_data_2020,benhamida_pyff_2021,zhang_privacy-by-design_2022,jones_profile_2020}. \emph{Data-transformation} PETs altered or masked inputs, for example with biometric template protection or Differential Privacy (DP) for unstructured media~\cite{melzi_overview_2024,zhao_survey_2022,wang_security_2025}. Several works coupled PETs with configuration interfaces, consent or contract logic and provenance hooks to align system behavior with policy constraints and to trace access~\cite{daoudagh_data_2021,wu_privacy-preserved_2022,lodge_privacy_2019,carmichael_personal_2024,jordan_selecting_2022}. Evaluations most often reported privacy--utility trade-offs and, less consistently, robustness under attack or reproducibility artifacts~\cite{yang_approaching_2024,weng_faster_2024,mazmudar_cache_2022,liu_privacy_2021,mo_security_2024,wagner_technical_2019}.

Evidence showed that adoption concentrated at the data-processing boundary (training/inference pipelines, query systems and analytics platforms) and used differential privacy or cryptography as default enforcement mechanisms~\cite{lo_systematic_2022,nguyen_federated_2023,zhao_scenario-based_2024,marcolla_survey_2022,witt_decentral_2023,yu_dop-sql_2024}. \textit{Finding 1:} Architectural PETs that embedded provenance, audit, or verifiability shifted part of privacy assurance from policy to executable infrastructure~\cite{daoudagh_data_2021,wu_privacy-preserved_2022,zhang_privacy-by-design_2022,carmichael_personal_2024,liang_identity_2024,alghuried_blockchain_2025}. \textit{Finding 2:} Instrumentation with configuration and logging interfaces supported traceability into governance dimensions but rarely assessed long-term maintainability in operational settings~\cite{iwaya_privacy_2024,iwaya_privacy_2023,daoudagh_data_2021,wu_privacy-preserved_2022,elkourdi_exploring_2024}.

PET studies repeatedly co-occurred with \emph{Privacy Metrics} when authors stated error bounds or quantitative guarantees~\cite{wagner_technical_2019,xia_towards_2024} and with \emph{Verification and Testing} when reporting attack-resilience or reproducible artifacts~\cite{liu_privacy_2021,mo_security_2024,jha_internet_2022,li_longitudinal_2023}. They also co-occurred with \emph{Governance and Accountability} when authors tied PET behavior to access control, provenance, or contract logic~\cite{daoudagh_data_2021,sun_data_2020,wu_privacy-preserved_2022,carmichael_personal_2024,rommetveit_privacy_2022,saksena_rebooting_2021}.

\subsubsection{Verification \& Testing.}
\begin{figure}[!h]
    \centering
    \includegraphics[width=0.50\linewidth]{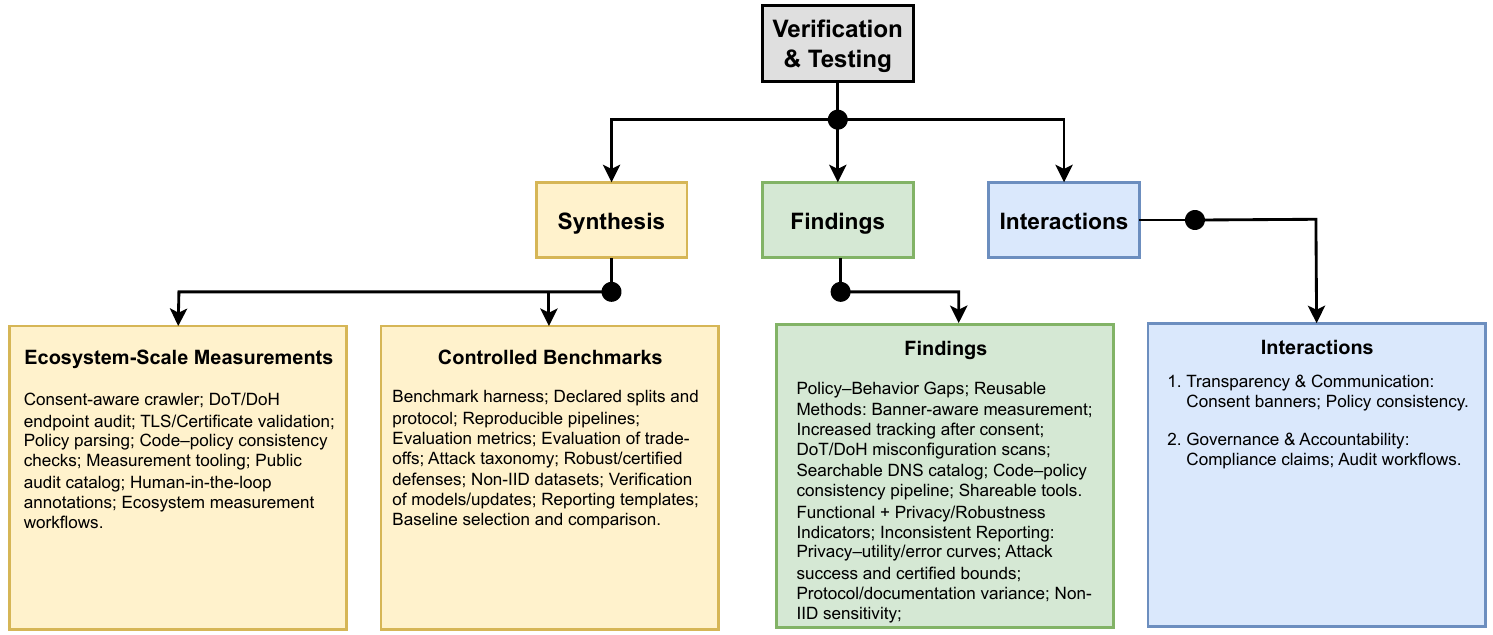}
    \caption{Overview of the insights for the Verification \& Testing dimension.}
    \label{fig:vt_insights}
\end{figure}

This dimension captured methods and evidence artifacts that test privacy properties and conformance claims, including ecosystem-scale measurements of deployed components and controlled benchmarks of proposed mechanisms under explicit baselines and assumptions. The topics of the selected studies include deployments on the web, network stacks, or apps to test whether executed behavior matched policy documents, In addition, it include works that benchmarked algorithms against explicit baselines~\cite{jha_internet_2022,li_longitudinal_2023,wang_as_2024,gebauer_human---loop_2023}. 
Contributing DTs included DTV1 (Benchmarks, measurement studies and ecosystem evaluations)~\cite{sun_data_2020,iwaya_security_2020,zhang_security_2020,daoudagh_data_2021,hammoudeh_service-oriented_2021,benhamida_pyff_2021,liu_privacy_2021,saksena_rebooting_2021,noauthor_guest_2020,beg_data_2022,jordan_selecting_2022,lo_systematic_2022,marcolla_survey_2022,zhang_privacy-by-design_2022,jha_internet_2022,wu_privacy-preserved_2022,witt_decentral_2023,issa_blockchain-based_2023,nguyen_federated_2023,pramod_privacy-preserving_2023,khalid_enhancing_2023,li_longitudinal_2023,zukaib_blockchain_2023,almarshoud_security_2024,liang_identity_2024,liu_privacy-preserving_2024,melzi_overview_2024,mo_security_2024,solis_exploring_2024,weng_faster_2024,yu_insights_2024,zhang_ppfed_2024,odema_privynas_2024,wang_pp-csa_2024,yu_dop-sql_2024,chen_advancements_2025,ali_privacy-preserved_2025,alghuried_blockchain_2025,wang_security_2025,du_privategaze_2024,zhao_visual_2025},
DTV2 (Testbeds, reproducible pipelines and verification harnesses)~\cite{sun_data_2020,issa_blockchain-based_2023,liang_identity_2024,yang_secudb_2024,ali_privacy-preserved_2025,chen_advancements_2025},
and DTV3 (User- and human-in-the-loop evaluations)~\cite{zhao_survey_2022,melzi_overview_2024,du_privategaze_2024,zhao_visual_2025}. Figure~\ref{fig:vt_insights} summarizes the insights for this dimension.

Two strands dominated. First, \emph{ecosystem-scale measurements} assessed the behavior of deployed artifacts: consent banners and post-consent tracking on popular sites, strict-privacy DNS deployments across resolvers, authorities and browsers and mini-app code–policy consistency at scale~\cite{jha_internet_2022,li_longitudinal_2023,wang_as_2024}. These studies engineered purpose-built tooling (a consent-aware crawler; a DNS-over-TLS (DoT)/DNS-over-HTTPS (DoH) search and recommender; a static/data-flow pipeline with prompt-based policy parsing) to enable repeatable tests and public reporting~\cite{jha_internet_2022,li_longitudinal_2023,wang_as_2024}. Related work on human-in-the-loop extraction released annotated artifacts to support independent verification of privacy-policy statements~\cite{gebauer_human---loop_2023}. Second, \emph{controlled benchmarks} evaluated PETs or architectures against comparative baselines, combining functional metrics (accuracy, latency) with privacy or robustness indicators in differentially private analytics, federated/edge learning and privacy-aware model design~\cite{yu_dop-sql_2024,mazmudar_cache_2022,weng_faster_2024,zhang_ppfed_2024,yang_approaching_2024,odema_privynas_2024}. Reviews from software engineering and FL perspectives contextualized evaluation practice and identified gaps in reporting and reproducibility~\cite{lo_systematic_2022,liu_privacy_2021,mo_security_2024}.

\emph{Finding 1:} Ecosystem studies revealed measurable gaps between intended policy and observed behavior and supplied reproducible methods that others could reuse. After accepting consent banners, web tracking increased and pages grew heavier, so ignoring banners biased web measurements~\cite{jha_internet_2022}. For strict DNS privacy, longitudinal scans showed substantial misconfiguration (e.g. \(\sim\!60\%\) invalid DoT and \(\sim\!44\%\) invalid DoH recursive resolver certificates) and centralization risks and delivered a searchable catalog to support ongoing audits~\cite{li_longitudinal_2023}. Mini-app analyses surfaced widespread code–policy inconsistencies at scale and packaged a pipeline that other auditors could replicate~\cite{wang_as_2024}. \emph{Finding 2:} Mechanism evaluations increasingly combined functional metrics with privacy or robustness indicators (DP \(\varepsilon\)/error curves, attack success, certified bounds), but reporting remained inconsistent across domains and stacks, which limited cross-paper comparison~\cite{yu_dop-sql_2024,mazmudar_cache_2022,weng_faster_2024,zhang_ppfed_2024,mo_security_2024}.

Verification co-occurred with \emph{Transparency and Communication} in measurements of user-facing mechanisms (consent banners, privacy policies) and with \emph{Governance and Accountability} when studies tied results to compliance claims, auditing workflows, or organizational practices (e.g., PIAs “in the wild,” health-data PET selection)~\cite{iwaya_privacy_2024,jordan_selecting_2022,jha_internet_2022,li_longitudinal_2023}.

\subsubsection{Transparency \& Communication.}
\begin{figure}[!h]
    \centering
    \includegraphics[width=0.55\linewidth]{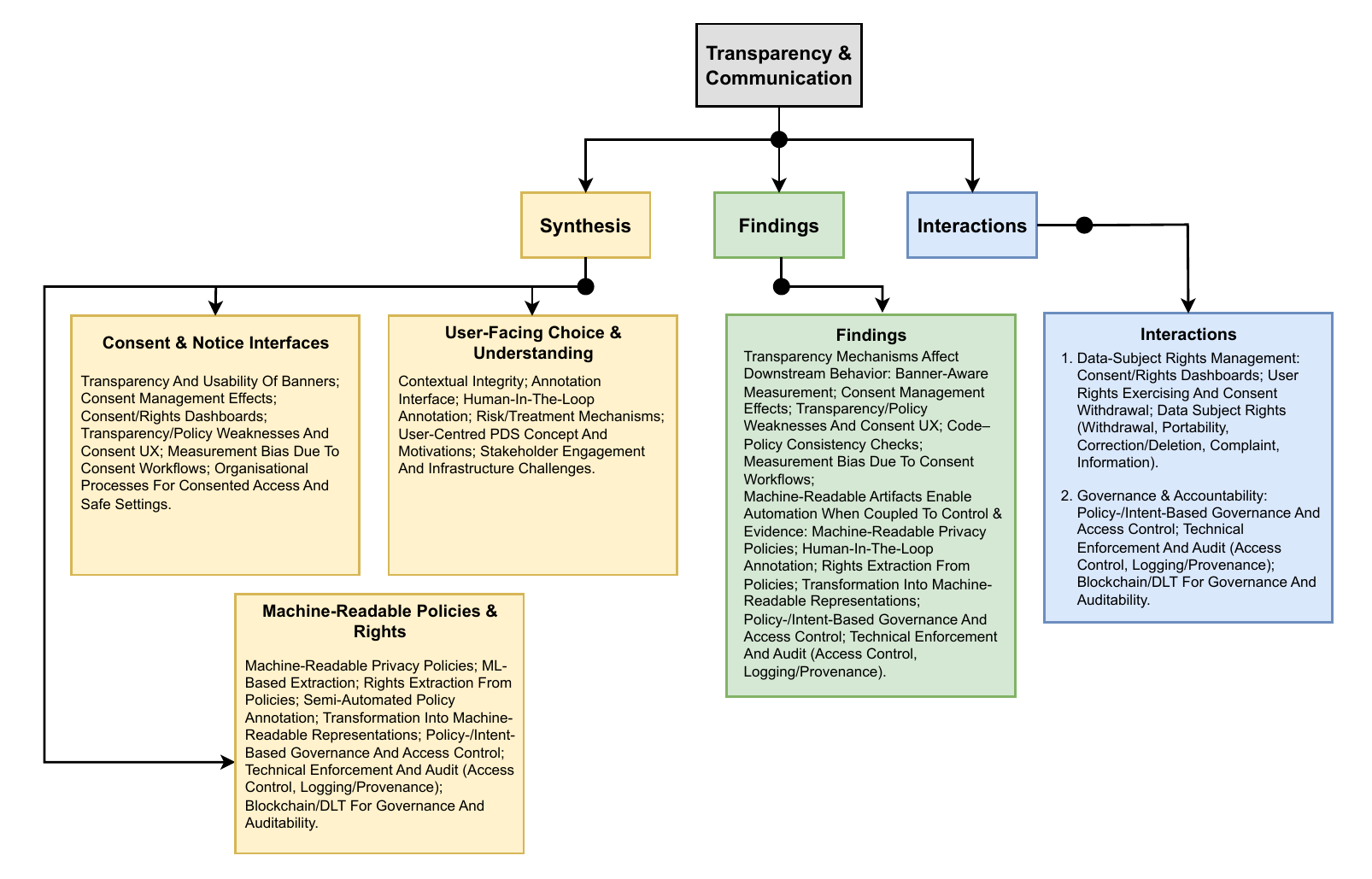}
    \caption{Overview of the insights for the Transparency \& Communication dimension.}
    \label{fig:tc_insights}
\end{figure}

We scoped this dimension to the engineering of transparency artifacts and their operational integration: how systems communicate data practices and control choices to individuals and how notices and policies become machine-readable inputs that drive enforcement, logging and audit. Contributing DTs included DTT1 (consent and transparency communication), DTT2 (machine-readable policies) and DTT3 (user- and human-in-the-loop communication). We covered consent and notice interfaces whose implementation affected downstream data collection, studies that engineered policy artifacts for automation and approaches that linked communication to enforcement or provenance. Contributing DTs included DTT1 (Consent and transparency communication (UX, explainability, banners))~\cite{islam_assurance_2018,barhamgi_user-centric_2018,hoel2019privacy,jones_profile_2020,iwaya_security_2020,alkhariji_synthesising_2021,daoudagh_data_2021,benhamida_pyff_2021,saksena_rebooting_2021,jandl_reasons_2021,jha_internet_2022,andrade_privacy_2022,beg_data_2022,zhang_privacy-by-design_2022,rommetveit_privacy_2022,gebauer_human---loop_2023,de_chaves_privacy_2023,hu_dark_2023,wilkowska_interdisciplinary_2023,carboni_privacy_2023,iwaya_privacy_2023-1,iwaya_privacy_2024,carmichael_personal_2024,wang_as_2024,elkourdi_exploring_2024,pallas2024privacy,cejas2024compai,ali_privacy-preserved_2025,wang_security_2025,de_chaves_user-centred_2025}, DTT2 (Machine-readable privacy policies and information extraction)~\cite{alshammari_privacy_2018,iwaya_security_2020,daoudagh_data_2021,gebauer_human---loop_2023,carmichael_personal_2024,pallas2024privacy,cejas2024compai} and DTT3 (Operational/sensory transparency in systems and robotics)~\cite{daoudagh_data_2021,carboni_privacy_2023,carmichael_personal_2024,de_chaves_user-centred_2025,grabler_privacy_2025}. Figure~\ref{fig:tc_insights} summarizes the insights for this dimension. 

We identified three strands. First, authors analyzed consent banners and related UI flows and measured their operational effect on tracking and configuration after user interaction; purpose-built crawlers and pipelines enabled consistent instrumentation at scale~\cite{jha_internet_2022,wang_as_2024}. Second, several works proposed interfaces that exposed data uses, trade-offs and choices in domains such as assisted living and lifelogging, emphasizing user comprehension and acceptance~\cite{carboni_privacy_2023,wilkowska_interdisciplinary_2023,barhamgi_user-centric_2018}. Third, studies introduced machine-readable representations of policies and rights to support automation (e.g., human-in-the-loop extraction of transparency statements, structured consent) and connected these artifacts to access control, identity/credential schemes, or contract logic so that declared policies influenced runtime behavior~\cite{gebauer_human---loop_2023,daoudagh_data_2021,khalid_enhancing_2023,zhang_privacy-by-design_2022}. In health and research settings, platform designs and personal data stores combined communication artifacts with audit logs and governance workflows to provide traceability for access, linkage and reuse~\cite{jones_profile_2020,carmichael_personal_2024,zhang_privacy-by-design_2022}.

\emph{Finding 1:} User-facing transparency mechanisms affected downstream tracking and configuration in practice; incorrect or incomplete implementations altered privacy outcomes despite nominal compliance. Accepting consent banners changed page composition and increased observable tracking, so ignoring banners biased measurements~\cite{jha_internet_2022}. At the application scale, code–policy inconsistencies were widespread in mini-app ecosystems and empirical analyses of mHealth apps identified policy gaps alongside data-sharing and permission issues~\cite{wang_as_2024,iwaya_privacy_2023-1}. \emph{Finding 2:} Machine-readable artifacts reduced ambiguity and enabled partial automation of rights handling and compliance checks, but they required tight coupling to access-control, credential and logging components to remain effective over time~\cite{gebauer_human---loop_2023,daoudagh_data_2021,khalid_enhancing_2023}. Research platforms and PDS ecosystems illustrated this coupling by binding consent and policy objects to auditable execution paths and provenance records~\cite{jones_profile_2020,carmichael_personal_2024,zhang_privacy-by-design_2022}.

Transparency \& Communication co-occurred with \emph{Data Subject Rights Management} when interfaces and artifacts operationalized access, deletion, or consent withdrawal (e.g., consent managers, dynamic consent, PDS-mediated sharing)~\cite{daoudagh_data_2021,khalid_enhancing_2023,carmichael_personal_2024}. It co-occurred with \emph{Governance and Accountability} when studies linked communication artifacts to audit, provenance and compliance traceability in secure research environments and organizational processes~\cite{iwaya_privacy_2024,zhang_privacy-by-design_2022,jones_profile_2020,jordan_selecting_2022}.

\subsubsection{Governance \& Accountability.}
\begin{figure}[!h]
    \centering
    \includegraphics[width=0.55\linewidth]{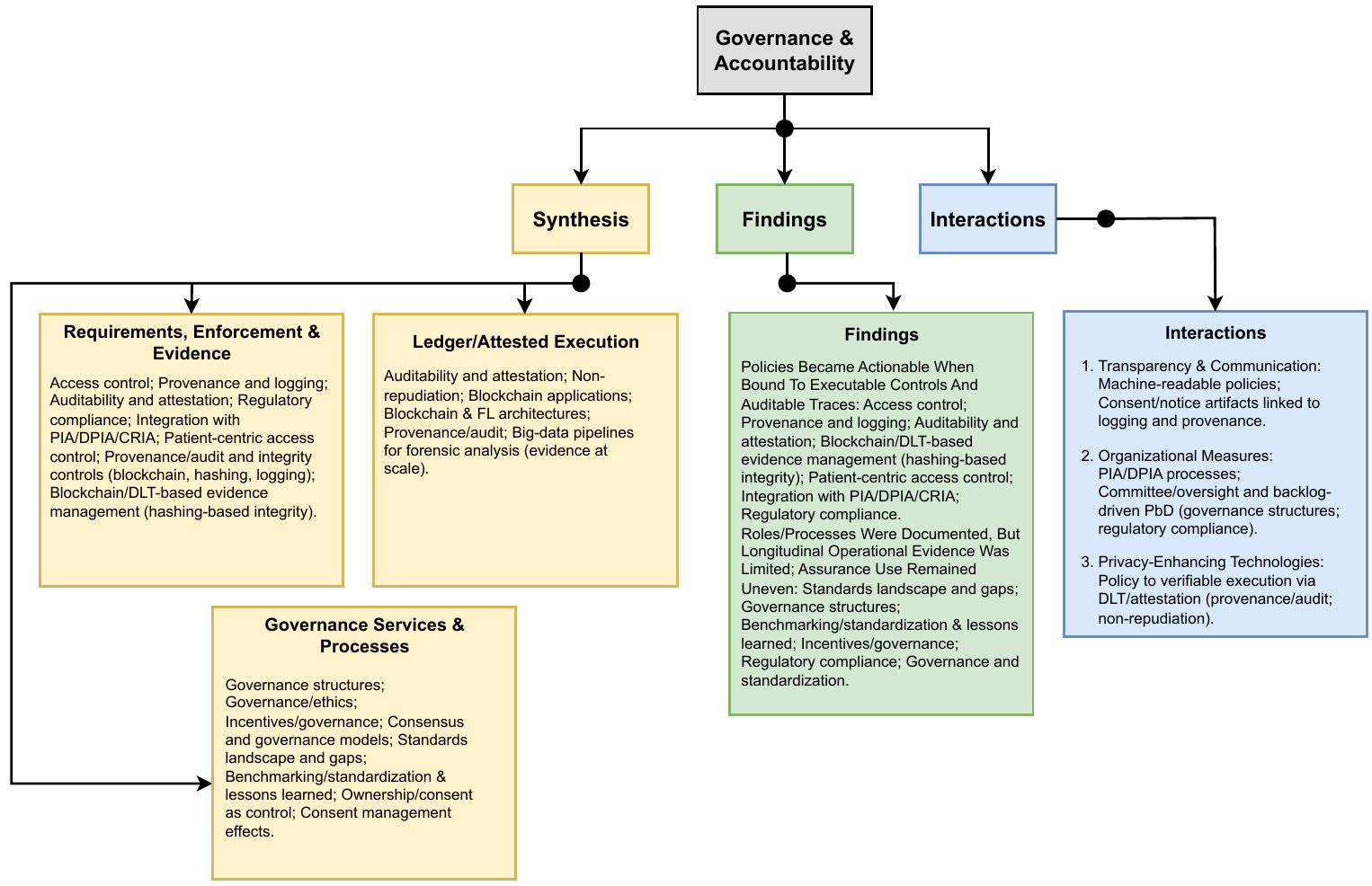}
    \caption{Overview of the insights for the Governance \& Accountability dimension.}
    \label{fig:ga_insights}
\end{figure}

This dimension covered governance controls that translate policies, standards and regulatory duties into accountable system behavior by assigning responsibilities and binding decisions to enforcement points and verifiable evidence, such as access control, provenance and audit logging. We aligned the scope with DTG1 (access control, provenance and auditing), DTG2 (governance patterns, roles and accountability) and DTG3 (standards, certifications and compliance processes). We included mechanisms that enforced policy conformance at runtime or build time, artifacts and roles used to allocate responsibilities and studies that operationalized standards into verifiable checks. Contributing DTs included DTG1 (Access control, provenance, traceability and audit logging)~\cite{lo_systematic_2022,ermakova_security_2020,yang_secudb_2024,sun_data_2020,daoudagh_data_2021,mastrolembo_ventura_enhancing_2023,khalid_enhancing_2023,zhang_privacy-by-design_2022,carmichael_personal_2024,benhamida_pyff_2021,saksena_rebooting_2021,jandl_reasons_2021,salem_comprehensive_2023,semantha_conceptual_2021,islam_assurance_2018,elkourdi_exploring_2024,wu_privacy-preserved_2022,liu_privacy-preserving_2024,iwaya_security_2020,pallas2024privacy}, DTG2 (Incentives, reputation and free-riding mitigation in distributed settings)~\cite{lo_systematic_2022,issa_blockchain-based_2023,nguyen_federated_2023,ali_privacy-preserved_2025,wang_security_2025,witt_decentral_2023,liu_privacy-preserving_2024} and DTG3 (Roles, governance services, standards and accountability structures)~\cite{alghuried_blockchain_2025,jha_internet_2022,mo_security_2024,gharib2020ontology,grabler_privacy_2025,lim_toward_2023,zhang_security_2020,nguyen_federated_2023,chen_advancements_2025,de_chaves_privacy_2023,andrade_privacy_2022,alshammari_privacy_2018,liang_identity_2024,ali_privacy-preserved_2025,wang_security_2025,alhirabi_security_2021,almarshoud_security_2024,senarath_will_2019,iwaya_privacy_2024,sun_data_2020,daoudagh_data_2021,mastrolembo_ventura_enhancing_2023,khalid_enhancing_2023,wilkowska_interdisciplinary_2023,carboni_privacy_2023,zhang_privacy-by-design_2022,carmichael_personal_2024,jordan_selecting_2022,iwaya_privacy_2023-1,benhamida_pyff_2021,saksena_rebooting_2021,jones_profile_2020,jandl_reasons_2021,rommetveit_privacy_2022,salem_comprehensive_2023,semantha_conceptual_2021,mazeli_framework_2022,islam_assurance_2018,zukaib_blockchain_2023,elkourdi_exploring_2024,spiekermann_inside_2019,iwaya_organizational_2022,yu_insights_2024,iwaya_privacy_2023,caiza_reusable_2019,marcolla_survey_2022,herwanto_toward_2024,xia_towards_2024,pallas2024privacy,hoel2019privacy,cejas2024compai}.
In practice, governance artifacts specified roles and responsibilities, lawful bases and purposes, assurance targets and audit requirements and then bound these to technical enforcement points (e.g., access control, logging, provenance) so that compliance became testable and repeatable~\cite{herwanto_leveraging_2024,daoudagh_data_2021,islam_assurance_2018} Organizational and sociotechnical studies framed how privacy engineering operates under regulatory pressure (e.g., GDPR), articulating governance committees, privacy champions and change processes that keep policy and implementation in sync across releases~\cite{spiekermann_inside_2019,rommetveit_privacy_2022,elkourdi_exploring_2024}. Domain platforms (e.g., PDS and TRE/PHT environments) added structural controls—approvals, secure settings, disclosure checks and public reporting—so that research or service access remained accountable to stakeholders and regulators~\cite{carmichael_personal_2024,jones_profile_2020,zhang_privacy-by-design_2022}. Figure~\ref{fig:ga_insights} summarizes the insights for the Governance \& Accountability dimension.

Prior studies mapped legal requirements to access control and provenance mechanisms, defined roles and responsibilities and introduced audit logging and evidence generation~\cite{herwanto_leveraging_2024,daoudagh_data_2021,zhang_privacy-by-design_2022,jones_profile_2020}. User-story and policy-rule approaches translated statutory provisions (e.g., consent withdrawal, purpose limitation) into enforceable decisions and they logged each decision with the basis, requester and scope so that auditors could reconstruct events~\cite{herwanto_leveraging_2024,daoudagh_data_2021}. Several works proposed governance workflows (for example, DPIA (Data Processing Impact Assessment), backlog-driven privacy by design, or committee oversight) and connected these to technical artifacts so that audits could follow recorded decisions (risk, control, log)~\cite{iwaya_privacy_2024,lim_toward_2023,spiekermann_inside_2019}. Some studies embedded verifiability and non-repudiation through ledgers or attested components, using smart contracts for consented access, event trails for disclosures and attestation chains or temporal tables for tamper-evidence~\cite{wu_privacy-preserved_2022,liang_identity_2024,alghuried_blockchain_2025,zhang_security_2020, yang_secudb_2024} Assurance-oriented work linked these enforcement points to explicit objectives and quality attributes (e.g., completeness, auditability coverage, depth of incident reporting) to guide deployment choices and evaluations~\cite{islam_assurance_2018,xia_towards_2024}.

\emph{Finding 1:} Governance claims became actionable when authors bound policies to executable controls (access control, logging, provenance) and produced auditable traces~\cite{daoudagh_data_2021,zhang_privacy-by-design_2022,wu_privacy-preserved_2022,liang_identity_2024}. Smart-city consent and access-control pipelines recorded declared purposes and legal bases alongside decisions; TRE/PHT platforms logged project approvals, researcher identities and output checks; blockchain-based EMR exchange captured consented disclosures as immutable events; and enclave/attested DBMS designs emitted verifiable histories—all of which enabled ex post review and redress~\cite{daoudagh_data_2021,jones_profile_2020,zhang_privacy-by-design_2022,wu_privacy-preserved_2022,yang_secudb_2024}. \emph{Finding 2:} Studies documented roles and processes but reported fewer longitudinal evaluations of accountability in operation: SLRs and scoping reviews called for evidence on durability of governance arrangements over time, operational friction and stability under real workloads, even where frameworks and committees were well defined~\cite{iwaya_privacy_2024,elkourdi_exploring_2024,spiekermann_inside_2019,lim_toward_2023}. Assurance models proposed measurement scaffolds (e.g., coverage and rigor of audits) but empirical use at scale remained uneven, which limited direct comparability across organizations~\cite{islam_assurance_2018,xia_towards_2024}.

Governance and Accountability co-occurred with \emph{Transparency and Communication} via machine-readable policies and interfaces that drove consent, notices and explanations into enforcement and logs; ecosystem measurements and code–policy consistency checks showed how transparency quality influenced governance outcomes~\cite{jha_internet_2022,wang_as_2024,gebauer_human---loop_2023,zhang_privacy-by-design_2022}. Governance and Accountability also co-occurred with \emph{Organizational Measures} through formal processes (e.g., DPIA templates, disclosure control, committee oversight), procurement/vendor controls and training artifacts that embedded accountability into routine engineering practices~\cite{iwaya_privacy_2024,lim_toward_2023,carmichael_personal_2024,jones_profile_2020,spiekermann_inside_2019}. In ledger- and enclave-backed designs, governance intersected with \emph{PETs} by coupling policy to verifiable execution (contracts, attestations, temporal tables) and by treating evidence generation as a first-class engineering goal~\cite{wu_privacy-preserved_2022,liang_identity_2024,alghuried_blockchain_2025,yang_secudb_2024}.

\subsubsection{Data Subject Rights Management.}
\begin{figure}[!h]
    \centering
    \includegraphics[width=0.55\linewidth]{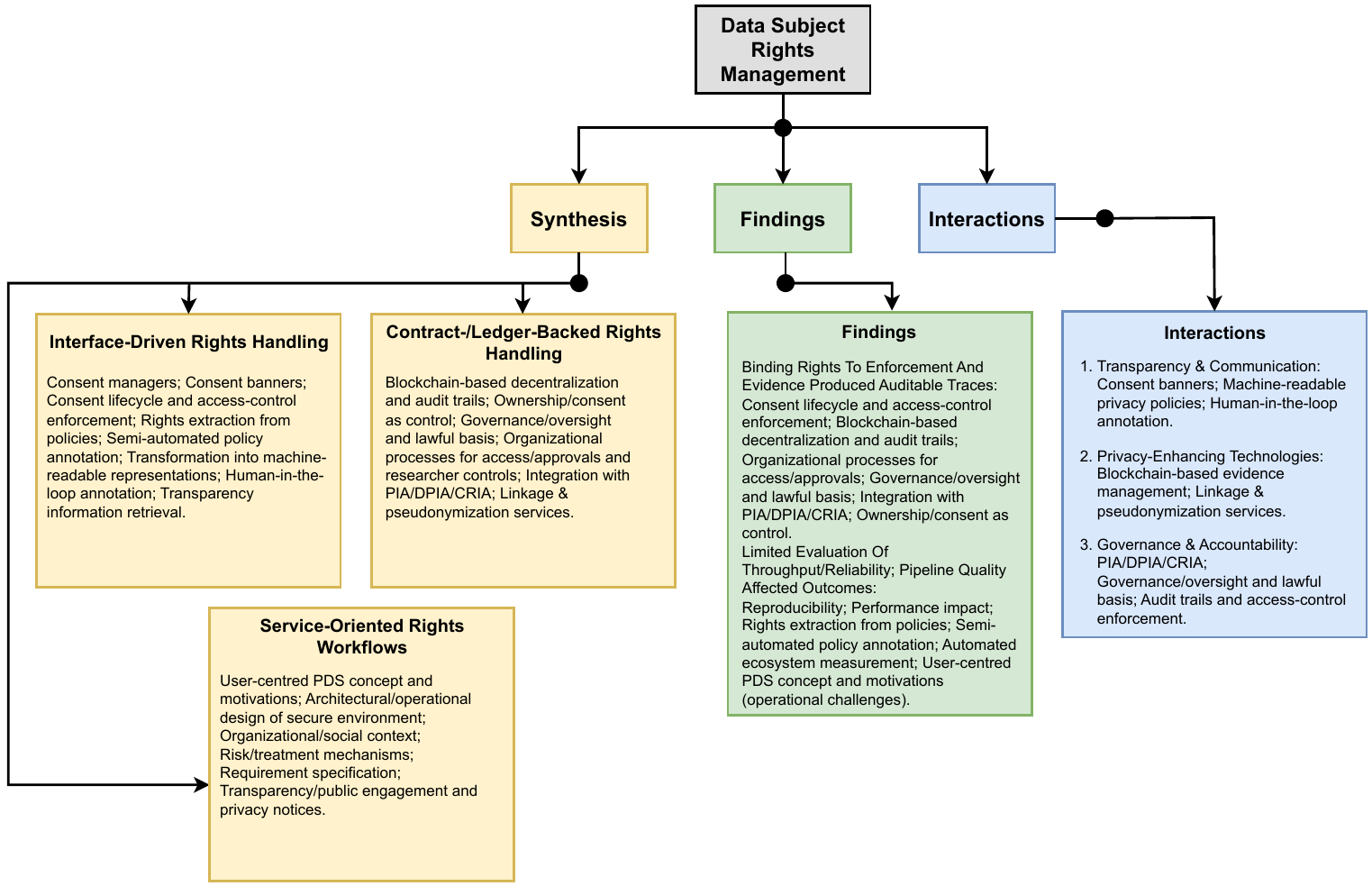}
    \caption{Overview of the insights for the Data Subject Rights Management dimension.}
    \label{fig:dsrm_insights}
\end{figure}

This dimension covered end-to-end rights operations that receive, authenticate, execute and evidence data subject requests, including how systems coordinate fulfillment across services and retain an auditable state for accountability and redress. Contributing DTs included DTR1 (Rights extraction/management (consent, access, deletion, portability))~\cite{barhamgi_user-centric_2018,ermakova_security_2020,gharib2020ontology,jones_profile_2020,alkhariji_synthesising_2021,daoudagh_data_2021,saksena_rebooting_2021,jha_internet_2022,wu_privacy-preserved_2022,zhang_privacy-by-design_2022,gebauer_human---loop_2023,khalid_enhancing_2023,lim_toward_2023,zukaib_blockchain_2023,carmichael_personal_2024,herwanto_leveraging_2024,liu_privacy-preserving_2024,pallas2024privacy,de_chaves_user-centred_2025}, DTR2 (Rolling consent and user control modalities)~\cite{gharib2020ontology,daoudagh_data_2021,jha_internet_2022,iwaya_privacy_2023,de_chaves_user-centred_2025,grabler_privacy_2025} and DTR3 (Data subject rights tooling \& dashboards)~\cite{ermakova_security_2020,jones_profile_2020,daoudagh_data_2021,saksena_rebooting_2021,jha_internet_2022,gebauer_human---loop_2023,khalid_enhancing_2023,carmichael_personal_2024,de_chaves_user-centred_2025}. In practice, studies implemented consent managers and GDPR-aligned access control mechanisms that link stated purposes to enforceable decisions on personal data, thus making the execution of rights concrete in operating systems \cite{daoudagh_data_2021,herwanto_leveraging_2024}. Personal Data Store (PDS) ecosystems and secure research environments positioned individuals as active controllers of sharing and re-use, offering patterns for access, portability and revocation that travel with the data across organizations \cite{carmichael_personal_2024,jones_profile_2020}. Tooling for extracting machine-readable obligations from privacy policies (and for maintaining those annotations) was used to operationalize rights requests at scale and to reduce ambiguity in downstream enforcement \cite{gebauer_human---loop_2023,herwanto_leveraging_2024}. User-centered privacy engineering in the Internet of Things (IoT) and broader Human–Computer Interaction (HCI) reviews stressed that rights mechanisms must be discoverable, actionable and integrated with accounts and consent flows to be effective in practice \cite{barhamgi_user-centric_2018,de_chaves_user-centred_2025}. Figure~\ref{fig:dsrm_insights} summarizes the information for this dimension.

Evidence clustered around two approaches: interface-driven rights handling that integrated with consent and account management and contract- or ledger-based handling that linked requests to verifiable access decisions and logs. Interface-driven lines included consent managers coupled to purpose-specific access control and provenance policies; these systems capture user intent, attach it to processing contexts and expose request/response state to users and administrators \cite{daoudagh_data_2021,jha_internet_2022}. Rights automation frequently relied on structured representations—policy ontologies, machine-extracted clauses and NLP-assisted dataflow models—to map user stories, data categories and processing purposes onto concrete enforcement and evidence artifacts \cite{gharib2021copri,gebauer_human---loop_2023,herwanto_leveraging_2024}. Contract- and ledger-based designs bound rights (e.g., access, erasure, disclosure logging) to smart-contract logic and auditable events, combining attribute-based access, differential privacy and cryptographic proofs to implement verifiable compliance in health and industrial data sharing \cite{wu_privacy-preserved_2022,khalid_enhancing_2023,zukaib_blockchain_2023}. Several architectures embedded rights workflows as callable services—within TRE/PHT-style environments, PDS ecosystems and industrial big-data platforms—so that other components can invoke access/portability/erasure with traceable outcomes and platform-wide audit trails \cite{jones_profile_2020,carmichael_personal_2024,liu_privacy-preserving_2024}. Across both strands, normative guidance on Data Protection by Design and accountability was used to align technical artifacts (logs, proofs, DPIA records) with legal obligations and organizational processes \cite{herwanto_leveraging_2024,iwaya_privacy_2024,lim_toward_2023}.

\emph{Finding 1:} Implementations that bound rights operations to provenance and access-control layers produced auditable traces and clearer accountability. In smart-city deployments, consent managers coupled to GDPR-based access control recorded purposes and legal bases alongside data flows, enabling subsequent audits of who accessed what, under which entitlement and when \cite{daoudagh_data_2021}. Blockchain-backed EMR exchange and EHR sharing frameworks recorded consented disclosures and policy checks as immutable events. When combined with local differential privacy or attribute-level controls, these designs produced both individual-centric configuration and system-level traceability \cite{wu_privacy-preserved_2022,zukaib_blockchain_2023}. Secure research platforms and PDS ecosystems demonstrated that tying rights requests to data-lineage and governance layers (e.g., project approvals, dataset linkages, access committees) supports end-to-end evidencing of fulfillment and reduces ambiguity in redress \cite{jones_profile_2020,carmichael_personal_2024}. Platform proposals for industrial big data similarly layered function, security and system architectures so that rights actions trigger verifiable logging and policy enforcement across services \cite{liu_privacy-preserving_2024}. Where database and service designs surfaced fine-grained provenance with rights hooks, reviewers could reconcile subject requests with concrete processing episodes and generate auditable reports \cite{yang_secudb_2024,ermakova_security_2020}. \emph{Finding 2:} Studies rarely evaluated the end-to-end latency and reliability of rights workflows at scale, which limited claims about operational readiness. HCI and user-centred syntheses reported uneven user involvement and limited empirical validation of the usability and throughput of rights in live settings \cite{de_chaves_user-centred_2025,barhamgi_user-centric_2018}. Scoping and SLR works in health and industrial domains demonstrated open issues around scalability, governance handoffs and cross-organizational fulfillment, even when cryptographic or ledger primitives were in place \cite{zukaib_blockchain_2023,liu_privacy-preserving_2024}. Commentary and ecosystem studies highlighted that consent and notice implementations materially affect downstream behavior and that automation pipelines for rights depend on robust, consistent policy extraction and integration—areas where field measurements still expose gaps \cite{jha_internet_2022,gebauer_human---loop_2023}. Privacy-by-design guidance and PIA reviews urged stronger empirical evidence on timeliness, failure modes and exception handling to move from compliant designs on paper to reliable rights operations in production \cite{herwanto_leveraging_2024,iwaya_privacy_2024,lim_toward_2023}.

Data Subject Rights Management management co-occurred with \emph{Transparency and Communication} when interfaces operationalized access, deletion, or consent withdrawal and when policy artifacts became machine-readable inputs for enforcement pipelines; measurements and human-in-the-loop extraction directly linked transparency quality to the correctness of subsequent rights execution \cite{jha_internet_2022,gebauer_human---loop_2023,de_chaves_user-centred_2025}. It co-occurred with \emph{Governance and Accountability} when DPIAs, policy ontologies and accountability metrics anchored technical workflows to organizational controls and regulatory duties \cite{herwanto_leveraging_2024,iwaya_privacy_2024,lim_toward_2023,gharib2021copri}. In ledger-backed systems, it also co-occurred with \emph{PETs} through contract enforcement and verifiable logging, where cryptographic budgets, proofs and event trails jointly supported subject rights and system attestations across health and industrial ecosystems \cite{wu_privacy-preserved_2022,khalid_enhancing_2023,zukaib_blockchain_2023,liu_privacy-preserving_2024}.

\subsubsection{Modeling \& Specification.}
\begin{figure}[!h]
    \centering
    \includegraphics[width=0.55\linewidth]{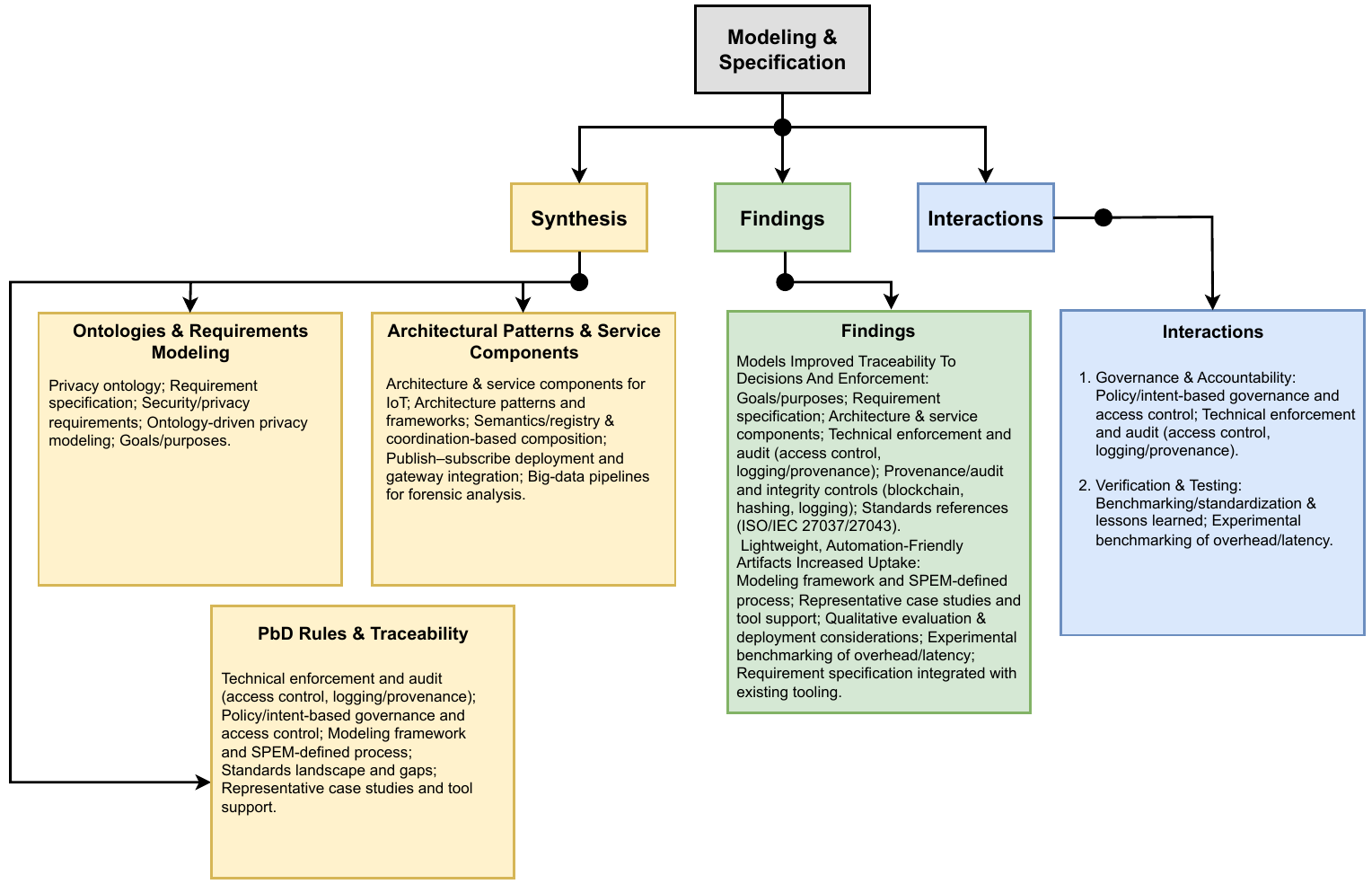}
    \caption{Overview of the insights for the Modeling \& Specification dimension.}
    \label{fig:ms_insights}
\end{figure}

This dimension captured specification artifacts that make privacy requirements, threats and design decisions explicit and traceable across development and operation and that serve as inputs to enforcement, verification and governance artifacts. Contributing DTs included DTS1 (Ontologies and requirements modeling for privacy)~\cite{alshammari_privacy_2018,islam_assurance_2018,caiza_reusable_2019,hoel2019privacy,gharib2020ontology,sun_data_2020,alkhariji_synthesising_2021,alhirabi_security_2021,daoudagh_data_2021,benhamida_pyff_2021,semantha_conceptual_2021,andrade_privacy_2022,de_chaves_privacy_2023,iwaya_privacy_2023-1,salem_comprehensive_2023,elkourdi_exploring_2024,herwanto_leveraging_2024,herwanto_toward_2024,cejas2024compai,de_chaves_user-centred_2025}, DTS2 (Architectural patterns and service components (Service-Oriented Architecture(SOA), Data Governance Service (DGS)/Personal Data Accounts (PDAs))~\cite{alshammari_privacy_2018,barhamgi_user-centric_2018,caiza_reusable_2019,hoel2019privacy,sun_data_2020,jones_profile_2020,chen_holistic_2021,daoudagh_data_2021,benhamida_pyff_2021,semantha_conceptual_2021,hammoudeh_service-oriented_2021,andrade_privacy_2022,wu_privacy-preserved_2022,de_chaves_privacy_2023,mastrolembo_ventura_enhancing_2023,wilkowska_interdisciplinary_2023,salem_comprehensive_2023,elkourdi_exploring_2024,solis_exploring_2024,liu_privacy-preserving_2024,herwanto_toward_2024,pallas2024privacy} and DTS3 (Privacy-by-Design architectural tactics methodology)~\cite{alshammari_privacy_2018,zhang_security_2020,alhirabi_security_2021,jha_internet_2022,herwanto_leveraging_2024,herwanto_toward_2024,pallas2024privacy}. In practice, these artifacts spanned formal ontologies for privacy concepts, structured requirement models and threat catalogs that informed both design choices and validation activities~\cite{gharib2021copri,alhirabi_security_2021,de_chaves_privacy_2023}. Studies also grounded privacy by design as a \emph{traceable} engineering concern by mapping modeled requirements to specific architectural views and decision points, enabling review and change management across the lifecycle~\cite{alshammari_privacy_2018,alkhariji_synthesising_2021,caiza_reusable_2019}.
Figure~\ref{fig:ms_insights} summarizes the insights for this dimension.

Papers reported privacy ontologies, goal and asset models and data-flow or misuse diagrams to structure requirements and threats~\cite{gharib2021copri,alhirabi_security_2021,alkhariji_synthesising_2021,iwaya_privacy_2023-1}. Across these accounts, ontologies captured domain concepts and relations (e.g., data subjects, purposes, processing operations), while goal asset models and misuse Data Flow Diagram (DFD) diagrams linked threats to assets and flows in a way that could be checked against implementation choices~\cite{gharib2021copri,alhirabi_security_2021,iwaya_privacy_2023-1}. Surveys and pattern catalogs consolidated recurring modeling choices, clarifying when to prefer tactic-level patterns (e.g., minimization, separation) versus higher-level strategies to satisfy sets of privacy goals~\cite{alshammari_privacy_2018,alkhariji_synthesising_2021,de_chaves_privacy_2023}.  
Prior studies used architectural patterns and tactics to encode recurring solutions and linked these to concrete components and interfaces~\cite{alshammari_privacy_2018,caiza_reusable_2019,alkhariji_synthesising_2021,de_chaves_privacy_2023}. Pattern repositories and mapping studies showed how “privacy tactics” (e.g., anonymize, aggregate, log) could be bound to components such as gateways, data stores and API facades, with trace links that supported reviews and change impact analysis~\cite{alshammari_privacy_2018,caiza_reusable_2019,de_chaves_privacy_2023}. These works emphasized decision records and forces (e.g., utility, cost, compliance) so that pattern selection remained auditable and reproducible across projects~\cite{alkhariji_synthesising_2021,andrade_privacy_2022,de_chaves_privacy_2023}.  
Several works expressed legal provisions as requirements (for example, user stories or policy rules) and connected them to enforcement mechanisms such as access control or logging~\cite{herwanto_leveraging_2024,herwanto_toward_2024,daoudagh_data_2021,wu_privacy-preserved_2022}. Natural Language Processing (NLP) and template-driven approaches extracted GDPR-like constraints from user stories or policy texts and produced actionable rules/acceptance criteria, which could then be tied to authorization hooks, consent checks, or provenance logging~\cite{herwanto_leveraging_2024,herwanto_toward_2024,gebauer_human---loop_2023}. Domain architectures (e.g., smart cities, e-health) illustrated how policy-bound rules propagate into access-control models and ledgered/audited workflows for later accountability~\cite{daoudagh_data_2021,wu_privacy-preserved_2022,zhang_privacy-by-design_2022}.  
Tool-supported approaches generated checklists, traced requirements to code or configuration, or produced compliance evidence~\cite{iwaya_privacy_2024,mazeli_framework_2022,andrade_privacy_2022,de_chaves_privacy_2023}. DPIA-centric toolchains yielded structured assessment templates, artifact checklists and trace matrices linking risks/controls to modules and configurations; several studies reported these as reusable assets to support audits~\cite{iwaya_privacy_2024,de_chaves_privacy_2023,andrade_privacy_2022}. Frameworks for developers emphasized lightweight guidance and repository integration (e.g., issue trackers, CI steps) so that privacy models remained synchronized with code and documentation~\cite{mazeli_framework_2022,herwanto_leveraging_2024,herwanto_toward_2024}.

\emph{Finding 1:} Modeling improved traceability when authors connected requirements and threats to concrete architectural decisions and enforcement points~\cite{alshammari_privacy_2018,de_chaves_privacy_2023,caiza_reusable_2019}. Case studies and domain frameworks demonstrated end-to-end traces from goals and misuse cases to components that implemented access control, minimization, or provenance, enabling testable justifications of why a given design satisfied a stated privacy objective~\cite{alshammari_privacy_2018,caiza_reusable_2019,zhang_privacy-by-design_2022}. Where enforcement relied on smart contracts or policy engines, models acted as the source of truth for rule deployment and audit log configuration, strengthening post hoc accountability~\cite{wu_privacy-preserved_2022,daoudagh_data_2021,zhang_privacy-by-design_2022}.  
\emph{Finding 2:} Methods that integrated with agile workflows and code repositories increased uptake but required lightweight artifacts and automation~\cite{herwanto_leveraging_2024,herwanto_toward_2024,andrade_privacy_2022}. Studies reported higher practitioner engagement when modeling outputs were small, updateable (e.g., checklists, story-level criteria) and surfaced in existing tooling (IDEs, issue boards, CI), whereas heavyweight models without automation saw limited adoption~\cite{mazeli_framework_2022,de_chaves_privacy_2023,andrade_privacy_2022}. Pipelines that generated tests, linters, or configuration stubs directly from models further reduced friction and improved consistency across releases~\cite{herwanto_leveraging_2024,mazeli_framework_2022,de_chaves_privacy_2023}.

Modeling co-occurred with Governance and Accountability through compliance tracing and with Verification and Testing when authors used models to drive checks or tests~\cite{iwaya_privacy_2024,de_chaves_privacy_2023,andrade_privacy_2022}. PIA-driven models and policy rule sets yielded evidence packages (requirements–control–log traces) for audits, while metric- and threat-driven models fed benchmark design and regression checks that validated privacy behavior over time~\cite{iwaya_privacy_2024,wagner_technical_2019,xia_towards_2024}. Domain exemplars tied model elements (purposes, consents, lawful bases) to access decisions and verifiable logs, closing the loop between specification, enforcement and independent verification~\cite{daoudagh_data_2021,wu_privacy-preserved_2022,zhang_privacy-by-design_2022}.

\subsubsection{Data Minimization \& Purpose Limitation.}
\begin{figure}[!h]
    \centering
    \includegraphics[width=0.55\linewidth]{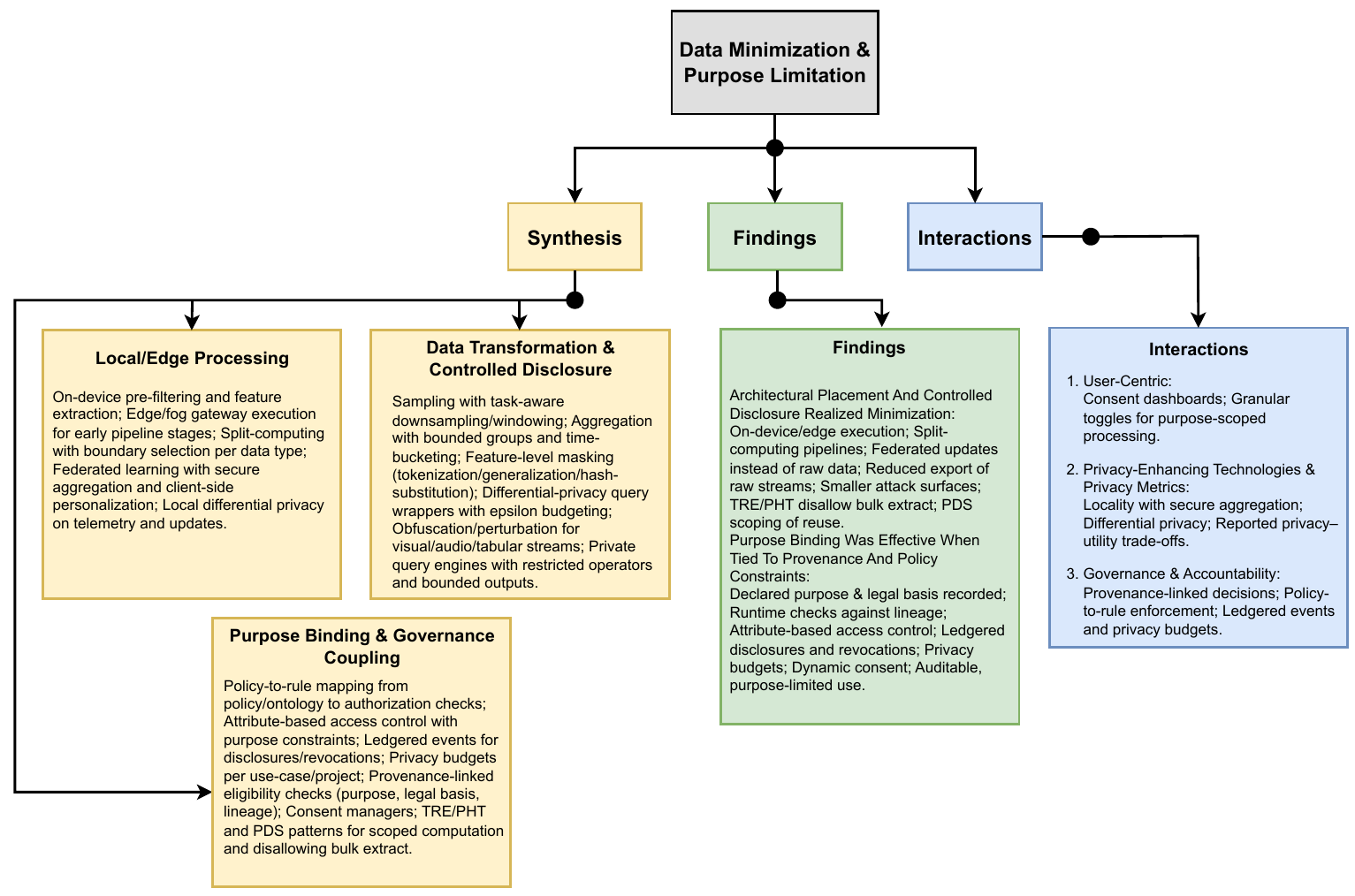}
    \caption{Overview of the insights for the Data Minimization \& Purpose Limitation dimension.}
    \label{fig:dmpl_insights}
\end{figure}

This dimension captured engineering choices that restrict data collection and processing to what supports declared purposes, by placing computation to reduce data movement, constraining secondary use through purpose-bound rules and linking these constraints to enforcement and evidence. Contributing DTs included DTM1 (Data minimization via on-device processing, LDP, FL)~\cite{hoel2019privacy,jones_profile_2020,alkhariji_synthesising_2021,daoudagh_data_2021,saksena_rebooting_2021,jandl_reasons_2021,iwaya_privacy_2023-1,nguyen_federated_2023,pallas2024privacy,ali_privacy-preserved_2025,chen_advancements_2025} and DTM2 (Purpose binding and policy-to-rule mapping)~\cite{hoel2019privacy,gharib2020ontology,alkhariji_synthesising_2021,andrade_privacy_2022,de_chaves_privacy_2023,lim_toward_2023}.
In privacy-preserving health research and federated analysis, secure research environments (e.g., TREs/PHTs) and personal data stores exemplify minimization by “bringing compute to the data,” constraining extraction and secondary use to declared projects and purposes \cite{jones_profile_2020,zhang_privacy-by-design_2022,carmichael_personal_2024}. User-centred privacy engineering for IoT similarly frames minimization as a first-class design goal—collect only what is necessary, as locally as possible and expose purpose to the user at decision time \cite{barhamgi_user-centric_2018,de_chaves_user-centred_2025}. Empirical audits of mHealth and lifelogging technologies further underscore the need for strict minimization and purpose limitation by documenting unnecessary permissions, excessive third-party sharing and ambiguous notices in deployed ecosystems \cite{iwaya_privacy_2023-1,wilkowska_interdisciplinary_2023,carboni_privacy_2023}.
 Figure~\ref{fig:dmpl_insights} summarizes the insights for this dimension.

Studies implemented minimization by processing data on device or at the edge and by limiting data movement to central services. Federated and split/edge computing approaches embody this pattern: FL avoids raw-data centralization, while fog/edge and split-computing pipelines limit exposure by keeping early or sensitive stages local \cite{lo_systematic_2022,nguyen_federated_2023,chen_advancements_2025,issa_blockchain-based_2023,solis_exploring_2024,benhamida_pyff_2021,odema_privynas_2024,hammoudeh_service-oriented_2021}. Several works degraded or summarized data before release (for example, sampling, aggregation, or feature-level masking) and bound access to declared purposes through configuration or contract logic. Differential-privacy systems, information-theoretic obfuscation and private query engines provided controllable disclosure while reporting explicit accuracy/privacy budgets \cite{zhao_scenario-based_2024,zhao_survey_2022,yu_dop-sql_2024,mazmudar_cache_2022,yang_approaching_2024,sun_toward_2022,weng_faster_2024,pal_privacy_2019}. Architectures often combined minimization with rights workflows (for example, erasure and revocation) and with provenance so that components could verify intended purposes at decision time. Smart-city consent managers, blockchain-backed EMR exchange and dynamic consent frameworks demonstrate purpose binding via attribute-based access control, ledged events and privacy budgets attached to specific use-cases \cite{daoudagh_data_2021,wu_privacy-preserved_2022,khalid_enhancing_2023}. Evaluations reported task-performance impacts from reduced data and described configuration ranges that preserved acceptable utility. Empirical and survey work in FL and split/edge designs repeatedly quantified accuracy/latency impacts under data locality and noise, outlining regimes where utility remains competitive despite minimized data movement \cite{nguyen_federated_2023,chen_advancements_2025,odema_privynas_2024,weng_faster_2024,yu_dop-sql_2024}.

\emph{Finding 1:} Authors realized minimization primarily through architectural placement (local processing/edge) and controlled disclosure rather than post-hoc sanitization. Fog/edge architectures (e.g., PyFF, SOA) and split-computing pipelines relocate sensitive processing to constrained devices or private gateways, reducing the need to export raw streams and shrinking attack surfaces \cite{benhamida_pyff_2021,hammoudeh_service-oriented_2021,odema_privynas_2024}. Federated and decentralized learning likewise keep data in situ and share updates under aggregation or perturbation, substituting design-time locality and protocol-level controls for later anonymization \cite{lo_systematic_2022,nguyen_federated_2023,chen_advancements_2025,issa_blockchain-based_2023}. Secure research environments (TREs/PHTs) and personal data stores operationalize minimization by disallowing bulk extract and by constraining computations to approved, project-scoped purposes \cite{jones_profile_2020,zhang_privacy-by-design_2022,carmichael_personal_2024}. \emph{Finding 2:} Purpose binding became effective when authors linked access decisions to provenance records and policy constraints instead of treating purpose as a static label. GDPR-aligned consent and access-control components that co-record legal basis, declared purpose and data lineage enable runtime checks and post-hoc audits of whether a request matches an authorized purpose \cite{daoudagh_data_2021,wu_privacy-preserved_2022}. Dynamic-consent and contract-based designs further tied disclosures and revocations to verifiable events (e.g., smart-contract logs, privacy budgets), improving traceability of purpose-limited use \cite{khalid_enhancing_2023,wu_privacy-preserved_2022}. Conversely, ecosystem measurements and code-policy consistency analyses show that when purpose is handled only textually (e.g., banners/policies) and not bound to enforcement/provenance, downstream behavior diverges from stated intent \cite{jha_internet_2022,wang_as_2024,gebauer_human---loop_2023}. TRE/PHT governance records and project scoping illustrate how embedding provenance at the platform layer supports verifiable purpose restriction over time \cite{jones_profile_2020,zhang_privacy-by-design_2022}.

Data Minimization \& Purpose Limitation co-occurred with \emph{User-Centric} when interfaces exposed data-sharing choices and made purpose-scoped processing intelligible to end-users (e.g., consent dashboards, granular toggles), a theme emphasized in IoT/HCI syntheses and assisted-living deployments \cite{barhamgi_user-centric_2018,de_chaves_user-centred_2025,wilkowska_interdisciplinary_2023,carboni_privacy_2023}. It co-occurred with \emph{Governance and Accountability} when systems enforced purpose through access control and logging at the platform layer—smart-city consent managers, TRE/PHT controls, PDS governance and industrial data platforms all couple minimization with auditable records and policy checks \cite{daoudagh_data_2021,jones_profile_2020,zhang_privacy-by-design_2022,carmichael_personal_2024,liu_privacy-preserving_2024,sun_data_2020}. In federated/edge pipelines, minimization also intersected with \emph{PETs} and \emph{Privacy Metrics}, where locality and controlled disclosure were quantified via privacy-utility trade-offs and protocol guarantees \cite{lo_systematic_2022,nguyen_federated_2023,weng_faster_2024,yu_dop-sql_2024,mazmudar_cache_2022}.

\subsubsection{Organizational Measures.}
\begin{figure}[!h]
    \centering
    \includegraphics[width=0.55\linewidth]{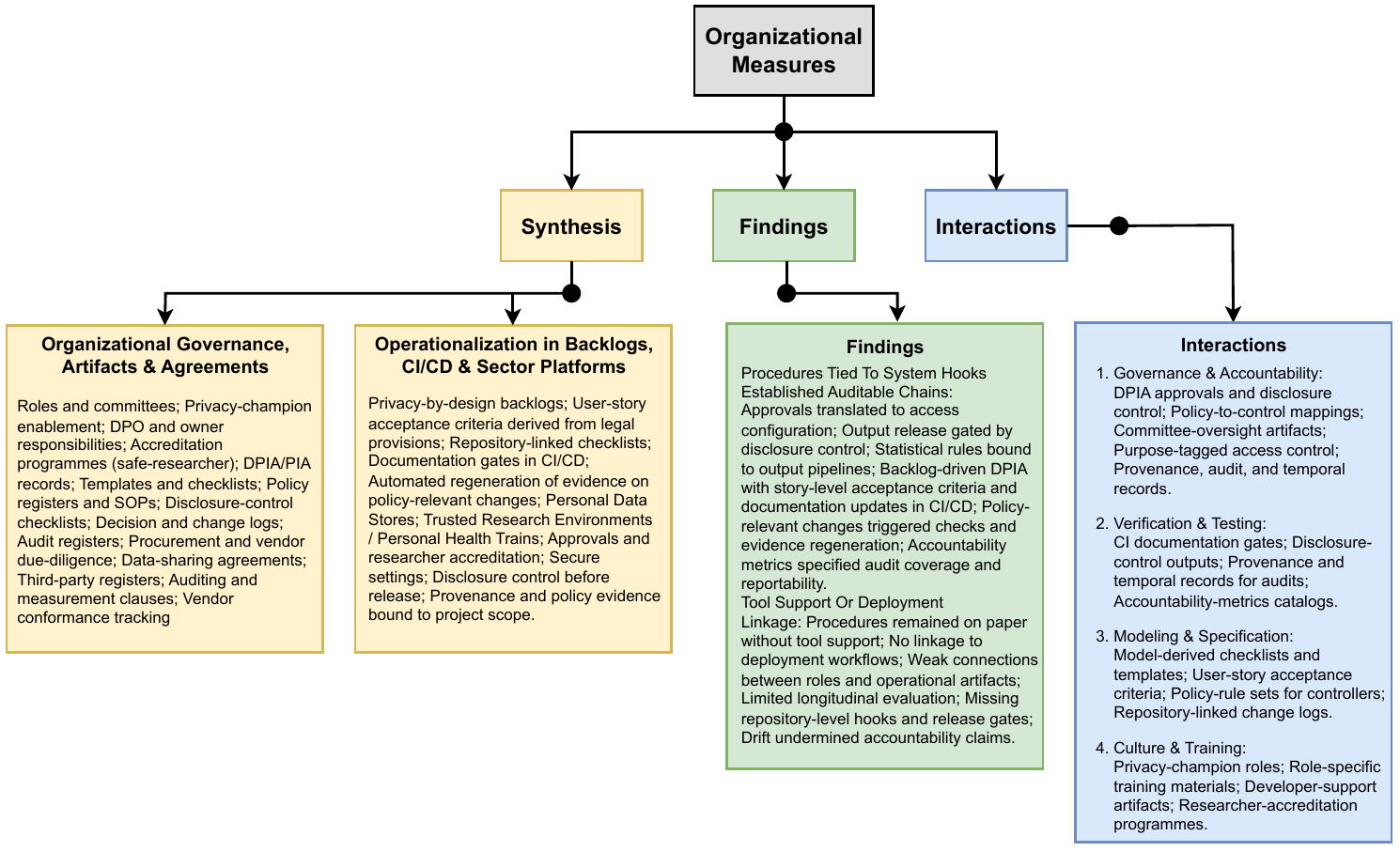}
    \caption{Overview of the insights for the Organizational Measures dimension.}
    \label{fig:om_insights}
\end{figure}

This dimension captured organizational practices that operationalize privacy-by-design within engineering workflows and operations, including assessments, review gates, vendor controls and backlog integration that produce repeatable evidence. Contributing DTs included DTO1 (Organizational processes, privacy-by-design backlogs and practices)~\cite{spiekermann_inside_2019,hoel2019privacy,islam_assurance_2018,jones_profile_2020,alhirabi_security_2021,saksena_rebooting_2021,jandl_reasons_2021,rommetveit_privacy_2022,mazeli_framework_2022,zhang_privacy-by-design_2022,andrade_privacy_2022,iwaya_organizational_2022,ermakova_security_2020,iwaya_privacy_2023-1,carboni_privacy_2023,de_chaves_privacy_2023,mastrolembo_ventura_enhancing_2023,lim_toward_2023,carmichael_personal_2024,almarshoud_security_2024,elkourdi_exploring_2024,herwanto_toward_2024,xia_towards_2024,iwaya_privacy_2024,de_chaves_user-centred_2025} and DTO2 (Training \& culture enablement)~\cite{spiekermann_inside_2019,andrade_privacy_2022,iwaya_organizational_2022,de_chaves_privacy_2023,iwaya_privacy_2023,lim_toward_2023}. In practice, works defined governance structures (e.g., roles, committees, privacy champions), described how  prepared and maintained privacy artifacts (e.g., DPIA/PIA records, policy registers, SOPs) and showed how those artifacts constrained day-to-day operations across projects and releases~\cite{spiekermann_inside_2019,iwaya_privacy_2024,de_chaves_privacy_2023}. Sector-specific platforms and ecosystems (e.g., PDS, TRE/PHT) situated organizational controls alongside technical controls, so approvals, researcher accreditation and disclosure checks remained visible and auditable to all stakeholders~\cite{carmichael_personal_2024,jones_profile_2020,zhang_privacy-by-design_2022}. Figure~\ref{fig:om_insights} illustrates the insights for this dimension.

Papers described privacy impact assessments, procurement and vendor controls, data-sharing agreements and operational checklists~\cite{iwaya_privacy_2024,elkourdi_exploring_2024,lim_toward_2023}. PIA/ScR syntheses cataloged how organizations planned assessments, appraised methodological quality and maintained change logs for decisions and controls, while HIPAA-focused studies mapped SE activities to safeguard families and highlighted gaps in traceability and evaluation in practice~\cite{iwaya_privacy_2024,elkourdi_exploring_2024}. Secure research environments and data trusts combined procedural safeguards (review boards, secure settings, disclosure control) with technical controls so that approvals and outputs remained documented; operators recorded project justifications, researcher training and output checks alongside provenance and policy evidence~\cite{jones_profile_2020,zhang_privacy-by-design_2022}. Several works integrated privacy tasks into development backlogs and release processes—for example, translating legal provisions into user stories and access-control templates, adding DPIA checks and documentation gates to CI and using lightweight checklists linked to code repositories to keep models and artifacts aligned with changes~\cite{herwanto_leveraging_2024,andrade_privacy_2022,de_chaves_privacy_2023,mazeli_framework_2022}. In public-sector and Small and Medium Enterprises (SME) settings, operational guides connected procurement/vendor constraints and site-specific safety/privacy analysis to concrete deployment choices (e.g., role design, anonymized telemetry, infrastructure-less modes), showing how organizational constraints shape architectures and data flows~\cite{mastrolembo_ventura_enhancing_2023,chen_holistic_2021,alhirabi_security_2021}. Industry surveys and SLRs further reported how organizations documented third-party sharing, negotiated data-sharing agreements and structured due-diligence around measurement and auditing, but also noted uneven evidence for sustained evaluation and vendor conformance at scale~\cite{jandl_reasons_2021,salem_comprehensive_2023,elkourdi_exploring_2024}.

\emph{Finding 1:} Organizational measures became effective when authors tied procedures to concrete system hooks (for example, approvals to configuration changes and outputs to disclosure checks). In TRE/PHT and PDS ecosystems, approvals and researcher accreditation translated to access configuration and output release gates and disclosure control bound statistical rules to output pipelines; the resulting packages (decision, basis, artifacts) established an auditable chain from request to release~\cite{jones_profile_2020,zhang_privacy-by-design_2022,carmichael_personal_2024}. Backlog-driven PbD integrated DPIA tasks, story-level acceptance criteria and documentation updates with CI/CD so that policy-relevant changes (e.g., new data uses) triggered checks and evidence regeneration~\cite{herwanto_leveraging_2024,andrade_privacy_2022,de_chaves_privacy_2023}.Accountability-metrics catalogs complemented these procedures with measurable expectations (e.g., audit coverage, reportability) to track whether processes produced inspectable evidence over time~\cite{xia_towards_2024}. 
\emph{Finding 2:} Studies reported gaps when procedures existed on paper but lacked tool support or linkage to deployment workflows. Reviews of HIPAA-aligned SE practice and OPCC literature observed limited longitudinal evaluation and weak connections between declared roles/processes and operational artifacts, which reduced the ability to verify that procedures remained effective as systems evolved~\cite{elkourdi_exploring_2024,iwaya_privacy_2024,spiekermann_inside_2019}. Case reports and audits in sector settings similarly noted that without pipeline integration and repository-level hooks (e.g., automated evidence regeneration, release gates), procedures drifted from implementation and undermined accountability claims~\cite{de_chaves_privacy_2023,jandl_reasons_2021}.

Organizational Measures co-occurred with \emph{Governance and Accountability} through formal processes (DPIA, approvals, disclosure control), committee oversight and policy-to-control mappings and with \emph{Culture and Training} through privacy-champion roles, developer support and training artifacts embedded in routine engineering activities~\cite{iwaya_privacy_2024,spiekermann_inside_2019,de_chaves_user-centred_2025,lim_toward_2023}. In platforms where organizations coupled procedures to verifiable technical hooks, organizational measures also intersected with \emph{Modeling and Specification} and \emph{Verification and Testing}, as modeled constraints produced checklists and tests and measurement artifacts fed audits and release decisions~\cite{de_chaves_privacy_2023,zhang_privacy-by-design_2022,jandl_reasons_2021}.

\subsubsection{Privacy Metrics.}
\begin{figure}[!h]
    \centering
    \includegraphics[width=0.55\linewidth]{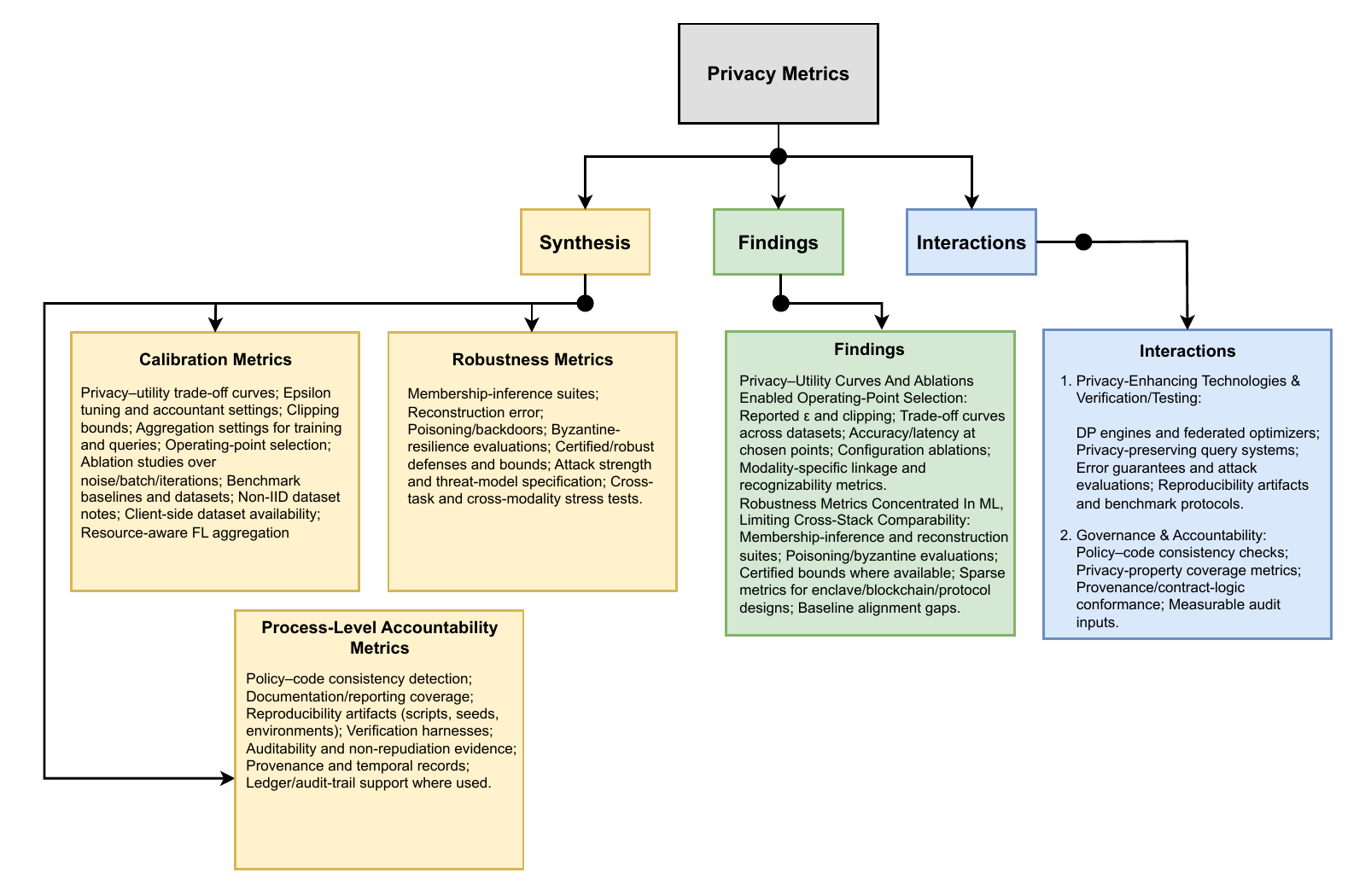}
    \caption{Overview of the insights for the Privacy Metrics dimension.}
    \label{fig:pm_insights}
\end{figure}

We scoped this dimension to quantitative measures that characterize privacy guarantees and privacy risk in relation to utility, performance and robustness, including calibration metrics for disclosure control and indicators of attack resistance and conformance. Contributing DTs included DPM1 (privacy–utility calibration, epsilon tuning, error/accuracy metrics) and DPM2 (robustness metrics, attack-surface measurement, certification). Contributing DTs included DPM1 (privacy–utility calibration, epsilon tuning, error/accuracy metrics)~\cite{pal_privacy_2019,ermakova_security_2020,mazmudar_cache_2022,zhao_survey_2022,giordano_use_2022,beg_data_2022,jordan_selecting_2022,marcolla_survey_2022,sun_toward_2022,jha_internet_2022,pramod_privacy-preserving_2023,khalid_enhancing_2023,yu_dop-sql_2024,wang_pp-csa_2024,du_privategaze_2024,melzi_overview_2024,liang_identity_2024,zhao_scenario-based_2024,yang_approaching_2024,weng_faster_2024,odema_privynas_2024,zhao_visual_2025,ali_privacy-preserved_2025,wang_security_2025} and DPM2 (robustness metrics, attack-surface measurement, certification)~\cite{noauthor_guest_2020,liu_privacy_2021,lo_systematic_2022,jha_internet_2022,mo_security_2024,melzi_overview_2024,liang_identity_2024,wang_security_2025}.
Figure~\ref{fig:pm_insights} illustrates the insights for this dimension.

We identified three recurrent metric families. First, \emph{calibration metrics} balanced privacy budgets against accuracy or latency in differentially private training and query systems (e.g., $\varepsilon$, clipping bounds, aggregation settings), typically reported as trade-off curves rather than single operating points~\cite{yu_dop-sql_2024,mazmudar_cache_2022,weng_faster_2024,yang_approaching_2024,zhang_ppfed_2024}. Second, \emph{robustness metrics} quantified resistance to inference, reconstruction, poisoning and extraction in ML pipelines, often via attack success rate, reconstruction error, or certified bounds~\cite{liu_privacy_2021,mo_security_2024,ali_privacy-preserved_2025,pramod_privacy-preserving_2023}. Third, \emph{process-level accountability metrics} operationalized auditability and conformance (e.g., detection and matching rates for policy–code consistency; coverage of documentation and reporting artifacts)\cite{wang_as_2024,jordan_selecting_2022,giordano_use_2022}. Across these families, studies exposed explicit configuration “knobs” and reported ablations to calibrate utility loss against privacy parameters~\cite{wagner_technical_2019,yu_dop-sql_2024,mazmudar_cache_2022}. Works in biometrics and content obfuscation complemented DP-centric reporting with modality-specific fidelity and linkage measures~\cite{melzi_overview_2024,du_privategaze_2024}.

\emph{Finding 1:} Authors preferred privacy–utility curves and ablation studies when they evaluated DP mechanisms or obfuscation; this practice enabled reasoned selection of operating points across datasets and tasks~\cite{zhao_survey_2022,zhao_scenario-based_2024,weng_faster_2024,zhang_ppfed_2024,odema_privynas_2024}. Biometric and visual pipelines extended these curves with linkage risk and recognizability metrics to capture modality-specific leakage~\cite{melzi_overview_2024,du_privategaze_2024}. \emph{Finding 2:} Robustness metrics concentrated in ML-focused studies and appeared less frequently in systems and architectural PET papers, which limited comparability across enclave-, blockchain-, or protocol-level approaches~\cite{liu_privacy_2021,mo_security_2024,witt_decentral_2023,ali_privacy-preserved_2025}.

Metric reporting co-occurred with \emph{PETs} and \emph{Verification and Testing}: DP engines, federated optimizers and privacy-preserving query systems attached explicit error guarantees, empirical attack evaluations, or reproducibility artifacts~\cite{chen_advancements_2025,lo_systematic_2022,yu_dop-sql_2024,mazmudar_cache_2022}. Governance-oriented works referenced measurable conformance when they operationalized auditability, provenance, or contract logic (e.g. policy code consistency checks; privacy property coverage)~\cite{wang_as_2024,jordan_selecting_2022,liang_identity_2024}.

\subsubsection{Culture \& Training.}
\begin{figure}[!h]
    \centering
    \includegraphics[width=0.55\linewidth]{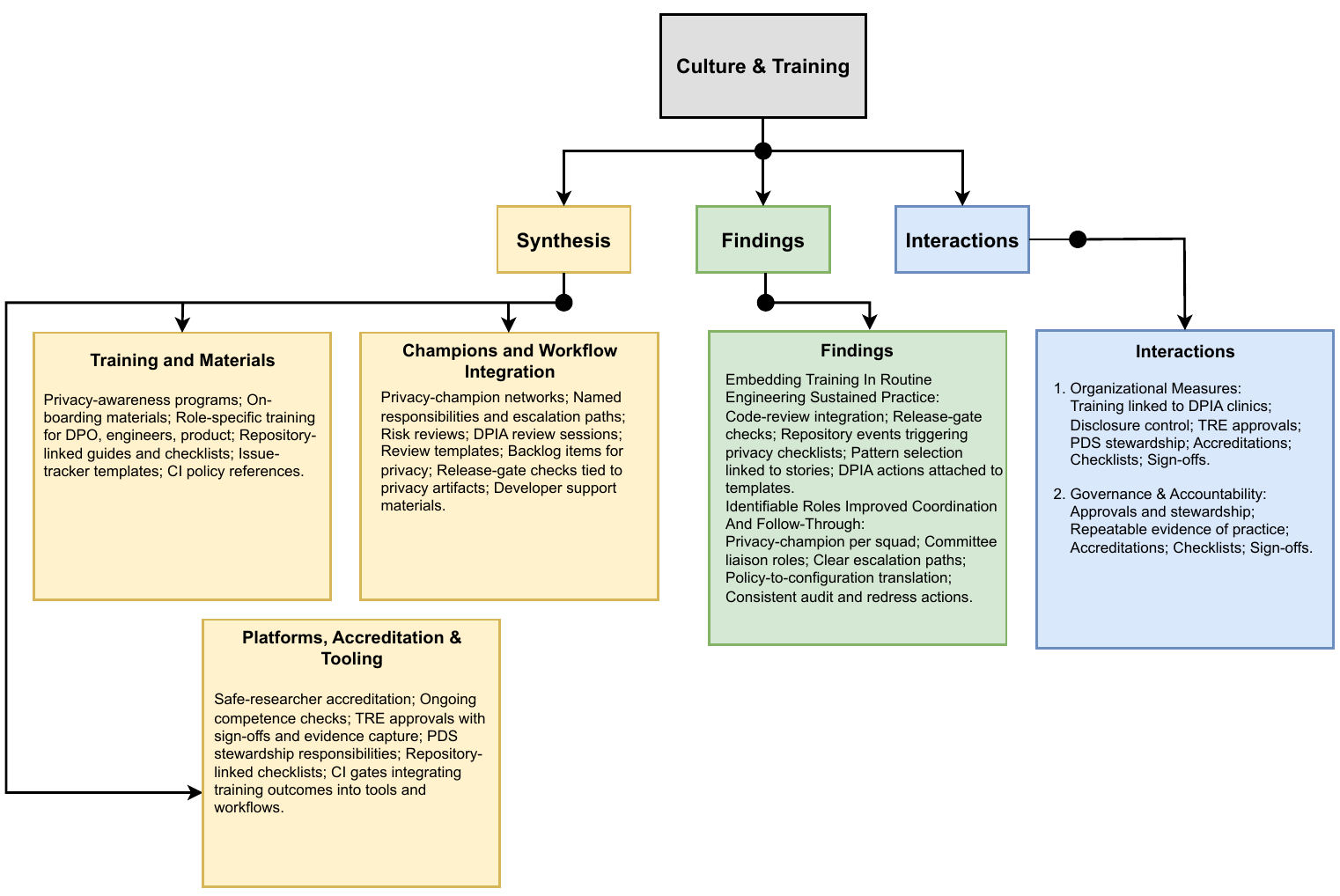}
    \caption{Overview of the insights for the Culture \& Training dimension.}
    \label{fig:ct_insights}
\end{figure}

This dimension covered workforce practices that build and sustain privacy competence and norms, including role-based training, champion networks and organizational climate factors that influence the adoption of privacy controls and evidence-based practices. Contributing DTs included DTC1 (Privacy awareness \& training)~\cite{senarath_will_2019,spiekermann_inside_2019,jones_profile_2020,alhirabi_security_2021,andrade_privacy_2022,mazeli_framework_2022,iwaya_organizational_2022,de_chaves_privacy_2023,hu_dark_2023,wilkowska_interdisciplinary_2023,carboni_privacy_2023,iwaya_privacy_2023-1,iwaya_privacy_2024} and DTC2 (Organizational Privacy Culture \& Climate)~\cite{hoel2019privacy,senarath_will_2019,spiekermann_inside_2019,rommetveit_privacy_2022,zhang_privacy-by-design_2022,iwaya_organizational_2022,de_chaves_privacy_2023,iwaya_privacy_2023,iwaya_privacy_2023-1,lim_toward_2023,elkourdi_exploring_2024}. Empirical and scoping studies described how leadership signals, norms and shared beliefs influenced whether teams integrated privacy tasks into day-to-day engineering and how staff interpreted obligations during design and operations~\cite{iwaya_organizational_2022,spiekermann_inside_2019}. HCI-oriented mappings and stakeholder analyses reported that culture affected the effectiveness of communication, consent and risk interpretation, particularly in settings with vulnerable users (e.g., assisted living and lifelogging)~\cite{de_chaves_user-centred_2025,carboni_privacy_2023,wilkowska_interdisciplinary_2023}. These results positioned culture/training as a first-order determinant of whether privacy processes remained active beyond initial roll-out~\cite{elkourdi_exploring_2024,iwaya_organizational_2022}.
Figure~\ref{fig:ct_insights} illustrates the insights for this dimension.

Studies reported privacy-awareness programs, on-boarding materials and the use of privacy champions within teams~\cite{iwaya_organizational_2022,spiekermann_inside_2019,de_chaves_user-centred_2025}. Organizational reviews documented awareness sessions, role-specific training (e.g. DPO, engineers, product) and internal networks of “champions” who supported backlog grooming, reviews and incident preparation; these interventions aimed to normalize privacy tasks in routine work rather than treating them as episodic exercises~\cite{iwaya_organizational_2022,spiekermann_inside_2019}. Several works associated culture with sustained adoption of privacy processes and with the ability to interpret policies and risks during design and operations, showing that teams with explicit privacy roles and recurring rituals (e.g., risk huddles, PIA clinics) maintained higher process fidelity~\cite{de_chaves_user-centred_2025,elkourdi_exploring_2024}. Evidence also indicated that training without integration into tools and workflows had limited effect: developer-facing frameworks and guides were most effective when embedded in repositories, issue trackers and CI policies so that training translated directly into actions and artifacts~\cite{mazeli_framework_2022,andrade_privacy_2022,de_chaves_privacy_2023}. Sector platforms illustrated complementary organizational mechanisms: secure research environments required “safe researcher” accreditation and ongoing competence checks, while personal data store ecosystems formalized stewardship responsibilities for different actors~\cite{jones_profile_2020,carmichael_personal_2024}.

\emph{Finding 1:} Programs that embedded training into routine engineering activities (for example, code reviews or release gates) sustained practice more consistently than stand-alone sessions~\cite{mazeli_framework_2022,andrade_privacy_2022,de_chaves_privacy_2023}. Developer-centred proposals that tied checklists, pattern selection and PIA actions to repository events and review templates reported higher uptake and fewer regressions than classroom-only initiatives, indicating that proximity to day-to-day work mattered for persistence~\cite{mazeli_framework_2022,de_chaves_privacy_2023}. \emph{Finding 2:} The presence of identifiable privacy roles or champions improved coordination across governance, engineering and product teams~\cite{spiekermann_inside_2019,iwaya_organizational_2022,carboni_privacy_2023}. Studies linked named responsibilities (e.g., privacy champion per squad; committee liaisons) to clearer escalation paths, better translation of policy into code/configuration and more consistent follow-through on audit and redress actions~\cite{spiekermann_inside_2019,iwaya_organizational_2022}.

Culture \& Training. co-occurred with \emph{Organizational Measures} and \emph{Governance and Accountability}~\cite{iwaya_organizational_2022,elkourdi_exploring_2024,lim_toward_2023}. In settings where organizations coupled training to documented processes (e.g., DPIA clinics, disclosure control) and to platform controls (e.g., TRE approvals, PDS stewardship), cultural interventions reinforced governance by producing repeatable evidence of practice (accreditations, checklists, sign-offs)~\cite{jones_profile_2020,carmichael_personal_2024,zhang_privacy-by-design_2022}.

\subsubsection{User-Centric.}
\begin{figure}[!h]
    \centering
    \includegraphics[width=0.55\linewidth]{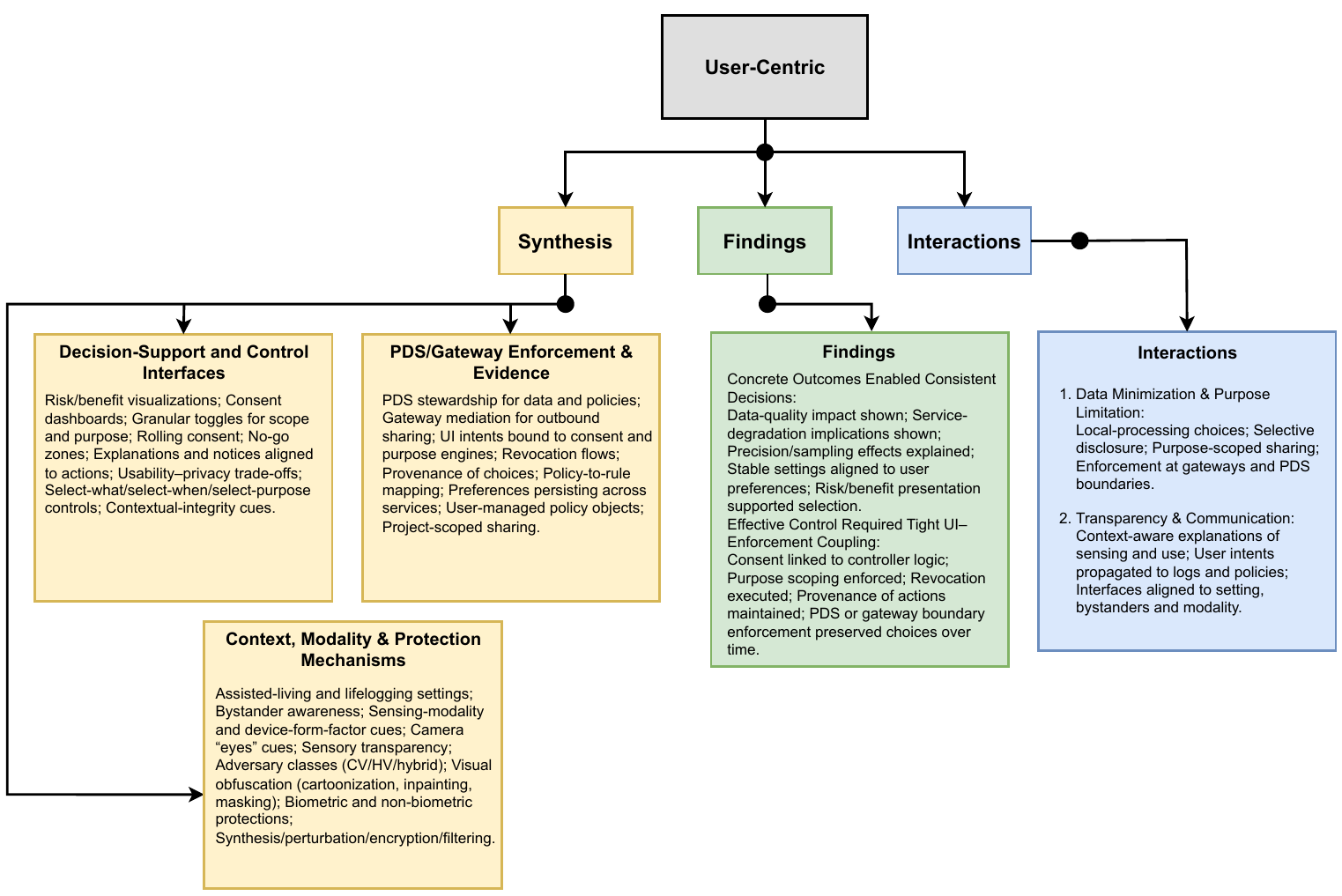}
    \caption{Overview of the insights for the User-Centric dimension.}
    \label{fig:user_centric_insights}
\end{figure}

This dimension addressed user-facing mechanisms that support informed, context-aware decisions and effective control and that bind user intent to enforcement and provenance so choices persist in execution. Contributing DTs included DTUC1 (User-centred privacy in context (contextual integrity, multidimensional privacy))~\cite{barhamgi_user-centric_2018,hoel2019privacy,benhamida_pyff_2021,alhirabi_security_2021,de_chaves_privacy_2023,hu_dark_2023,wilkowska_interdisciplinary_2023,carboni_privacy_2023,carmichael_personal_2024,grabler_privacy_2025,zhao_visual_2025,de_chaves_user-centred_2025}. User-centred designs presented privacy choices in ways that people could act on (e.g., selecting what to share, when and for which purpose) and framed those choices in terms of concrete effects on service behavior or output quality~\cite{barhamgi_user-centric_2018,de_chaves_user-centred_2025}. Personal Data Store (PDS) ecosystems operationalized self-management by giving individuals a managed space for data, policies and provenance, so that outbound sharing and reuse reflected user-defined preferences rather than platform defaults~\cite{carmichael_personal_2024} In lifelogging and assisted-living contexts, studies treated privacy as situational and multi-dimensional, showing that expectations depended on setting, bystanders, sensing modality and social norms; user-facing artifacts therefore had to match those contexts to be effective~\cite{wilkowska_interdisciplinary_2023,carboni_privacy_2023,grabler_privacy_2025}. Figure~\ref{fig:user_centric_insights} summarizes the insights for this dimension.

Authors built decision-support interfaces that estimated risks and benefits, exposed configuration choices and suggested actions~\cite{barhamgi_user-centric_2018,de_chaves_user-centred_2025}. These interfaces included risk/benefit visualizations, sliders and toggles for data-sharing scope and explanations of what a given choice would change, which helped users align configuration with their goals or constraints~\cite{de_chaves_user-centred_2025,barhamgi_user-centric_2018}. Some works implemented PDS or gateway components that mediated sharing and applied policies on behalf of the user, linking UI intents to enforcement points (e.g., consent, purpose scoping, revocation) and recording provenance so that choices remained effective over time~\cite{carmichael_personal_2024,benhamida_pyff_2021}. Studies in domains such as robotics and mobile health considered contextual cues (e.g., device form factor, presence of bystanders, visible camera “eyes”) and consent dynamics and they evaluated how interface and embodiment choices affected behavior and comfort in both public and private settings~\cite{grabler_privacy_2025,hu_dark_2023,wilkowska_interdisciplinary_2023}. Several reports connected user-facing artifacts to broader organizational processes by embedding guidance, templates and role support (e.g., privacy champions) so that user control remained a first-class engineering concern rather than an afterthought~\cite{de_chaves_user-centred_2025,carboni_privacy_2023}.

\emph{Finding 1:} Systems that presented concrete outcomes of choices (for example, data-quality impacts or service degradation) enabled more consistent decisions than those that showed only policy text~\cite{de_chaves_user-centred_2025,carmichael_personal_2024,barhamgi_user-centric_2018}. Evidence from user-centered surveys and platform case studies showed that when interfaces explained, for example, how reducing precision or sampling would affect recommendation quality or fidelity of analysis, users selected stable settings that matched their preferences, while abstract policy language alone did not support stable choices~\cite{de_chaves_user-centred_2025,barhamgi_user-centric_2018}. \emph{Finding 2:} User control required tight coupling between interfaces and enforcement layers; without that link, systems reverted to defaults that ignored user intent~\cite{carmichael_personal_2024,benhamida_pyff_2021,wilkowska_interdisciplinary_2023} PDS and gateway designs that bound UI actions (e.g., scoping, revocation) to policy engines and provenance logs maintained effect over time and across services, while studies in lifelogging and robotics showed that decoupled or purely presentational interfaces left downstream processing unchanged despite visible “consent” cues~\cite{carmichael_personal_2024,grabler_privacy_2025,hu_dark_2023}.

User-Centric co-occurred with \emph{Data Minimization and Purpose Limitation} when interfaces exposed local processing, selective disclosure and purpose-scoped sharing as first-order choices and system components enforced those choices at gateways or PDS boundaries~\cite{barhamgi_user-centric_2018,benhamida_pyff_2021,carmichael_personal_2024}. It co-occurred with \emph{Transparency and Communication} when interfaces provided concrete, context-aware explanations of sensing and use and when user intents propagated to logs and policies that other components could consume~\cite{wilkowska_interdisciplinary_2023,grabler_privacy_2025,hu_dark_2023}.

\subsubsection{Lifelong Management.}
\begin{figure}[!h]
    \centering
    \includegraphics[width=0.55\linewidth]{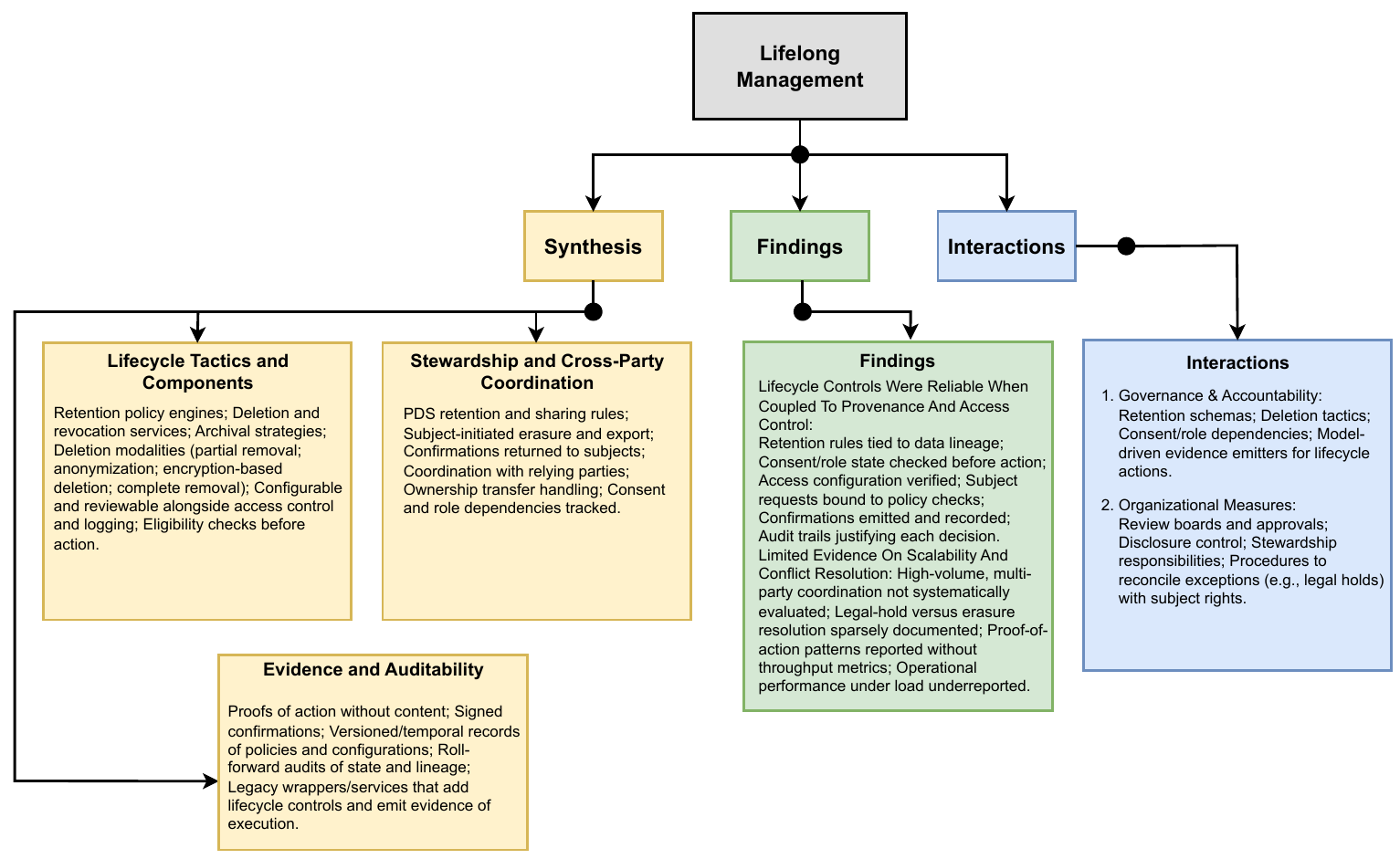}
    \caption{Overview of the insights for the Lifelong Management dimension.}
    \label{fig:lifelong_insights}
\end{figure}

This dimension captured lifecycle operations that manage retention, deletion, revocation, archival and transfer over time, including the system hooks that coordinate actions across parties and emit proofs or logs suitable for audit. Contributing DTs included DTL1 (Retention/deletion policies and lifecycle management)~\cite{senarath_will_2019,jandl_reasons_2021,andrade_privacy_2022,mazeli_framework_2022,lim_toward_2023,carmichael_personal_2024,pallas2024privacy}. In architectural accounts, authors specified how systems represent and enforce retention periods, deletion and revocation and archival constraints as first-class design elements rather than ad hoc scripts~\cite{pallas2024privacy}. In stewardship-oriented ecosystems, personal data stores positioned individuals to initiate and track lifecycle actions (e.g., export, retention change, erasure), with the platform mediating obligations across data holders and documenting the outcome for later inspection~\cite{carmichael_personal_2024}.
Figure~\ref{fig:lifelong_insights} summarizes the insights for this dimension.

Authors implemented data-retention schedules, deletion and revocation mechanisms and archival strategies~\cite{senarath_will_2019,mazeli_framework_2022}. Architectural guidance described explicit lifecycle tactics and components—e.g., retention policy engines, deletion services and evidence emitters—so that retention and erasure became configurable and reviewable alongside access control and logging; the guidance also distinguished deletion modalities (partial removal, anonymization, encryption-based deletion, complete removal) to match legal and operational constraints~\cite{lim_toward_2023}. Stewardship designs for personal data stores detailed how end users set retention and sharing rules, request erasure and obtain confirmations, while the ecosystem coordinated with relying parties and recorded proofs that actions occurred; this design aimed to keep lifecycle decisions transparent and traceable over time across organizations~\cite{senarath_will_2019} Some architectures supported versioned or temporal records (e.g., time-stamped configuration and policy histories) and recorded ownership transfer, enabling roll-forward audits that reconcile what data existed, where and under which policy or consent at a given point in time; these histories underpinned later verification that lifecycle actions complied with current policy and role bindings~\cite{jandl_reasons_2021}. Several systems combined deletion with persistent audit by storing proofs of action (e.g., signed confirmations) alongside content removal, allowing teams to demonstrate compliance without retaining personal data indefinitely; this pattern balanced erasure with accountability artifacts required for redress or dispute resolution~\cite{andrade_privacy_2022}. Legacy-modernization reports described wrappers and services that added lifecycle controls to existing systems, so administrators could enforce retention and erasure without full re-architecture and still produce evidence of execution~\cite{pallas2024privacy}.

\emph{Finding 1:} Lifecycle controls became reliable when authors integrated retention and deletion with provenance and access control so that systems could verify eligibility before action~\cite{senarath_will_2019,carmichael_personal_2024}. Architectural specifications that tied retention rules to data lineage and role/consent state prevented accidental removal or unlawful retention and they produced an audit trail that justified each lifecycle decision at the time it was executed~\cite{andrade_privacy_2022}. PDS stewardship flows that bound user requests (e.g., erasure or export) to policy checks and emitted confirmations helped relying parties coordinate consistent actions and gave subjects inspectable records of what happened~\cite{carmichael_personal_2024}. \emph{Finding 2:} Evidence on scalability and on recovery from conflicting obligations (for example, legal hold versus erasure) remained limited; while patterns for “proof-of-action without content” addressed accountability, few reports evaluated performance under high-volume, multi-party coordination or documented resolution strategies when retention exceptions overrode user-initiated requests~\cite{carmichael_personal_2024,jandl_reasons_2021}.

Lifelong Management co-occurred with \emph{Governance and Accountability} when policy and rule representations (e.g., retention schemas, deletion tactics, consent/role dependencies) drove enforcement and evidence generation and with \emph{Culture \& Training} when procedures (e.g., review boards, disclosure control, stewardship responsibilities) audited lifecycle actions and reconciled exceptions (e.g., legal holds) with subject rights~\cite{pallas2024privacy,lim_toward_2023}.

\subsubsection{Incident Response \& Management.}
\begin{figure}[!h]
    \centering
    \includegraphics[width=0.55\linewidth]{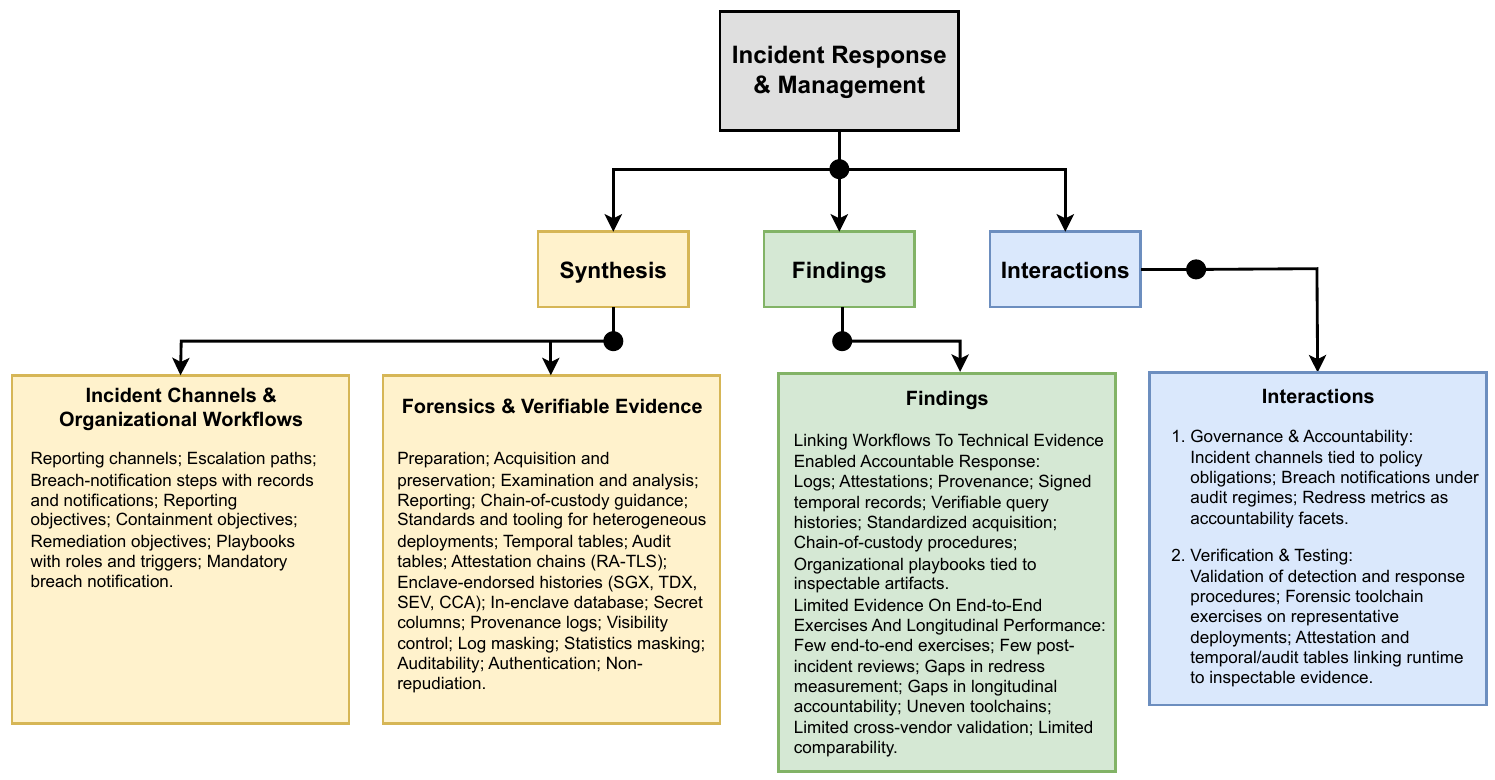}
    \caption{Overview of the insights for the Incident Response \& Management dimension.}
    \label{fig:irm_insights}
\end{figure}

This dimension addressed incident-handling mechanisms and workflows that support detection, containment, investigation and notification by linking response procedures to tamper-evident logs, attestations, provenance and forensic artifacts. Contributing DTs included DTI1 (Tamper-evidence, auditability and forensics (temporal/audit tables))~\cite{senarath_will_2019,ermakova_security_2020,alkhariji_synthesising_2021,jandl_reasons_2021,andrade_privacy_2022,salem_comprehensive_2023,khalid_enhancing_2023,yang_secudb_2024,xia_towards_2024,pallas2024privacy}. In practice, works defined incident-handling objectives (reporting, escalation, containment, remediation) and linked them to organizational and technical artifacts so that teams could move from detection to auditable response~\cite{xia_towards_2024,alkhariji_synthesising_2021}. Responsible AI metrics catalogs treated redress and incident reporting as measurable accountability facets, making expectations explicit for evidence capture and review~\cite{xia_towards_2024}. Pattern-based PbD syntheses for healthcare included breach-notification and response patterns among the reusable elements available to practitioners, clarifying the roles, triggers and artifacts needed to execute response in regulated settings~\cite{khalid_enhancing_2023}.
Figure~\ref{fig:irm_insights} summarizes the insights for this dimension.

Papers described incident reporting channels, escalation paths and breach-notification workflows and they associated each step with concrete artifacts (records, logs, notifications) that enabled post hoc auditing~\cite{senarath_will_2019,pallas2024privacy}. Several works connected response to logging, provenance and forensics so that teams could reconstruct events and apply containment or remediation: IoT digital-forensics frameworks detailed preparation, acquisition/preservation, examination/analysis and reporting phases with guidance on evidence handling, standards and tooling; these phases form the backbone of post-incident investigations in heterogeneous deployments~\cite{salem_comprehensive_2023}. Systems papers embedded verifiability and non-repudiation in the data layer—e.g., enclave-endorsed temporal tables, attestation chains (RA-TLS) and audit streams—so response teams could correlate anomalies with tamper-evident histories and produce verifiable reports~\cite{yang_secudb_2024}. Metrics catalogs specified redress and incident-process indicators (e.g., incident reporting completeness, auditability coverage), offering targets for organizations to monitor and improve over time~\cite{ermakova_security_2020,xia_towards_2024}.

\emph{Finding 1:} Systems that linked incident workflows to technical evidence (logs, attestations and provenance) supported clearer accountability and faster response~\cite{yang_secudb_2024,salem_comprehensive_2023}. Enclave-backed database designs provided attested execution, signed temporal records and verifiable query histories that shortened triage and enabled defensible post-incident reports; digital-forensics frameworks prescribed standardized acquisition and chain-of-custody procedures that preserved evidentiary value across devices, networks and cloud backends~\cite{yang_secudb_2024,salem_comprehensive_2023}. When combined with organizational playbooks (who reports, who escalates and when), these evidentiary hooks turned ad hoc handling into a repeatable process tied to artifacts that auditors could inspect~\cite{alkhariji_synthesising_2021,xia_towards_2024}. \emph{Finding 2:} Studies rarely reported end-to-end exercises or post-incident reviews, which limited claims about operational effectiveness~\cite{xia_towards_2024,salem_comprehensive_2023}. Bibliometric and metrics-oriented work highlighted gaps in redress measurement and longitudinal accountability, while IoT forensics surveys pointed to uneven tool chains and limited cross-vendor validation, both factors that reduce comparability and make it difficult to demonstrate sustained operational performance in incident response~\cite{xia_towards_2024,salem_comprehensive_2023}.

Incident Response and Management co-occurred with \emph{Governance and Accountability} when incident channels, breach notifications and redress metrics were tied to policy obligations and audit regimes and with \emph{Verification and Testing} when teams validated detection and response procedures or exercised forensic toolchains against representative deployments~\cite{alkhariji_synthesising_2021,xia_towards_2024,salem_comprehensive_2023}. Platform designs that embed attestation and temporal/audit tables further link responses to verifiable execution, creating a direct bridge from governance claims to inspectable technical evidence during and after incidents~\cite{yang_secudb_2024,pallas2024privacy}.

\subsection{Cross-cutting Patterns}\label{sec:cross_cut}
In this section, we report cross-cutting patterns that organize how the dimensions relate, where they align along engineering and data lifecycles and how emphasis varies by domain. First, we present co-occurrence among dimensions to characterize recurrent pairings and clusters. Then, we examine lifecycle placement across the software development and personal data lifecycles to identify typical phases and handoffs. Finally, we inspect how dimensions interact across application domains. These findings advance the paper’s goal by providing an integrated map that links intentions to enforcement, measurement and evidence, directly informing RQ2, RQ3 and RQ4.

\subsubsection{Co-occurrence of Dimensions}\label{sec:cooccur}
\begin{figure}[!h]
    \centering
    \includegraphics[width=0.60\linewidth]{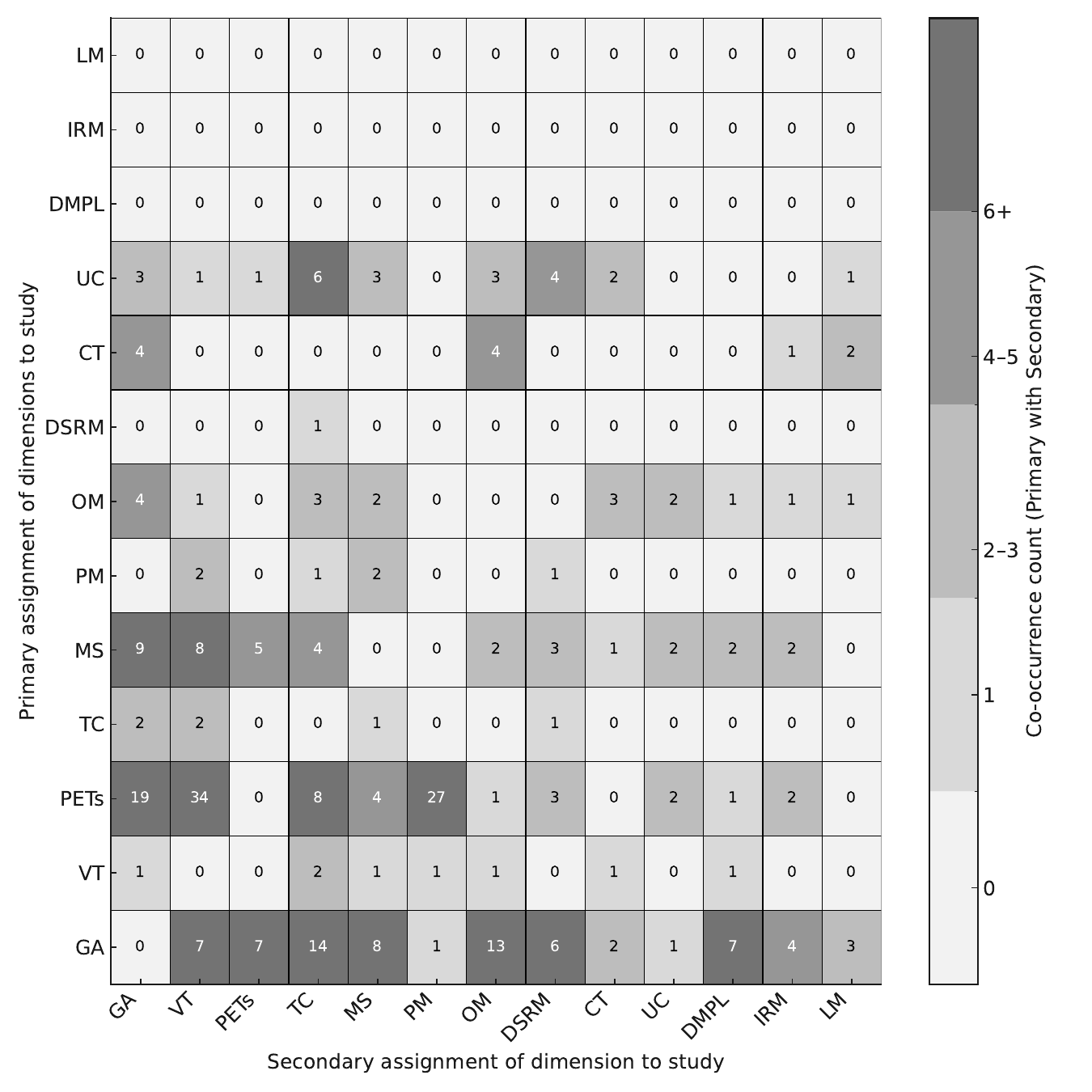}
    \caption{Co-occurrence matrix of privacy-engineering dimensions across the included studies. Y-axis show the primary assignment per study, x-axis show the secondary assignment and each cell reports the count of primary-to-secondary pairings. Abbreviations: GA = Governance \& Accountability; VT = Verification \& Testing; PETs = Privacy-Enhancing Technologies; TC = Transparency \& Communication; MS = Modeling \& Specification; PM = Privacy Metrics; OM = Organizational Measures; DSRM = Data Subject Rights Management; CT = Culture \& Training; UC = User-Centric; DMPL = Data Minimization \& Purpose Limitation; IRM = Incident Response \& Management; LM = Lifelong Management.}
    \label{fig:co_occur}
\end{figure}

We analyzed co-occurrence by constructing a directional matrix of primary-secondary dimension pairings at study level. For each paper, we recorded every pairing that linked its designated primary dimension to its secondary dimensions. Figure~\ref{fig:co_occur} illustrates the resulting co-occurrence counts across the included studies.

\begin{figure}[!h]
    \centering
    \includegraphics[width=0.45\linewidth]{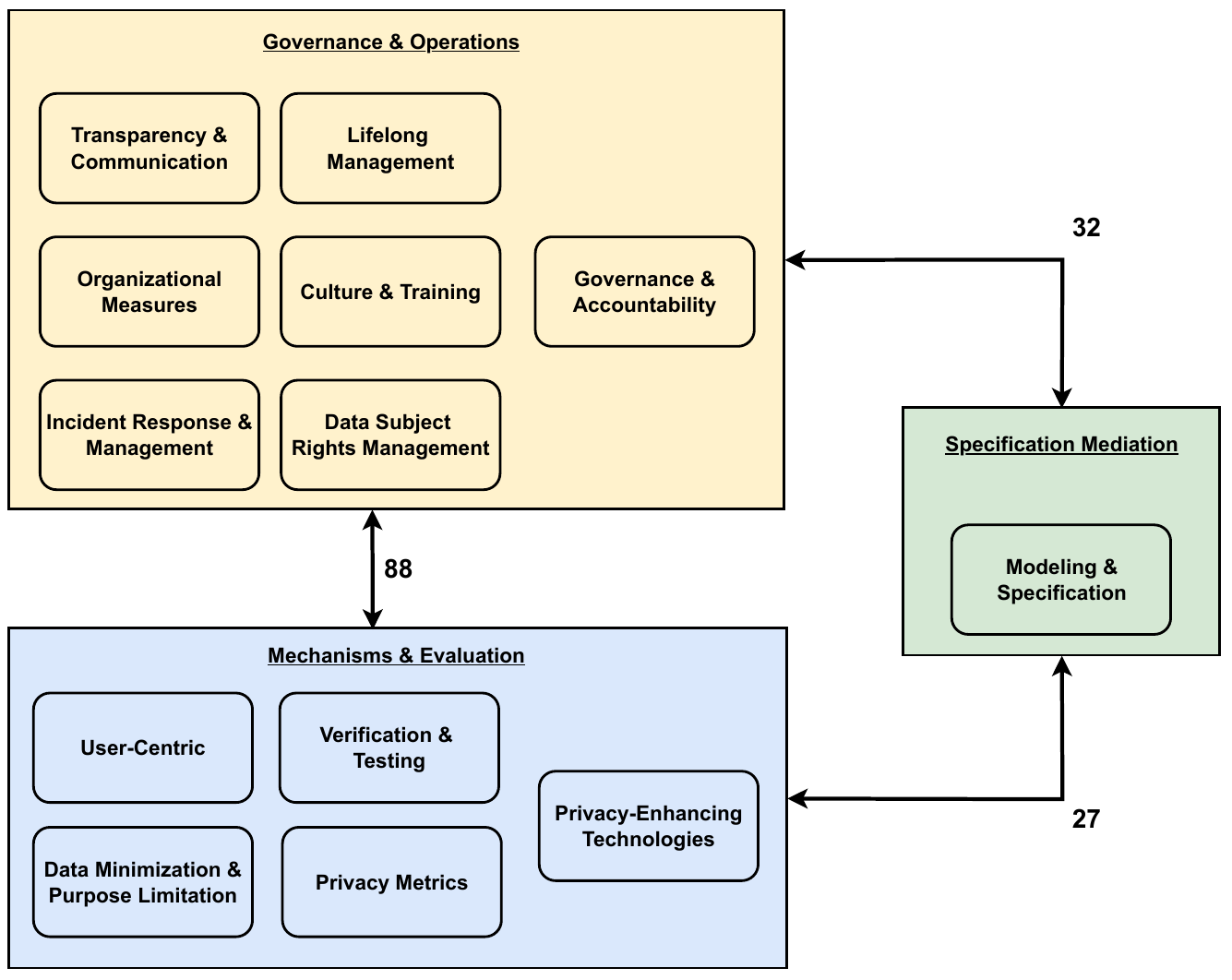}
    \caption{Group-level directional co-occurrence of dimensions. Mechanisms and Evaluation and Governance and Operations form two interacting cores. Modeling and Specification are shown as Specification Mediation, linking to both cores without belonging to either. The numbers on arrows denote the total primary to secondary co-occurrences for each pair.}
    \label{fig:co_occur_group}
\end{figure}

The matrix revealed two cores. The mechanism-evaluation core linked Privacy-Enhancing Technologies with Verification \& Testing and Privacy Metrics and attached User-Centric and Data Minimization \& Purpose Limitation to this side. The governance-operations core linked Governance \& Accountability with Transparency \& Communication and Organizational Measures and attached Data Subject Rights Management, Incident Response \& Management, Culture \& Training and Lifelong Management to this side. Modeling \& Specification sat outside both cores in a Specification Mediation role that connected specification artifacts to checks on the mechanism side and to governance and enforcement on the operations side; the corpus also supported direct links between the cores and the schematic emphasized Specification Mediation rather than a unique path. Figure~\ref{fig:co_occur_group} summarizes the two cores, their attached dimensions and the Specification Mediation connections~\cite{wagner_technical_2019,daoudagh_data_2021,mazmudar_cache_2022,jha_internet_2022,li_longitudinal_2023,gebauer_human---loop_2023,yu_dop-sql_2024,herwanto_leveraging_2024}. 

The two cores reflected how studies substantiated claims. Mechanism-oriented papers typically paired mechanisms with calibration or robustness metrics and evaluated them through controlled experiments or ecosystem measurements, while policy- and operations-oriented papers documented consent, notices and rights workflows and tied them to access control, logging, or audit processes~\cite{wagner_technical_2019,daoudagh_data_2021,lo_systematic_2022,jha_internet_2022,mo_security_2024,herwanto_leveraging_2024}. The corpus also supported bridging roles across cores~\cite{gharib2021copri,andrade_privacy_2022,de_chaves_privacy_2023}: Organizational Measures connected governance constructs to day-to-day development and operations and Modeling \& Specification connected high-level intentions to concrete components and checks~\cite{alshammari_privacy_2018,andrade_privacy_2022,de_chaves_privacy_2023}. We also observed domain effects~\cite{jones_profile_2020,benhamida_pyff_2021,zhang_privacy-by-design_2022,jha_internet_2022,odema_privynas_2024}. Healthcare papers often combined Governance \& Accountability with Organizational Measures and Rights, reflecting regulated environments and secure data platforms. IoT and edge papers paired PETs with Minimization and User-Centric controls, reflecting on-device processing and constrained resources. Web measurement papers paired Verification \& Testing with Transparency \& Communication, reflecting evaluations of banners, tracking and policy conformance at scale~\cite{jones_profile_2020,benhamida_pyff_2021,zhang_privacy-by-design_2022,jha_internet_2022,carmichael_personal_2024,odema_privynas_2024}. These patterns did not follow frequency alone. We traced them to recurring architectural choices (e.g., ledger-backed access and provenance, enclave-backed databases, federated pipelines) and process choices (e.g., privacy-by-design backlogs, incident workflows) that connected code paths and organizational workflows~\cite{wu_privacy-preserved_2022,lo_systematic_2022,nguyen_federated_2023,de_chaves_privacy_2023,yang_secudb_2024,iwaya_privacy_2024}. Where PET designs embedded provenance or attestation, papers also discussed governance implications and compliance evidence; where governance frameworks defined roles and responsibilities, papers also introduced machine-readable artifacts or interfaces connected to enforcement points~\cite{daoudagh_data_2021,wu_privacy-preserved_2022,gebauer_human---loop_2023,yang_secudb_2024,herwanto_leveraging_2024}. Conversely, the corpus showed weak co-occurrence between Lifelong Management and Verification \& Testing because few studies exercised retention and deletion controls under test conditions and weak co-occurrence between User-Centric and Incident Response because incident handling often remained an organizational process disconnected from end-user interactions~\cite{barhamgi_user-centric_2018,hu_dark_2023,pallas2024privacy,carmichael_personal_2024,xia_towards_2024}.

We interpreted the strongest triads and sequence patterns as common engineering paths. PETs and Privacy Metrics often co-occurred with Verification in a build-measure loop: authors implemented a mechanism, reported privacy-utility or robustness curves and evaluated the result against baselines; when mechanisms shifted data distributions or model behavior (e.g., DP in federated learning, obfuscation for visual data), evaluations often extended to downstream tasks and linked to Modeling \& Specification when authors traced requirements or attack surfaces~\cite{wagner_technical_2019,mazmudar_cache_2022,lo_systematic_2022,zhao_survey_2022,nguyen_federated_2023,yu_dop-sql_2024,weng_faster_2024,melzi_overview_2024,du_privategaze_2024}. When authors deployed architectural PETs (e.g., enclave-backed databases, contract-governed access), studies introduced provenance, attestation, or audit tables and linked to Governance \& Accountability and Organizational Measures because these artifacts changed operational processes (e.g., approval of releases, interpretation of logs, handling of erasure)~\cite{jones_profile_2020,wu_privacy-preserved_2022,de_chaves_privacy_2023,yang_secudb_2024}. Transparency \& Communication often co-occurred with Verification \& Testing because user-facing artifacts (notices, banners, consent interfaces, machine-readable policies) mediated access, deletion, correction, or portability; when authors expressed policies and rights as code or structured rules, they also linked to Modeling \& Specification and, when logs or attestations entered the design, to Governance \& Accountability~\cite{daoudagh_data_2021,jha_internet_2022,zhang_privacy-by-design_2022,gebauer_human---loop_2023,herwanto_leveraging_2024}. Incident Response \& Management connected to Governance \& Accountability when papers framed incident reporting, redress, or breach notification as accountability elements~\cite{salem_comprehensive_2023,xia_towards_2024,yang_secudb_2024}. Culture \& Training rarely appeared alone; it clustered with Organizational Measures and Governance through privacy champions, training and committee oversight embedded in backlogs or change-management gates~\cite{spiekermann_inside_2019,iwaya_organizational_2022,elkourdi_exploring_2024}. Lifelong Management co-occurred with Culture \& Training and Organizational Measures in studies that implemented retention schedules and erasure in secure platforms; designs that kept immutable audit trails often recorded proofs of action rather than content, linking back to Governance \& Accountability~\cite{pallas2024privacy,carmichael_personal_2024,yang_secudb_2024}. Across domains, the corpus repeated these couplings with different emphasis: healthcare highlighted governance-oriented triads around Rights, while autonomous or mobile scenarios highlighted PET triads around Metrics and Verification~\cite{jones_profile_2020,benhamida_pyff_2021,zhang_privacy-by-design_2022,odema_privynas_2024}. This stability supported a structural interpretation: authors strengthened privacy claims by linking mechanisms to measurements and tests and they strengthened policy claims by linking processes and artifacts to enforcement and evidence~\cite{wagner_technical_2019,daoudagh_data_2021,lo_systematic_2022,herwanto_leveraging_2024}.

We checked outliers and negative evidence to avoid over-interpreting frequent pairs. Some mechanism-focused papers reported accuracy without privacy metrics; we treated these cases as PETs-Verification or PETs-only and recorded missing metrics during extraction~\cite{lo_systematic_2022,mo_security_2024}. Some governance-focused papers described roles and processes without connecting them to executable controls; we marked these as Governance-only or Governance-Organizational and avoided inferring unsupported links~\cite{spiekermann_inside_2019,elkourdi_exploring_2024,iwaya_privacy_2024}. In edge and IoT settings, we observed PETs and Data Minimization without User-Centric controls when systems processed data locally without exposing decisions to users and we recorded this absence because it affected the interpretation of on-device processing patterns~\cite{benhamida_pyff_2021,odema_privynas_2024}. We also examined cases where Transparency appeared without Rights, which typically involved machine-readable policy representations designed for audit or automation without an end-user interface~\cite{gebauer_human---loop_2023,herwanto_leveraging_2024}. Finally, we controlled for co-occurrence driven by common authorship or survey scope by checking whether papers implemented or evaluated the linkage, rather than merely mentioning it~\cite{lo_systematic_2022,de_chaves_privacy_2023,mo_security_2024}. After these checks, the conclusions held: PETs coupled with Metrics and Verification in mechanism-centric work; Governance \& Accountability coupled with Transparency \& Communication and Organizational Measures in policy-centric work; Modeling \& Specification acted as specification mediation; Incident Response attached to Governance \& Accountability and sometimes to Modeling \& Specification only as a secondary assignment~\cite{wagner_technical_2019,daoudagh_data_2021,jha_internet_2022,zhang_privacy-by-design_2022,xia_towards_2024}.

\subsubsection{Lifecycle Placement}\label{sec:lifecycle}
\begin{figure}[!h]
    \centering
    \includegraphics[width=0.50\linewidth]{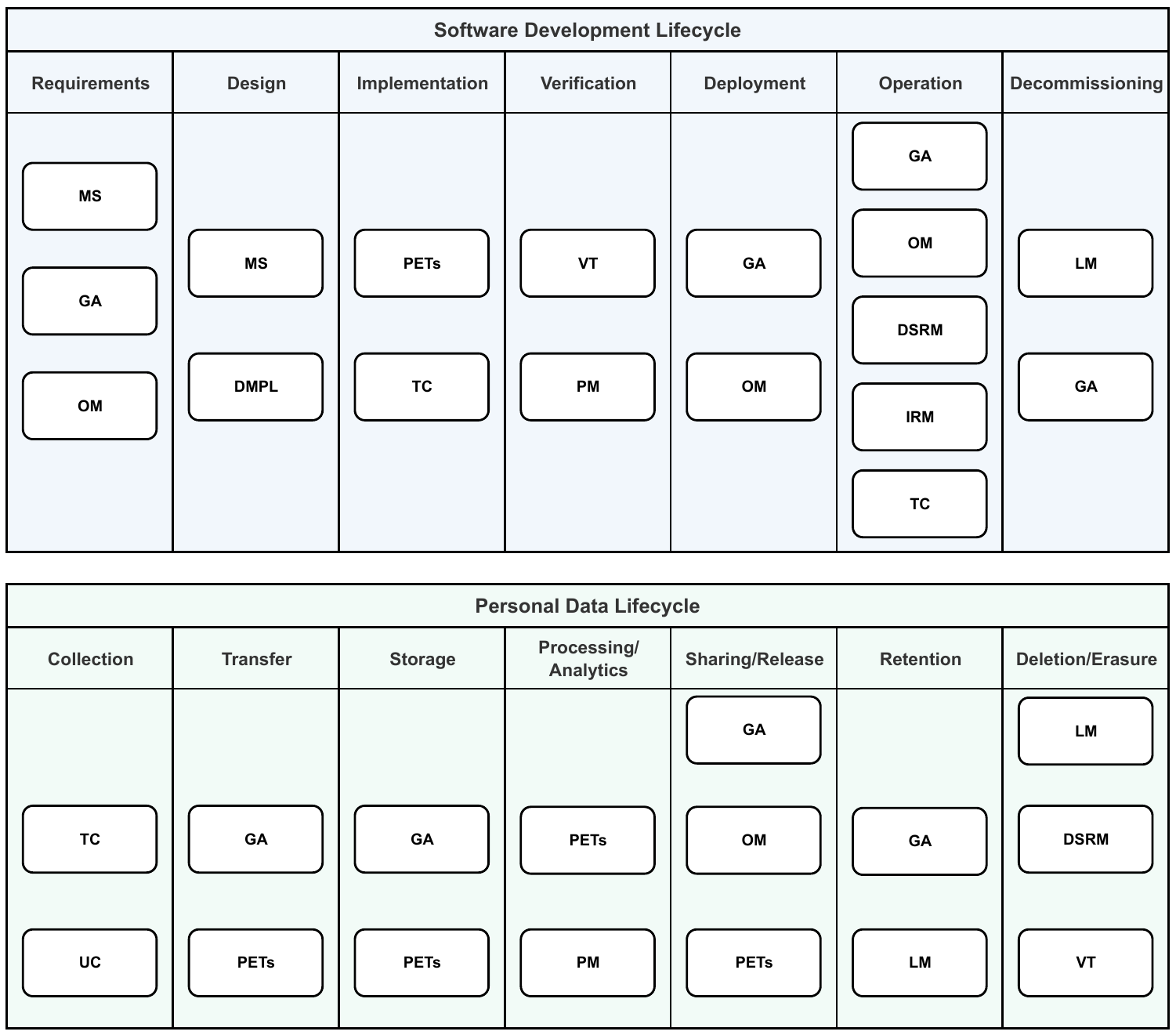}
    \caption{Lifecycle placement of dimensions across two frames. The upper panel shows where dimensions concentrate along the software development lifecycle  and the lower panel shows placements along the personal data lifecycle. Abbreviations: GA = Governance \& Accountability; VT = Verification \& Testing; PETs = Privacy-Enhancing Technologies; TC = Transparency \& Communication; MS = Modeling \& Specification; PM = Privacy Metrics; OM = Organizational Measures; DSRM = Data Subject Rights Management; CT = Culture \& Training; UC = User-Centric; DMPL = Data Minimization \& Purpose Limitation; IRM = Incident Response \& Management; LM = Lifelong Management.}
    \label{fig:lifecycles}
\end{figure}

We mapped evidence to lifecycle phases in two parallel frames: the software development lifecycle (requirements, design, implementation, verification, deployment, operation, decommissioning) and the personal-data lifecycle (collection, transfer, storage, processing/analytics, sharing/release, retention, deletion/erasure). During extraction we annotated each DT occurrence with a primary phase in each frame and, where the paper explicitly covered a hand-off or cross-phase dependency, a secondary phase. We based placements on what the study actually implemented, evaluated, or measured rather than on aspirational statements. For example, when a study described a consent interface bound to controller logic and logs, we placed Transparency and Communication (DTT*) at collection and recorded a secondary placement in operation. We aggregated these DT annotations by dimension to produce a dimension-by-phase matrix for both frames. We then reviewed the underlying papers for the densest cells to confirm the mechanisms that produced those concentrations and identify systematic hand-off points. We used this matrix to order the per-dimension results and to interpret co-occurrence as structural couplings along a lifecycle rather than as mere co-mention.
Figure~\ref{fig:lifecycles} illustrates the placement of all thirteen dimensions across the two lifecycles and highlights the recurrent hand-offs between phases that emerged from the evidence.

Along the software development lifecycle, Modeling and Specification dominated the requirements and early design phases~\cite{alshammari_privacy_2018,gharib2021copri,de_chaves_privacy_2023}. Ontologies, goal/asset models, misuse/abuse cases and pattern/tactic catalogs specified privacy requirements, threats and design intents and several papers connected these artifacts to later enforcement (for example, generating access-control templates, policy rules, or architectural skeletons)~\cite{alshammari_privacy_2018,caiza_reusable_2019,gharib2021copri}. Governance and Accountability and Organizational Measures appeared early when authors translated regulatory provisions into user stories, backlogs, or review workflows and assigned roles and responsibilities; these early placements often persisted into deployment through audit logging, evidence capture and sign-off procedures~\cite{spiekermann_inside_2019,herwanto_leveraging_2024,iwaya_privacy_2024}. Data Minimization and Purpose Limitation typically entered during design, where authors decided data-flow boundaries (on-device vs service-side), aggregation granularity and purpose-binding mechanisms; designs that favored local processing or edge aggregation reduced downstream exposure and constrained subsequent implementation choices~\cite{benhamida_pyff_2021,nguyen_federated_2023,odema_privynas_2024}. Implementation concentrated Privacy-Enhancing Technologies: differential privacy mechanisms in pipelines and query engines; cryptographic protections (secure aggregation, homomorphic encryption, multiparty computation) or trusted execution for data‐at-use; and architectural PETs such as enclave-backed databases or contract-mediated access~\cite{mazmudar_cache_2022,marcolla_survey_2022,wu_privacy-preserved_2022,zhao_scenario-based_2024,yu_dop-sql_2024,yang_secudb_2024}. Transparency and Communication appeared in implementation when studies built consent widgets, policy explainers, or machine-readable policy components and connected them to controller logic~\cite{daoudagh_data_2021,gebauer_human---loop_2023,herwanto_leveraging_2024}. Verification and Testing and Privacy Metrics dominated the verification phase: papers benchmarked PETs, measured privacy–utility trade-offs, quantified robustness to attacks, validated banner behavior or DNS privacy at ecosystem scale and released tooling to support reproduction~\cite{wagner_technical_2019,jha_internet_2022,li_longitudinal_2023,mo_security_2024}. Deployment and operation concentrated Governance and Accountability, Organizational Measures, Data Subject Rights Management and Incident Response and Management: access control and provenance logs went live; rights workflows (access, erasure, portability, complaint) integrated with ticketing or self-service portals; incident channels and escalation paths attached to evidence sources (logs, attestations, provenance) for forensics and redress; and some systems enforced contractual or ledger-recorded policies at runtime~\cite{jones_profile_2020,zhang_privacy-by-design_2022,wu_privacy-preserved_2022,xia_towards_2024}. Lifelong Management and parts of Governance and Accountability dominated decommissioning: studies implemented retention schedules, temporal tables and erasure workflows; some designs reconciled erasure with immutable audit by recording proofs-of-action instead of content~\cite{pallas2024privacy,carmichael_personal_2024,yang_secudb_2024}. Across the Software Development Life Cycle (SDLC), we observed stable hand-offs~\cite{lo_systematic_2022,de_chaves_privacy_2023}. Requirements/models that produced artifacts (rules, patterns, components) influenced design and implementation without re-specification; PET implementations that emitted metrics and logs increased verifiability downstream; deployment choices (for example, enclaves or ledgers) changed how governance produced evidence; and lifecycle services (retention/erasure) required tight coupling to provenance to verify eligibility~\cite{wagner_technical_2019,marcolla_survey_2022,yang_secudb_2024,pallas2024privacy}.

Across the personal-data lifecycle, the matrix showed a complementary structure~\cite{jones_profile_2020,de_chaves_privacy_2023}. Collection concentrated Transparency and Communication and User-Centric: consent, notices, risk–benefit interfaces and personal data stores mediated the initial intake~\cite{barhamgi_user-centric_2018,jha_internet_2022,carmichael_personal_2024}. Papers that tied interfaces to controller logic and logs propagated user intent downstream; where interfaces remained decoupled, defaults reasserted control and later phases could not reconstruct original intent~\cite{daoudagh_data_2021,gebauer_human---loop_2023,herwanto_leveraging_2024}. Transfer and storage concentrated Governance and Accountability and PET deployments for data-at-rest and data-in-transit (for example, encrypted DNS, certificate validation and transport protections) and for data-at-use via TEEs; provenance and access-control configurations recorded the basis for later audits, redress, or rights handling~\cite{zhang_privacy-by-design_2022,li_longitudinal_2023,yang_secudb_2024}. Processing/analytics concentrated PETs and Privacy Metrics: differential privacy in batch and interactive systems; federated/distributed learning; obfuscation and attribute-inference resistance for visual/biometric/tabular data; and calibration or robustness measurements to select operating points~\cite{wagner_technical_2019,zhao_survey_2022,lo_systematic_2022,melzi_overview_2024,du_privategaze_2024}. Sharing/release combined Governance and Accountability with Organizational Measures and sometimes PETs: disclosure control in secure research environments, policy-driven release gates and verifiable logging or contract enforcement; studies that used immutable ledgers captured non-repudiation and enclave/attestation chains recorded trust roots~\cite{jones_profile_2020,zhang_privacy-by-design_2022,wu_privacy-preserved_2022,liang_identity_2024}. Retention brought Governance and Accountability together with Lifelong Management: authors encoded schedules and legal bases, sometimes with temporal tables and signed records to support audits~\cite{pallas2024privacy,carmichael_personal_2024}. Deletion/erasure concentrated Lifelong Management, Rights and Verification when authors attempted to validate effective erasure under constraints such as backups, distributed replicas and append-only logs; a frequent pattern replaced content with proofs while preserving accountability~\cite{wu_privacy-preserved_2022,pallas2024privacy,carmichael_personal_2024}. Two consistent cross-cuts traversed the data lifecycle. First, Minimization and Purpose Limitation acted at collection and processing by reducing or transforming data before sharing, thereby reducing reliance on downstream controls and changing which PETs were necessary~\cite{benhamida_pyff_2021,nguyen_federated_2023,odema_privynas_2024}. Second, Rights workflows traversed all phases: access and portability required joined-up storage and processing views; rectification required provenance and versioning; erasure required consistency across storage engines, indices and logs; complaint and redress required governance hooks and evidence~\cite{jones_profile_2020,daoudagh_data_2021,zhang_privacy-by-design_2022}. Domain lenses expressed these lifecycle couplings differently: healthcare emphasized retention, disclosure control and auditability; IoT and edge emphasized on-device processing and constrained sharing; web ecosystems emphasized banner behavior, tracking after consent and transport-layer privacy~\cite{jones_profile_2020,benhamida_pyff_2021,zhang_privacy-by-design_2022,jha_internet_2022,li_longitudinal_2023}. In all cases, placements depended on whether artifacts bound policy to execution. Where authors connected interfaces, models and rules to code and logs, the placements formed continuous chains; where artifacts stopped at documentation, placements fragmented and later phases could not rely on upstream intent~\cite{daoudagh_data_2021,zhang_privacy-by-design_2022,herwanto_leveraging_2024}.

\subsubsection{Application Domains}\label{sec:doms}
\begin{figure}[!h]
    \centering
    \includegraphics[width=0.60\linewidth]{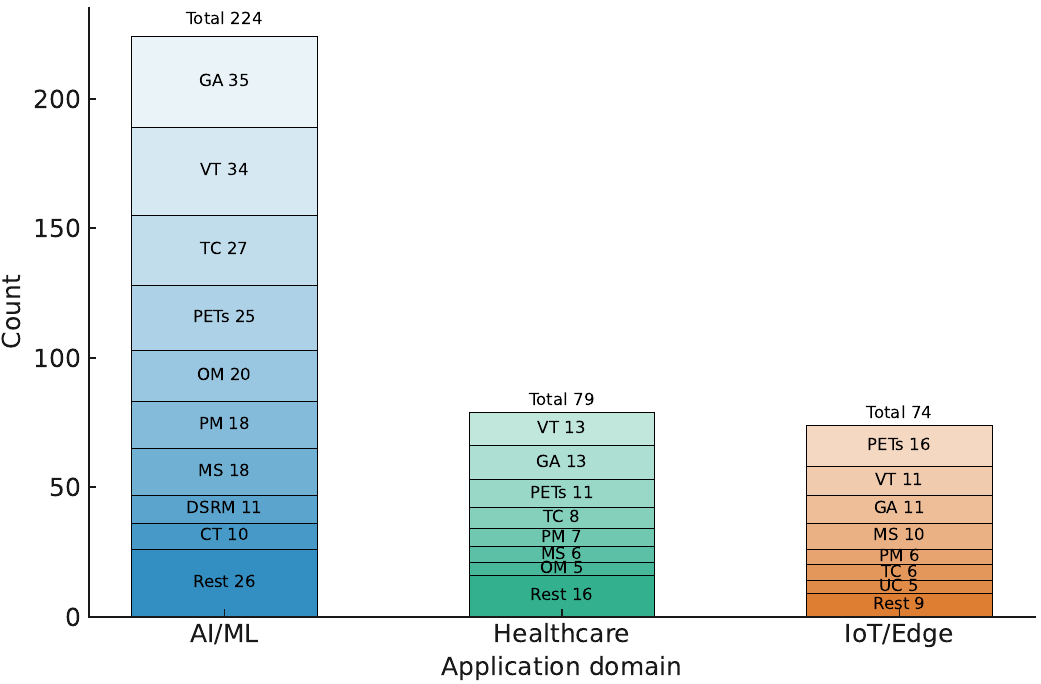}
    \caption{Stacked counts of dimension occurrences per application domain. Bars show AI/ML, Healthcare and IoT/Edge; segments display major dimensions within each domain, with low-frequency dimensions grouped as “Rest.” Counts aggregate primary and secondary assignments. Abbreviations: GA = Governance \& Accountability; VT = Verification \& Testing; PETs = Privacy-Enhancing Technologies; TC = Transparency \& Communication; MS = Modeling \& Specification; PM = Privacy Metrics; OM = Organizational Measures; DSRM = Data Subject Rights Management; CT = Culture \& Training; UC = User-Centric; DMPL = Data Minimization \& Purpose Limitation; IRM = Incident Response \& Management; LM = Lifelong Management.}
    \label{fig:domain_stacked}
\end{figure}

\begin{figure}[!h]
    \centering
    \includegraphics[width=0.60\linewidth]{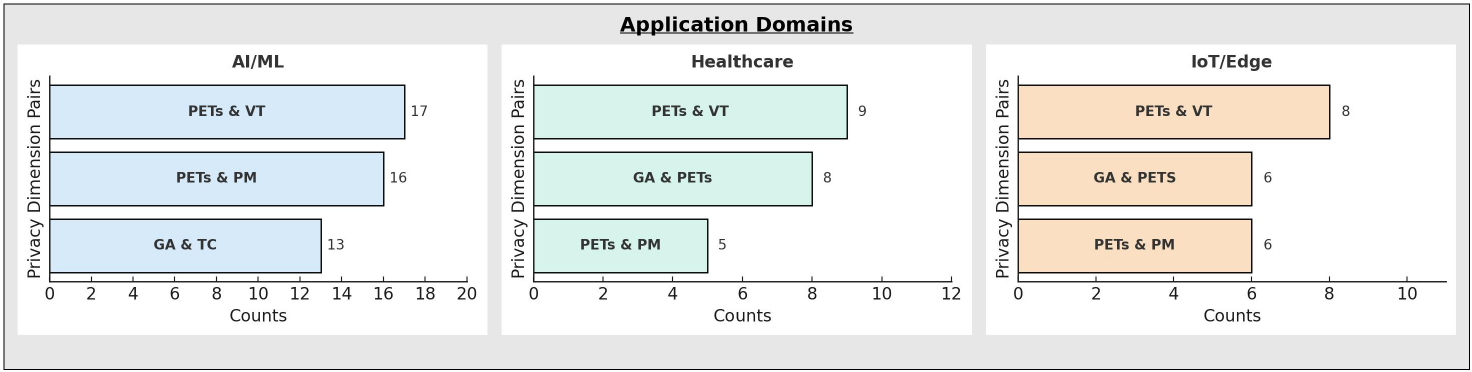}
    \caption{Top co-occurring dimension pairs by domain (undirected sums). Bars show the three largest pair totals for Healthcare, IoT/Edge and AI/ML; counts aggregate primary and secondary assignments and colors indicate domain. Abbreviations: PETs = Privacy-Enhancing Technologies; VT = Verification \& Testing; PM = Privacy Metrics;GA = Governance \& Accountability; TC = Transparency \& Communication; }
    \label{fig:domain_pairs}
\end{figure}

We examined how evidence is distributed across application domains by aggregating DT assignments and inspecting how each domain places evidence across the thirteen dimensions and the two lifecycle frames. Figure~\ref{fig:domain_stacked} reports stacked counts of dimension occurrences per domain (AI/ML, Healthcare, IoT/Edge), aggregating primary and secondary assignments. Figure~\ref{fig:domain_pairs} reports the strongest co-occurrence pairs per domain by showing the top three pair sums for Healthcare, IoT/Edge and AI/ML, again aggregating primary and secondary assignments.

Healthcare concentrated evidence in Governance \& Accountability together with Verification \& Testing and Privacy-Enhancing Technologies (PETs); Transparency \& Communication and Privacy Metrics also appeared, while Organizational Measures remained less prominent and Data Subject Rights Management and Lifelong Management appeared at lower levels~\cite{jones_profile_2020,ermakova_security_2020,zhang_privacy-by-design_2022,carmichael_personal_2024}. Studies translated regulatory provisions into operational controls, tied access decisions to provenance and audit logs and implemented disclosure control in secure research environments and data trusts~\cite{jones_profile_2020,zhang_privacy-by-design_2022,herwanto_leveraging_2024}. Modeling \& Specification appeared early through ontologies, goal/asset models and rule sets that generated access-control templates or policy checks and these artifacts persisted into deployment through signed decisions and release justifications~\cite{alshammari_privacy_2018,gharib2021copri,de_chaves_privacy_2023}. PET adoption emphasized enclave-backed processing, secure aggregation for multi-institution analytics and, to a lesser extent, Differential Privacy for interactive or batch analysis~\cite{lo_systematic_2022,nguyen_federated_2023,zukaib_blockchain_2023,yang_secudb_2024}. Authors often combined PETs with approvals, contracts and redaction pipelines so technical and procedural safeguards jointly determined release outcomes~\cite{jones_profile_2020,zhang_privacy-by-design_2022,jordan_selecting_2022}. Transparency \& Communication and User-Centric mechanisms appeared where consent management, patient portals, or risk-benefit interfaces mediated access and sharing and their effectiveness depended on binding interfaces to controller logic and logs~\cite{carmichael_personal_2024,de_chaves_user-centred_2025}. Verification \& Testing appeared as ecosystem audits for policy conformance in secure environments and as controlled evaluations of PETs against clinical or administrative tasks~\cite{wagner_technical_2019,jha_internet_2022,li_longitudinal_2023}. Privacy Metrics addressed disclosure risk and perturbation effects but remained heterogeneous, limiting comparability~\cite{wagner_technical_2019,xia_towards_2024}. Lifecycle placement formed continuous chains from requirements and design (models, roles, safeguards) to deployment and operation (evidence capture, release gates, rights workflows) and to decommissioning (retention schedules, erasure)~\cite{jones_profile_2020,zhang_privacy-by-design_2022,pallas2024privacy}. This chain formed a triad around Governance \& Accountability with Organizational Measures and Data Subject Rights Management, attached Lifelong Management through temporal records and attached PETs and Privacy Metrics when analytics extended beyond a single controller or required measurable guarantees~\cite{wagner_technical_2019,jones_profile_2020,zhang_privacy-by-design_2022}.

IoT, edge, robotics, vehicular networks and smart-city settings emphasized PETs with Verification \& Testing and Privacy Metrics; Minimization \& Purpose Limitation appeared without dominating and User-Centric controls varied by context~\cite{benhamida_pyff_2021,alhirabi_security_2021,odema_privynas_2024}. Studies moved computation to devices or edge gateways, limited export and transformed data before sharing (aggregation, feature masking, selective disclosure)~\cite{benhamida_pyff_2021,odema_privynas_2024}. Federated and distributed learning dominated PET adoption, with secure aggregation and related cryptographic mechanisms as default enablers; when systems required provenance or trust across nodes, designs added blockchain-backed audit and access or hardware attestation~\cite{lo_systematic_2022,nguyen_federated_2023,weng_faster_2024,liang_identity_2024,yang_secudb_2024,alghuried_blockchain_2025}. Transparency \& Communication appeared through dashboards, policy explainers and consent prompts in urban platforms or assisted-living contexts and depended on binding interface choices to enforcement points~\cite{barhamgi_user-centric_2018,carboni_privacy_2023,wilkowska_interdisciplinary_2023}. Verification \& Testing remained uneven: some papers reported end-to-end experiments with real hardware, traces, or city datasets, while others relied on simulations with limited external validity~\cite{hammoudeh_service-oriented_2021,nguyen_federated_2023,solis_exploring_2024}. Privacy Metrics concentrated on privacy--utility curves, communication or energy costs in federated settings and robustness to poisoning, byzantine behavior, or attribute inference; incident-oriented metrics and redress reporting appeared rarely~\cite{wagner_technical_2019,lo_systematic_2022,mo_security_2024}. Organizational Measures and Governance \& Accountability emerged when cross-agency or multi-stakeholder settings required roles, review boards, or contracts, but several studies described the governance conceptually without executable controls, thereby reducing alignment with Data Subject Rights Management and Transparency \& Communication~\cite{spiekermann_inside_2019,alhirabi_security_2021}. Lifecycle placement differed from healthcare: collection and processing concentrated PETs (and, when present, Minimization \& Purpose Limitation and User-Centric) on device or gateway; deployment centered on trust roots (attestation, keys) and coordination infrastructure; operation emphasized drift and updates, with limited evidence on decommissioning and erasure across heterogeneous nodes~\cite{benhamida_pyff_2021,odema_privynas_2024,yang_secudb_2024}. Dominant co-occurrences therefore involved PETs with Verification \& Testing and Privacy Metrics; governance co-occurred where platforms integrated data marketplaces, consent services, or contract enforcement~\cite{daoudagh_data_2021,lo_systematic_2022,nguyen_federated_2023,weng_faster_2024,solis_exploring_2024}.

Web and mobile ecosystems, large-scale measurement and AI/ML pipelines showed different emphases~\cite{jha_internet_2022,li_longitudinal_2023}. Web measurement paired Transparency \& Communication with Verification \& Testing by auditing banners, post-consent tracking and transport-layer privacy; studies built crawlers and harnesses, reported reproducible methods and published artifacts, with Governance \& Accountability relevance where results informed compliance assessment or enforcement~\cite{jha_internet_2022,li_longitudinal_2023,wang_as_2024}. AI/ML pipelines paired PETs with Privacy Metrics and used Verification \& Testing for attacks, transferability tests and stress scenarios; Governance \& Accountability co-occurred when studies tied lineage to logging and audit or used contracts and policies to control training-data access~\cite{liu_privacy_2021,zhang_privacy-by-design_2022,mo_security_2024,carmichael_personal_2024}. User-Centric mechanisms appeared sporadically as interfaces that explained risks, adjustable parameters, or consent and depended on pipeline enforcement of user constraints downstream~\cite{barhamgi_user-centric_2018,wilkowska_interdisciplinary_2023,de_chaves_user-centred_2025}. Blockchain-backed systems concentrated Governance \& Accountability with PETs via smart contracts and ledgers for access and provenance and addressed tensions between immutability and erasure through off-chain storage, redactable logs, or proofs-of-action~\cite{wu_privacy-preserved_2022,liang_identity_2024,pallas2024privacy,alghuried_blockchain_2025}. Lifelong Management appeared when systems implemented retention schedules, versioned tables, or temporal records; Verification \& Testing rarely exercised deletion at scale, keeping this intersection weak~\cite{pallas2024privacy,yang_secudb_2024}. Across domains, Modeling \& Specification bridged domains when authors converted policies, user stories, or threat models into code, rules, or components that influenced tests, deployments and audits~\cite{alshammari_privacy_2018,gharib2021copri,de_chaves_privacy_2023,herwanto_leveraging_2024}. Organizational Measures and Culture \& Training attached when teams integrated privacy tasks into backlogs, change-management gates, or role definitions; regulated settings strengthened this attachment, while mechanism-centric experiments weakened it~\cite{spiekermann_inside_2019,iwaya_organizational_2022,elkourdi_exploring_2024}. Overall, application domains did not alter the dimension structure; they shifted weights and typical entry points. Regulated contexts began with Governance \& Accountability and Organizational Measures and then extended to PETs and Verification \& Testing when analytics or data sharing required measurable guarantees. Resource-constrained or distributed contexts started with PETs (and Minimization \& Purpose Limitation where applicable) and propagated into Governance \& Accountability when coordination, provenance, or accountability became necessary. Web-scale evaluations started with Verification \& Testing and Transparency \& Communication and propagated into Governance \& Accountability when evidence informed policy or enforcement~\cite{jones_profile_2020,benhamida_pyff_2021,zhang_privacy-by-design_2022,lo_systematic_2022,jha_internet_2022,li_longitudinal_2023,nguyen_federated_2023}.

\subsection{Answers to Research Questions}
We formulated reporting research questions to reflect the evidence synthesized in Section~\ref{sec:per_dimension} and Section~\ref{sec:cross_cut}. For each RQ, we state a one-sentence answer and then explain how the corpus supported that answer with pointers to the relevant findings.

\noindent\textbf{Which privacy engineering dimensions emerged from the corpus and what do they cover?}

\vspace{2pt}
\noindent
\begin{tikzpicture}[baseline=(box.south)]
  \node[
    draw=black!40,
    fill=gray!15,
    rounded corners=2pt,
    inner sep=4pt,           
    text width=\columnwidth,
    align=left
  ] (box)
  {%
Thirteen dimensions emerged from corpus-wide constant comparison, covering mechanisms (PETs), assurance (metrics and verification), policy communication and rights, governance and organizational practice, user control with minimization, modeling and specification, incident handling and lifecycle operations.
  };
\end{tikzpicture}

We derived the dimensions by clustering Descriptive Themes (DTs) after stabilizing the DT catalog. Section~\ref{sec:per_dimension} reported one results block per dimension with scope, contributing DTs, synthesis and numbered findings. Mechanism-oriented evidence concentrated in Privacy-Enhancing Technologies, where studies operationalized differential privacy, cryptography and trusted execution, obfuscation for visual/biometric data, federated or distributed learning and blockchain-backed enforcement. Assurance evidence appeared in Privacy Metrics and in Verification and Testing, where authors calibrated privacy–utility trade-offs, measured robustness to attacks and benchmarked or audited systems and ecosystems. Policy-facing evidence appeared in Transparency and Communication and in Data Subject Rights Management, where interfaces and machine-readable artifacts mediated notices, consent and rights workflows tied to enforcement points. Governance and organizational evidence appeared in Governance and Accountability, Organizational Measures and Culture and Training; these results showed how roles, processes, audits and training connected policies to executable controls and evidence. User-facing technical and architectural choices appeared in User-Centric and in Minimization and Purpose Limitation, where designs reduced data exposure and offered control through local processing or mediated sharing. Operational evidence appeared in Incident Response and Management and in Lifelong Management, where systems linked incidents, retention and erasure to provenance, logging and access control. The Modeling and Specification dimension connected intentions to implementation through ontologies, goal/asset models, user stories, tactics and rules that generated or constrained configurations. Together, these results blocks covered the full span from mechanism and measurement to communication, governance, user control and operations.

\noindent\textbf{How did the dimensions interact across studies?}

\smallskip
\noindent
\begin{tikzpicture}[baseline=(box.south)]
  \node[
    draw=black!40,
    fill=gray!15,
    rounded corners=2pt,
    inner sep=4pt,
    text width=\columnwidth,
    align=left
  ] (box)
  {%
Dimensions co-occurred in two stable cores and their nearest pairs: PETs with Verification \& Testing and Privacy Metrics; Governance \& Accountability with Transparency \& Communication and Organizational Measures. Modeling \& Specification served as Specification Mediation, linking the two cores while remaining conceptually independent of both.
  };
\end{tikzpicture}
\par\smallskip

In Section~\ref{sec:cooccur}, we analyzed the directional primary--secondary matrix. The mechanism--evaluation core joined PETs with Verification \& Testing and Privacy Metrics, with User--Centric and Data Minimization \& Purpose Limitation attached to this side. The governance--operations core joined Governance \& Accountability with Transparency \& Communication and Organizational Measures, with Data Subject Rights Management, Incident Response \& Management, Culture \& Training and Lifelong Management attached to this side. Modeling \& Specification remained outside both cores and served as Specification Mediation, linking specification artifacts to verification on the mechanism side and to governance and enforcement on the operations side.

\noindent\textbf{Where in the software and data lifecycles did the dimensions concentrate and what hand-offs did the corpus support?}

\smallskip
\noindent
\begin{tikzpicture}[baseline=(box.south)]
  \node[
    draw=black!40,
    fill=gray!15,
    rounded corners=2pt,
    inner sep=4pt,
    text width=\columnwidth,
    align=left
  ] (box)
  {%
Dimensions concentrated at requirements/design (Modeling \& Specification; Governance \& Accountability; Organizational Measures; Data Minimization \& Purpose Limitation), at implementation/verification (Privacy-Enhancing Technologies; Verification \& Testing; Privacy Metrics; Transparency \& Communication) and at operation/decommissioning (Governance \& Accountability; Organizational Measures; Data Subject Rights Management; Incident Response \& Management; Lifelong Management). Data Minimization \& Purpose Limitation acted at collection and processing, Rights workflows traversed all phases and hand-offs relied on models and rules, PET outputs and metrics and provenance-linked configurations.%
  };
\end{tikzpicture}
\par\smallskip

In Section~\ref{sec:lifecycle}, we mapped Descriptive Themes to phases in the software development lifecycle and in the personal-data lifecycle and then aggregated them at the dimension level. Throughout the software lifecycle, requirements and design focused on Modeling \& Specification, Governance \& Accountability, Organizational Measures and Data Minimization \& Purpose Limitation, where the authors introduced models, roles, review workflows and minimization constraints. Implementation concentrated on Privacy-Enhancing Technologies and the construction of Transparency \& Communication artifacts bound to controller logic, while verification concentrated on Verification \& Testing and Privacy Metrics through benchmarks, robustness tests and ecosystem-scale audits. Deployment and operation concentrated Governance \& Accountability, Organizational Measures, Data Subject Rights Management and Incident Response \& Management as systems executed controls, captured evidence and handled requests and incidents; decommissioning concentrated Lifelong Management together with parts of Governance \& Accountability through retention schedules, temporal tables and erasure workflows. Across the personal-data lifecycle, collection concentrated Transparency \& Communication and User-Centric mechanisms, with Data Minimization \& Purpose Limitation already acting at collection and processing by reducing or transforming data before sharing. Transfer and storage combined Governance \& Accountability with Privacy-Enhancing Technologies, while analytics concentrated on Privacy-Enhancing Technologies and Privacy Metrics and continued to involve Data Minimization \& Purpose Limitation. Sharing and release combined Governance \& Accountability, Organizational Measures and sometimes Privacy-Enhancing Technologies; retention combined Governance \& Accountability and Lifelong Management; deletion and erasure combined Lifelong Management, Data Subject Rights Management and Verification \& Testing when studies exercised erasure under constraints such as backups or distributed replicas. Rights workflows traversed all phases and models, rules, PET outputs, metrics, logs, deployment choices and provenance records connected phases so that upstream artifacts constrained and evidenced downstream behaviour. These placements and hand-offs aligned with the per-dimension findings and the co-occurrence structure in Sections~\ref {sec:per_dimension} and~\ref {sec:cooccur}.

\noindent\textbf{How did domains modulate the prevalence and coupling of dimensions without changing the overall structure?}

\vspace{2pt}
\noindent
\begin{tikzpicture}[baseline=(box.south)]
  \node[
    draw=black!40,
    fill=gray!15,
    rounded corners=2pt,
    inner sep=4pt,
    text width=\columnwidth,
    align=left
  ] (box)
  {%
Domains shifted emphasis but preserved structure:
healthcare weighted Governance \& Accountability with Verification \& Testing and Privacy-Enhancing Technologies;
IoT/edge weighted PETs with Verification \& Testing and Privacy Metrics at device/edge;
web emphasized Transparency \& Communication with Verification \& Testing, while AI/ML emphasized Privacy-Enhancing Technologies with Privacy Metrics.
  };
\end{tikzpicture}

Section~\ref{sec:doms} contrasted healthcare, IoT/edge and robotics and web/AI ecosystems. Healthcare concentrated Governance \& Accountability together with Verification \& Testing and Privacy-Enhancing Technologies, often under secure research environments and data trusts; Modeling \& Specification generated enforceable artifacts and Privacy-Enhancing Technologies appeared in enclave-backed processing, secure aggregation and, selectively, Differential Privacy. IoT/edge and vehicular contexts emphasized Privacy-Enhancing Technologies with Verification \& Testing and Privacy Metrics—especially federated or distributed learning and secure aggregation; when systems required provenance or cross-actor trust, Governance \& Accountability co-occurred through ledgers or attestation. Web-scale measurement emphasized Transparency \& Communication with Verification \& Testing by auditing banners, tracking after consent and transport-layer privacy. In AI/ML pipelines, Privacy-Enhancing Technologies and Privacy Metrics dominated through calibration, ablation and robustness tests. Across all domains, Modeling \& Specification and Organizational Measures acted as bridges when authors converted policies and roles into rules, components or procedures. Incident Response \& Management attached where response and redress used logged evidence. These shifts changed weights and entry points but not the dimension set or the main interaction pairs and triads.

\section{Discussion}\label{sec:discussion}
This section consolidates the synthesis results and explains the specific research contributions of the study. First, we summarize the principal findings across the corpus, highlighting patterns, underrepresented areas and cross-cutting regularities. Second, we discuss the added value of this review—what is new and why it matters for privacy engineering research and practice. Finally, we demonstrate how are findinds align with the GDPR and ISO/IEC 27701:2025 international standard for establishing a Privacy Information Management System (PIMS).

\subsection{Summary of Findings from the Systematic Literature Review}
Across 90 primary studies, the synthesis showed a consistent two–core structure. Mechanism-centric work clustered around Privacy-Enhancing Technologies (PETs) with systematic use of Privacy Metrics and Verification \& Testing to calibrate errors, robustness and performance; policy-centric work clustered around Governance \& Accountability with Transparency \& Communication and Organizational Measures, where interfaces and machine-readable artifacts linked policies to executable controls (access control, provenance, logs). Modeling \& Specification and Organizational Measures acted as bridges that translated intentions and processes into code, configuration and audit evidence. Domain lenses reweighted this structure without changing it: healthcare emphasized Governance \& Accountability together with Verification \& Testing and PETs; IoT/edge emphasized PETs with Verification \& Testing and Privacy Metrics at device/edge; web-scale measurement emphasized Transparency \& Communication with Verification \& Testing, while AI/ML pipelines emphasized PETs with Privacy Metrics. These regularities were corroborated by the co-occurrence analysis and by lifecycle placements, where we observed stable hand-offs from specification artifacts to implementation, from PET outputs to verification evidence and from deployment choices (e.g., enclaves, ledgers) to governance evidence.

In Appendix~\ref{sec:studies_to_dimensions}, Table~\ref{tab:pe-dim-primary2} shows that the literature contributed to every dimension. In combination with the descriptive analysis in Figure~\ref{fig:dimension-counts}, these analyses demonstrate that some dimensions remained underrepresented as \emph{primary} foci: Incident Response \& Management, Lifelong Management and Minimization \& Purpose Limitation appeared predominantly as secondary assignments rather than as the main focus of the accepted studies. These gaps align with our qualitative reading: incident/redress procedures were typically treated within broader accountability discussions rather than evaluated end-to-end; lifecycle controls (retention/erasure) were usually specified architecturally or procedurally but rarely exercised at scale; and minimization was widely assumed (e.g., locality, selective disclosure) but less often validated as the \emph{primary} goal outside federated/edge settings. In addition, Verification \& Testing and User-Centric appeared robust in some domains (e.g., web measurements; HCI/assisted contexts) but thinner in others (e.g., industrial modernization), reinforcing the need for cross-domain evaluation practice and context-aware user control. These observations summarize where the field currently concentrates and where evidence is sparser, setting a focused agenda for subsequent work.

\subsection{Contributions and Implications}
Methodologically, the review delivers an auditable pathway from study-level evidence to field-level structure. Concretely: (i) it translates qualitative findings into Descriptive Themes (DTs) via constant comparison and inter-rater assessment; (ii) it derives thirteen \emph{emergent} dimensions by clustering DTs post hoc (no a priori taxonomy); and (iii) it models relations through a directional co-occurrence matrix and a dual lifecycle map (software and personal-data lifecycles). This yields a reproducible scaffold—codebook, DT catalog, paper–DT–dimension matrix, change log and mapping policy—that other researchers can re-use to update or specialize the synthesis. Substantively, the synthesis contributes three actionable results. First, it explains the field’s structure—two dense cores (PETs–Metrics–Verification and Governance–Transparency–Organizational Measures), a specification mediator (Modeling \& Specification) and domain-specific emphases—showing how credible privacy and policy claims depend on traceable couplings (mechanisms, measurement, tests; policies/roles, interfaces, enforcement, evidence). Second, it identifies underrepresented dimensions as primary foci (incident/redress, lifecycle, minimization) and specifies where evaluation practice is currently thin (longitudinal accountability in operation; deletion/retention at scale; end-to-end incident exercises). Third, it connects results to engineering practice through lifecycle placement: which artifacts should exist when, which evidence they must emit and how hand-offs (e.g., models to rules/tests; PETs to metrics/logs; deployment choices to governance evidence) reduce drift between declared policy and runtime behavior. Together, these contributions move privacy engineering beyond scattered exemplars toward a coherent, testable map that practitioners can use to plan artifacts and evidence and that researchers can use to target gaps with high leverage such as incident/redress benchmarking and data minimization-first designs.

\subsection{Results Alignment with the GDPR and ISO IEC 27701}\label{sec:gdpr_iso_alignment}
Applying thematic synthesis to the accepted studies, we identified thirteen engineering dimensions for privacy engineering. These results align with the GDPR and ISO/IEC 27701, which organizations use as a starting point for establishing and demonstrating compliance. ISO/IEC 27701 frames these alignments through Clauses 5 and 7 and 8 and Annexes A and B provide control content and implementation guidance. Privacy-Enhancing Technologies align with GDPR Article 5(1)(f) (integrity and confidentiality), Article 32 (security of processing) and Article 25 (Data Protection by Design and by Default); implementations that protect data at rest, in transit and in use and expose verifiable controls also support controller responsibility under Article 24. Verification \& Testing align with GDPR Article 5(2) (accountability), Article 24 (controller responsibility) and Article 32(1)(d) (regular testing and evaluation of measures); benchmarks, reproducible checks and attack-resilience evaluations operationalize these duties. Transparency \& Communication align with GDPR Article 5(1)(a) (lawfulness, fairness, transparency), Articles 12--14 (information duties), Article 7 (consent) and, where applicable, Article 22 (safeguards for automated decision-making); interfaces, notices and machine-readable policy artifacts implement these obligations and ISO/IEC 27701 reflects them through controller obligations in Clause 7 and Annexes A and B. Modeling \& Specification align with GDPR Article 25 (design and default), Article 30 (records of processing activities), Article 35 (data protection impact assessments) and Article 6 (lawfulness and purpose specification); requirements models, policy rules and architectural decision records provide traceable artifacts and ISO/IEC 27701 supports this through documented information and change control in Clause 5 and controller-side specification of purposes and lawful bases in Clause 7, with Annexes A and B. Governance \& Accountability align with GDPR Article 5(2) (accountability), Article 24 (controller duties), Articles 26--28 (joint-controller and processor arrangements), Article 30 (record-keeping) and, where applicable, Articles 44--49 (cross-border transfer rules); role definitions, contracts, provenance and audit logging implement these requirements. Organizational Measures align with GDPR Articles 24 (appropriate technical and organizational measures), 25 (design and default), 35 (DPIAs) and 37--39 (Data Protection Officer and role competence). Backlogs, operating procedures, approvals and review boards implement these duties.

Privacy Metrics align with GDPR Article 5(2) (accountability), Article 24 (controller responsibility) and Articles 25 and 32 where safeguards require calibration; quantitative objectives and monitoring enable demonstrable conformity and ongoing risk reduction. Data Subject Rights Management align with GDPR Articles 12--23 (access, rectification, erasure, restriction, portability, objection), Article 7(3) (consent withdrawal) and Article 19 (notification duties); rights intake, verification, tracking and provenance-backed fulfillment implement these obligations end to end and ISO/IEC 27701 mirrors them through controller obligations in Clause 7 and Annexes A and B.
Culture \& Training align with GDPR Article 24 (organizational measures), Article 32(4) (processing under instruction) and Articles 37--39 (DPO tasks, including staff awareness and advice); programs embedded into development and operations support these obligations. User-Centric align with GDPR Article 5(1)(a) (fairness and transparency), Article 7 (consent and withdrawal), Articles 12--14 (information duties), Article 21 (right to object) and, where applicable, Article 20 (portability); interfaces that surface choices and bind them to enforcement support these articles and ISO/IEC 27701 reflects them through Clause 7 and Annexes A and B. Data Minimization \& Purpose Limitation align with GDPR Article 5(1)(b) (purpose limitation), Article 5(1)(c) (data minimization) and Article 25 (design and default), because locality, selective disclosure and purpose-scoped access implement these constraints; ISO/IEC 27701 requires explicit purpose specification and limitation for controllers in Clause 7 and supports minimization and retention choices through Annexes A and B. Incident Response \& Management align with GDPR Articles 33--34 (breach notification) and Article 32 (security of processing), because engineered logging, forensics and notification procedures enable timely reporting and remediation; ISO/IEC 27701 embeds incident handling in operation, performance evaluation and improvement under Clause 5 and provides controller and processor guidance in Annexes A and B. Lifelong Management align with GDPR Article 5(1)(e) (storage limitation), Article 17 (erasure), Article 18 (restriction), Article 30 (record-keeping) and Articles 13--14 (duties to inform about retention); retention schedules, deletion services and proofs of action implement these articles across the data lifecycle and ISO/IEC 27701 supports the same behavior through design and lifecycle controls in Clause 5 and Clause 7, with retention and deletion guidance in Annexes A and B.

\section{THREATS TO VALIDITY}\label{sec:threads}

\subsection{Internal Validity.} We executed the review protocol systematically: we piloted and froze the engine–specific queries, logged exact queries and result counts and then screened titles/abstracts and full texts against pre–specified criteria with a pilot round and reconciliation before full execution; we preserved verbatim anchors in the extraction, kept an audit/change log and maintained a screening record to support traceability and replication. Two reviewers calibrated decisions on a balanced subset, applied constant comparison during open coding and ran inter–rater spot checks with reconciliation meetings; we declared saturation when additional papers no longer yielded new labels. These safeguards followed the stepwise guidance for thematic synthesis in SE (credibility, confirmability, dependability) and mirrored good practice in recent SLRs. At the same time, internal rigor could be further strengthened by expanding duplicate screening and duplicate coding beyond pilots and by adding formal sensitivity checks to inclusion decisions and theme boundaries.

\subsection{External Validity.} We searched the major digital libraries focusing on software engineering aspects for privacy (IEEE Xplore, ACM DL, Scopus, PubMed) and complemented them with snowballing, then synthesized evidence across multiple application domains; this provided broad coverage of published research but still concentrated the corpus on scholarly venues. As a result, our findings generalized across the included domains and study types, yet they may underrepresent practitioner experience and fast–moving industrial practice, which are reported only in the grey literature. As a result, new research directions arise, particularly focusing on methodologies and tools within the industrial spectrum which could provide empirical evidence for the under-represented dimensions of privacy engineering as identified in our study.

\subsection{Construct Validity.} We applied thematic synthesis with immersion, inductive open coding, reciprocal translation to descriptive themes and abstraction to higher–order structures. We recorded credibility/confirmability/dependability checks (coverage/saturation, audit trail, inter–rater calibration). These choices increased the fit between evidence and constructs while reducing the risk of forcing data into predefined categories. Still, construct alignment could benefit from additional replication artifacts that quantify coder agreement over larger samples, formal refutational checks per theme and targeted practitioner feedback sessions that probe whether the engineered dimensions and lifecycle placements match operational usage.

\section{Conclusion}\label{sec:conclusion}
We examined privacy engineering as it is practiced and evaluated in the software engineering literature and we translated the resulting evidence into a coherent view of the field. We applied thematic synthesis to 90 primary studies from 2018--2025, moved from open codes to cross-paper Descriptive Themes and then grouped these themes into thirteen higher-order dimensions. We modeled relations among dimensions using a directional co-occurrence matrix and mapped dimension placements to both the software development lifecycle and the personal data lifecycle. The results indicated two dense cores: one around Privacy-Enhancing Technologies with Privacy Metrics and Verification \& Testing and one around Governance \& Accountability with Transparency \& Communication and Organizational Measures. Modeling \& Specification acted as a specification mediator between these cores. Domain lenses shifted emphasis without changing the structure: healthcare weighted Governance \& Accountability with Verification \& Testing and Privacy-Enhancing Technologies; IoT/edge weighted Privacy-Enhancing Technologies with Verification \& Testing and Privacy Metrics at device/edge; web-scale measurement weighted Transparency \& Communication with Verification \& Testing; AI/ML pipelines weighted Privacy-Enhancing Technologies with Privacy Metrics. Across lifecycles, we observed consistent hand-offs: specification artifacts informed implementation, Privacy-Enhancing Technologies emitted measurement that supported verification and deployment choices (e.g., enclaves and ledgers) produced governance evidence. We also identified dimensions that appeared less frequently as primary foci, namely Incident Response \& Management, Lifelong Management and Data Minimization \& Purpose Limitation, suggesting areas where evaluation and benchmarking can mature further. Beyond these results, we provide a replication-ready scaffold (codebook, Descriptive Theme catalog, paper-theme-dimension matrix and change log) that supports updates and specializations as new evidence accumulates.

These observations carried practical and research implications. Section~\ref{sec:reporting} showed a trend toward emphasizing some dimensions more than others, with primary attention on Privacy-Enhancing Technologies and secondary attention on Governance \& Accountability and Verification \& Testing. This pattern indicated that much prior work addressed privacy as a system property through mechanisms (Privacy-Enhancing Technologies), organizational control and oversight (Governance \& Accountability) and procedures for verifying and testing privacy (Verification \& Testing). This focus remained important given the need for explicit privacy treatment in Large Language Models (LLMs) and other data-intensive applications. However, awareness alone did not demonstrate compliance. For example, a state-of-the-art federated learning method paired with differential privacy to achieve precise privacy budget allocation did not, by itself, ensure GDPR compliance. A system still needed engineered methods that demonstrated purpose-bounded processing, data minimization relative to agreed purposes, mechanisms for data subjects to access their data and an engineered approach to lifelong privacy management with auditable evidence. The dimensions that appeared less frequently as primary contributions in our dataset therefore identified concrete areas where future work can add value, including Data Minimization \& Purpose Limitation, Incident Response \& Management, Lifelong Management, Data Subject Rights Management, Privacy Metrics and privacy-first approaches for Verification \& Testing. For scholars, we identified three directions. First, scholars can design reproducible, end-to-end evaluations for incident handling and redress that combine governance interfaces, operational logs and measurable outcomes, enabling comparisons of preparedness and accountability across systems and domains. Second, scholars can establish longitudinal protocols for lifecycle operations at scale (retention and deletion with verifiable proofs) and report failure modes and operational costs alongside privacy guarantees, aligning mechanism design with deployment conditions. Third, scholars can treat data minimization as a primary goal outside federated and edge settings and introduce task-relevant metrics and baselines that test whether data and model choices reduced collection, exposure and linkage risk while preserving utility. Our scaffold supports these efforts by preserving traceability from segments to themes and by stabilizing identifiers, enabling cumulative synthesis and controlled taxonomy growth over time. For practitioners, the map provides a planning aid: teams can place artifacts along both lifecycles early, specify which evidence each artifact must emit (metrics, tests, logs) and tie governance decisions to executable controls and their provenance so that policies, interfaces, enforcement points and evidence remain coupled through delivery and operation. Teams can adopt recurring couplings supported by the corpus, such as mechanisms with measurement and verification and governance with transparency artifacts and logging and schedule periodic scenario-based exercises that use the same instrumentation to assess incident handling, redress and erasure at scale. Looking ahead, we expect regulation, platform architectures and workloads to evolve and our synthesis provides both a baseline and an auditable update path for future work to extend coverage, refine constructs and strengthen comparability across mechanism, verification and governance layers.

\bibliographystyle{ACM-Reference-Format}
\bibliography{sample-base}

\appendix

\section{Review Protocol: Search, Selection and Extraction Tables}\label{sec:protocol_tables}

Table~\ref{tab:facet-terms} enumerates the four query facets used in Eq~\eqref{eq:core-query} and their representative term sets, which we applied (with library-specific field scoping) to retrieve records that combined privacy-engineering terminology with a software-engineering context and either implementation/evaluation content or governance/operations content.

\begin{table}[!h]
\caption{Search facets and representative terms.}
\label{tab:facet-terms}
\centering
\footnotesize
\setlength{\tabcolsep}{3pt}
\begin{tabularx}{\columnwidth}{@{}l l >{\raggedright\arraybackslash}X@{}}
\hline
\textbf{Facet} & \textbf{Scope} & \textbf{Representative terms} \\
\hline
\textbf{P} & Privacy engineering & ''privacy engineering'', ''privacy by design'', ''data protection by design'', ''privacy preserving'', ''data minimization'', ''transparency'', ''data subject rights'', ''accountability'', ''GDPR'', ''CCPA'' \\
\textbf{SE} & Software engineering & ''software engineering'', ''software'', ''software system'', ''requirements engineering'', ''software architecture'', ''software design'', SDLC, DevOps \\
\textbf{E} & Empirical / evidence & ''implementation'', ''evaluation'', ''literature review'', ''case study'', ''benchmark'',  \\
\textbf{G} & Governance / operations & ''consent'', ''access control'', ''audit log'', ''DPIA'' \\
\hline
\end{tabularx}
\end{table}

Table~\ref{tab:incl-excl-privacy} specifies the eligibility criteria that guided title/abstract screening and full-text assessment and that supported the logged exclusion reasons referenced in Section~\ref{execution}.

\begin{table}[!h]
\centering
\caption{Inclusion and Exclusion Criteria.}
\label{tab:incl-excl-privacy}
\small
\setlength{\tabcolsep}{4pt} 
\begin{tabular}{@{}l l p{0.66\columnwidth}@{}}
\toprule
\textbf{Criteria} & \textbf{Code} & \textbf{Criteria Description} \\
\midrule
\multirow{5}{*}{Inclusion}
  & I1 & Papers must be in English and be readily accessible in full text. \\
  & I2 & Papers include clear validation or evaluation methods, with supporting references. \\
  & I3 & Papers report practices and challenges in implementing privacy engineering, including case studies, best practices, or guidelines. \\
  & I4 & Papers provide insights into the trends and evolution of privacy engineering practice. \\
  & I5 & Papers identify key contributors and diverse perspectives within privacy engineering. \\
\midrule
\multirow{8}{*}{Exclusion}
  & E1 & Short papers or brief papers (less than six pages). \\
  & E2 & Papers that are not written in English. \\
  & E3 & Papers that lack relevant focus on privacy engineering. \\
  & E4 & Papers that predominantly discuss physical infrastructure or purely hardware aspects without a privacy-engineering link. \\
  & E5 & Duplicate papers from any source. \\
  & E6 & Papers lacking empirical evidence or practical applications to support theoretical propositions. \\
  & E7 & Workshop papers. \\
  & E8 & Conference papers later extended to journal. \\
\bottomrule
\end{tabular}
\end{table}

Table~\ref{tab:extraction-questions} defines the Extraction Questions (EQs) and recorded items that we captured per included study to standardize extraction and to support subsequent coding, Descriptive Theme construction and lifecycle placement as reported in Section~\ref{execution}.

\begin{table}[!h]
\centering
\caption{Protocol extraction questions (EQs) and recorded data items.}
\label{tab:extraction-questions}
\small
\begin{tabularx}{\columnwidth}{p{0.9cm} >{\raggedright\arraybackslash}X >{\raggedright\arraybackslash}X}
\toprule
\textbf{EQ} & \textbf{Question} & \textbf{Recorded items} \\
\midrule
EQ1 & What engineering artifacts or activities did the study report? & Requirements and specification artifacts; architectural decisions; implementation techniques; interface and communication artifacts; verification and measurement methods; operational controls; lifecycle operations such as retention, erasure and redress. \\
EQ2 & Which lifecycle phases did the study address? & Software lifecycle: requirements, design, implementation, verification, deployment, operation, decommissioning. Personal data lifecycle: collection, transfer, storage, processing and analytics, sharing or release, retention, deletion or erasure. \\
EQ3 & What problem setting and domain did the study cover? & Application domains; system context; deployment constraints. \\
EQ4 & What mechanisms and controls did the study instantiate? & Techniques in data pipelines and platforms; authorization and consent; provenance and logging; configuration or contract logic tied to controls. \\
EQ5 & What evaluation or evidence did the study present? & Datasets and workloads; validation protocol such as splits or k-fold; metrics including privacy–utility, robustness and performance; availability of code, data, or tools. \\
EQ6 & What organizational or process elements did the study link to controls? & Roles and responsibilities; approvals and assessments; operational procedures; evidence produced by processes. \\
EQ7 & What assumptions, threats, or constraints did the study state? & Adversary and threat model; trust boundaries; environmental constraints. \\
EQ8 & What artifacts did the study make available? & Tools, code, datasets, policy templates and measurement scripts with persistent links. \\
\bottomrule
\end{tabularx}
\end{table}

\section{Descriptive Theme Catalog and Study-to-Dimension Mapping}\label{sec:studies_to_dimensions}

Table~\ref{tab:dts_dims2} presents a catalog of 36 cross-paper Descriptive Themes (DTs), including codes and titles, mapped to thirteen high-order dimensions for privacy engineering. These results were obtained through the thematic synthesis process employed in this study.

\begin{table*}[h!]
\centering
\setlength\tabcolsep{2pt}
\renewcommand{\arraystretch}{1.12}
\caption{The output catalog of cross-paper Descriptive Themes (DTs) paired with the resulting privacy engineering high-order dimension.}
\label{tab:dts_dims2}
\scriptsize
\begin{tabularx}{\textwidth}{p{0.10\textwidth}|p{0.64\textwidth}|p{0.22\textwidth}}
\hline
\textbf{DT Code} & \textbf{DT Title} & \textbf{Dimension} \\
\hline
DTP1 & Differential Privacy operationalization (interactive, batch, local; utility-aware calibration). & \multirow{9}{*}{Privacy Enhancing Technologies (PETs)} \\
DTP2 & Cryptographic PETs (HE/MPC/ABE/ZK/NIZK) and privacy-preserving computation. & \\
DTP3 & Enclave/TEE-based enforcement and verifiable computation. & \\
DTP4 & Obfuscation \& attribute-inference prevention (visual, biometric, or tabular). & \\
DTP5 & Federated/Distributed learning mechanisms (secure aggregation, personalization, communication efficiency). & \\
DTP6 & Blockchain-backed privacy and accountability (smart contracts, auditability, incentives). & \\
DTP7 & Threats \& Attacks on ML (adversarial, poisoning, extraction/inversion) \& defenses. & \\
DTP8 & Robust aggregation and attack-resilient FL algorithms. & \\
DTP9 & Augmentation PETs (synthetic data \& digital twins). & \\
\hline
DTG1 & Access control, provenance, traceability and audit logging. & \multirow{4}{*}{Governance \& Accountability} \\
DTG2 & Incentives, reputation and free-riding mitigation in distributed settings. & \\
DTG3 & Roles, governance services, standards and accountability structures. & \\
\hline
DTV1 & Benchmarks, measurement studies and ecosystem evaluations & \multirow{3}{*}{Verification \& Testing} \\
DTV2 & Verification of models/updates, smart-contract verification, reproducibility artifacts & \\
DTV3 & Formal/robustness testing and certification (attack/defense evaluation) & \\
\hline
DTT1 & Consent and transparency communication (UX, explainability, banners) & \multirow{3}{*}{Transparency \& Communication} \\
DTT2 & Machine-readable privacy policies and information extraction & \\
DTT3 & Operational/sensory transparency in systems and robotics & \\
\hline
DTS1 & Ontologies and requirements modeling for privacy & \multirow{3}{*}{Modeling \& Specification} \\
DTS2 & Architectural patterns and service components (SOA, DGS/PDAS) & \\
DTS3 & Privacy-by-Design architectural tactics methodology & \\
\hline
DTR1 & Rights extraction/management (consent, access, deletion, portability) & \multirow{3}{*} {Data Subject Rights Management} \\
DTR2 & Rolling consent \& user control modalities \\
DTR3 & Data subject rights tooling \& dashboards \\
\hline
DTM1 & Data minimization via on-device processing, LDP, FL & \multirow{2}{*}{Minimization \& Purpose Limitation} \\
DTM2 & Purpose binding and policy-to-rule mapping & \\
\hline
DTO1 & Organizational processes, privacy-by-design backlogs and practices & \multirow{2}{*}{Organizational Measures} \\
DTO2 & Training \& culture enablement & \\
\hline
DPM1 & Privacy–utility calibration, epsilon tuning, error/accuracy metrics & \multirow{2}{*}{Privacy Metrics} \\
DPM2 & Robustness metrics, attack-surface measurement, certification & \\
\hline
DTC1 & Privacy awareness \& training & \multirow{2}{*}{Culture \& Training} \\
DTC2 & Organizational Privacy Culture \& Climate. & \\
\hline

DTL1 & Retention/deletion policies and lifecycle management & Lifelong Management \\
\hline
DTI1 & Tamper-evidence, auditability and forensics (temporal/audit tables) & Incident Response \& Management \\
\hline
DTUC1 & User-centred privacy in context (contextual integrity, multidimensional privacy) & User-Centric \\
\hline
\end{tabularx}
\end{table*}

Table~\ref{tab:pe-dim-primary2} summarizes the mapping of included studies to the dimensions of privacy engineering. For each dimension, it reports the assignment of each study, distinguishing primary from secondary. 

\begin{table*}[h!]
\centering
\setlength\tabcolsep{4pt}
\renewcommand{\arraystretch}{1.18}
\caption{Included studies mapped to privacy engineering dimensions. For each dimension, studies are classified as primary, which reflects the study’s main contribution, or secondary, which denotes an additional, substantiated contribution. Secondary assignments are marked with an asterisk (*).}
\label{tab:pe-dim-primary2}
\scriptsize
\begin{tabularx}{\textwidth}{@{}p{0.26\textwidth} p{0.24\textwidth} p{0.26\textwidth} p{0.24\textwidth}@{}}
\toprule
\multicolumn{4}{c}{Mapping of included studies to privacy engineering dimensions} \\
\midrule
Dimension & References & Dimension & References \\
\midrule
Privacy Metrics &
\begin{tabular}[t]{@{}l@{}}
\cite{wagner_technical_2019} \cite{pal_privacy_2019}* \cite{ermakova_security_2020}* \cite{noauthor_guest_2020}* \cite{liu_privacy_2021}* \\
\cite{beg_data_2022}* \cite{giordano_use_2022}* \cite{jordan_selecting_2022}* \cite{lo_systematic_2022}* \cite{mazmudar_cache_2022}* \\
\cite{sun_toward_2022}* \cite{zhao_survey_2022}* \cite{khalid_enhancing_2023}* \cite{nguyen_federated_2023}* \cite{pramod_privacy-preserving_2023}* \\
\cite{witt_decentral_2023}* \cite{du_privategaze_2024}* \cite{liang_identity_2024}* \cite{melzi_overview_2024}* \cite{mo_security_2024}* \\
\cite{odema_privynas_2024}* \cite{wang_pp-csa_2024}* \cite{weng_faster_2024}* \cite{yang_approaching_2024}* \cite{yu_dop-sql_2024}* \\
\cite{zhang_ppfed_2024}* \cite{zhao_scenario-based_2024}* \cite{ali_privacy-preserved_2025}* \cite{chen_advancements_2025}* \cite{wang_security_2025}*
\end{tabular} &
User-Centric &
\begin{tabular}[t]{@{}l@{}}
\cite{barhamgi_user-centric_2018} \cite{hu_dark_2023} \cite{wilkowska_interdisciplinary_2023} \cite{carmichael_personal_2024} \cite{de_chaves_user-centred_2025} \\
\cite{grabler_privacy_2025} \cite{hoel2019privacy}* \cite{alhirabi_security_2021}* \cite{benhamida_pyff_2021}* \cite{zhao_survey_2022}* \\
\cite{carboni_privacy_2023}* \cite{de_chaves_privacy_2023}* \cite{zhao_visual_2025}*
\end{tabular}
\\
\addlinespace[0.8ex]
Verification \& Testing &
\begin{tabular}[t]{@{}l@{}}
\cite{giordano_use_2022} \cite{iwaya_privacy_2023-1} \cite{wang_as_2024} \cite{islam_assurance_2018}* \cite{caiza_reusable_2019}* \\
\cite{pal_privacy_2019}* \cite{wagner_technical_2019}* \cite{iwaya_security_2020}* \cite{zhang_security_2020}* \cite{alhirabi_security_2021}* \\
\cite{chen_holistic_2021}* \cite{daoudagh_data_2021}* \cite{hammoudeh_service-oriented_2021}* \cite{jandl_reasons_2021}* \cite{liu_privacy_2021}* \\
\cite{semantha_conceptual_2021}* \cite{beg_data_2022}* \cite{iwaya_organizational_2022}* \cite{jha_internet_2022}* \cite{jordan_selecting_2022}* \\
\cite{lo_systematic_2022}* \cite{marcolla_survey_2022}* \cite{mazmudar_cache_2022}* \cite{sun_toward_2022}* \cite{wu_privacy-preserved_2022}* \\
\cite{zhao_survey_2022}* \cite{issa_blockchain-based_2023}* \cite{li_longitudinal_2023}* \cite{nguyen_federated_2023}* \cite{pramod_privacy-preserving_2023}* \\
\cite{salem_comprehensive_2023}* \cite{witt_decentral_2023}* \cite{zukaib_blockchain_2023}* \cite{cejas2024compai}* \cite{du_privategaze_2024}* \\
\cite{elkourdi_exploring_2024}* \cite{herwanto_leveraging_2024}* \cite{herwanto_toward_2024}* \cite{liang_identity_2024}* \cite{melzi_overview_2024}* \\
\cite{mo_security_2024}* \cite{odema_privynas_2024}* \cite{pallas2024privacy}* \cite{wang_pp-csa_2024}* \cite{weng_faster_2024}* \\
\cite{xia_towards_2024}* \cite{yang_approaching_2024}* \cite{yang_secudb_2024}* \cite{yu_insights_2024}* \cite{zhang_ppfed_2024}* \\
\cite{zhao_scenario-based_2024}* \cite{alghuried_blockchain_2025}* \cite{ali_privacy-preserved_2025}* \cite{chen_advancements_2025}* \cite{de_chaves_user-centred_2025}* \\
\cite{wang_security_2025}* \cite{zhao_visual_2025}*
\end{tabular} &
Transparency \& Communication &
\begin{tabular}[t]{@{}l@{}}
\cite{jha_internet_2022} \cite{cejas2024compai} \cite{alshammari_privacy_2018}* \cite{barhamgi_user-centric_2018}* \cite{islam_assurance_2018}* \\
\cite{hoel2019privacy}* \cite{gharib2020ontology}* \cite{iwaya_security_2020}* \cite{jones_profile_2020}* \cite{zhang_security_2020}* \\
\cite{alkhariji_synthesising_2021}* \cite{benhamida_pyff_2021}* \cite{daoudagh_data_2021}* \cite{jandl_reasons_2021}* \cite{saksena_rebooting_2021}* \\
\cite{andrade_privacy_2022}* \cite{beg_data_2022}* \cite{rommetveit_privacy_2022}* \cite{zhang_privacy-by-design_2022}* \cite{carboni_privacy_2023}* \\
\cite{de_chaves_privacy_2023}* \cite{gebauer_human---loop_2023}* \cite{hu_dark_2023}* \cite{iwaya_privacy_2023-1}* \cite{li_longitudinal_2023}* \\
\cite{lim_toward_2023}* \cite{wilkowska_interdisciplinary_2023}* \cite{carmichael_personal_2024}* \cite{du_privategaze_2024}* \cite{elkourdi_exploring_2024}* \\
\cite{iwaya_privacy_2024}* \cite{liu_privacy-preserving_2024}* \cite{pallas2024privacy}* \cite{wang_as_2024}* \cite{xia_towards_2024}* \\
\cite{ali_privacy-preserved_2025}* \cite{de_chaves_user-centred_2025}* \cite{grabler_privacy_2025}* \cite{wang_security_2025}* \cite{zhao_visual_2025}*
\end{tabular}
\\
\addlinespace[0.8ex]
Incident Response \& Management &
\begin{tabular}[t]{@{}l@{}}
\cite{senarath_will_2019}* \cite{ermakova_security_2020}* \cite{alkhariji_synthesising_2021}* \cite{jandl_reasons_2021}* \cite{andrade_privacy_2022}* \\ 
\cite{khalid_enhancing_2023}* \cite{salem_comprehensive_2023}* \cite{xia_towards_2024}* \cite{yang_secudb_2024}* \cite{pallas2024privacy}*
\end{tabular} &
Data Subject Rights Management &
\begin{tabular}[t]{@{}l@{}}
\cite{gebauer_human---loop_2023} \cite{barhamgi_user-centric_2018}* \cite{ermakova_security_2020}* \cite{gharib2020ontology}* \cite{jones_profile_2020}* \\
\cite{alkhariji_synthesising_2021}* \cite{daoudagh_data_2021}* \cite{jha_internet_2022}* \cite{wu_privacy-preserved_2022}* \cite{khalid_enhancing_2023}* \\
\cite{lim_toward_2023}* \cite{zukaib_blockchain_2023}* \cite{carmichael_personal_2024}* \cite{herwanto_leveraging_2024}* \cite{liu_privacy-preserving_2024}* \\
\cite{pallas2024privacy}* \cite{de_chaves_user-centred_2025}* \cite{grabler_privacy_2025}*
\end{tabular}
\\
\addlinespace[0.8ex]
Lifelong Management &
\begin{tabular}[t]{@{}l@{}}
\cite{senarath_will_2019}* \cite{jandl_reasons_2021}* \cite{andrade_privacy_2022}* \cite{mazeli_framework_2022}* \cite{carmichael_personal_2024}* \\
\cite{lim_toward_2023}* \cite{pallas2024privacy}*
\end{tabular} &
Minimization \& Purpose Limitation &
\begin{tabular}[t]{@{}l@{}}
\cite{hoel2019privacy}* \cite{gharib2020ontology}* \cite{jones_profile_2020}* \cite{alkhariji_synthesising_2021}* \cite{daoudagh_data_2021}* \\
\cite{jandl_reasons_2021}* \cite{saksena_rebooting_2021}* \cite{andrade_privacy_2022}* \cite{iwaya_privacy_2023-1}* \cite{lim_toward_2023}* \\
\cite{pallas2024privacy}* \cite{ali_privacy-preserved_2025}*
\end{tabular}
\\
\addlinespace[0.8ex]
Governance \& Accountability &
\begin{tabular}[t]{@{}l@{}}
\cite{islam_assurance_2018} \cite{hoel2019privacy} \cite{ermakova_security_2020} \cite{jones_profile_2020} \cite{daoudagh_data_2021} \\
\cite{jandl_reasons_2021} \cite{saksena_rebooting_2021} \cite{rommetveit_privacy_2022} \cite{zhang_privacy-by-design_2022} \cite{lim_toward_2023} \\
\cite{mastrolembo_ventura_enhancing_2023} \cite{elkourdi_exploring_2024} \cite{iwaya_privacy_2024} \cite{liu_privacy-preserving_2024} \cite{pallas2024privacy} \\
\cite{xia_towards_2024} \cite{yu_insights_2024} \cite{alshammari_privacy_2018}* \cite{caiza_reusable_2019}* \cite{senarath_will_2019}* \\
\cite{spiekermann_inside_2019}* \cite{gharib2020ontology}* \cite{iwaya_security_2020}* \cite{sun_data_2020}* \cite{zhang_security_2020}* \\
\cite{alhirabi_security_2021}* \cite{benhamida_pyff_2021}* \cite{semantha_conceptual_2021}* \cite{andrade_privacy_2022}* \cite{iwaya_organizational_2022}* \\
\cite{jha_internet_2022}* \cite{jordan_selecting_2022}* \cite{lo_systematic_2022}* \cite{marcolla_survey_2022}* \cite{mazeli_framework_2022}* \\
\cite{wu_privacy-preserved_2022}* \cite{carboni_privacy_2023}* \cite{de_chaves_privacy_2023}* \cite{issa_blockchain-based_2023}* \cite{iwaya_privacy_2023}* \\
\cite{iwaya_privacy_2023-1}* \cite{khalid_enhancing_2023}* \cite{nguyen_federated_2023}* \cite{salem_comprehensive_2023}* \cite{wilkowska_interdisciplinary_2023}* \\
\cite{witt_decentral_2023}* \cite{zukaib_blockchain_2023}* \cite{almarshoud_security_2024}* \cite{carmichael_personal_2024}* \cite{cejas2024compai}* \\
\cite{herwanto_toward_2024}* \cite{liang_identity_2024}* \cite{mo_security_2024}* \cite{yang_secudb_2024}* \cite{alghuried_blockchain_2025}* \\
\cite{ali_privacy-preserved_2025}* \cite{chen_advancements_2025}* \cite{grabler_privacy_2025}* \cite{wang_security_2025}*
\end{tabular} &
Organizational Measures &
\begin{tabular}[t]{@{}l@{}}
\cite{andrade_privacy_2022} \cite{iwaya_organizational_2022} \cite{carboni_privacy_2023} \cite{de_chaves_privacy_2023} \cite{islam_assurance_2018}* \\
\cite{hoel2019privacy}* \cite{senarath_will_2019}* \cite{spiekermann_inside_2019}* \cite{ermakova_security_2020}* \cite{jones_profile_2020}* \\
\cite{alhirabi_security_2021}* \cite{jandl_reasons_2021}* \cite{saksena_rebooting_2021}* \cite{lo_systematic_2022}* \cite{mazeli_framework_2022}* \\
\cite{rommetveit_privacy_2022}* \cite{zhang_privacy-by-design_2022}* \cite{iwaya_privacy_2023}* \cite{iwaya_privacy_2023-1}* \cite{lim_toward_2023}* \\
\cite{mastrolembo_ventura_enhancing_2023}* \cite{carmichael_personal_2024}* \cite{elkourdi_exploring_2024}* \cite{herwanto_toward_2024}* \cite{iwaya_privacy_2024}* \\
\cite{pallas2024privacy}* \cite{de_chaves_user-centred_2025}* \cite{grabler_privacy_2025}*
\end{tabular}
\\
\addlinespace[0.8ex]
Culture \& Training &
\begin{tabular}[t]{@{}l@{}}
\cite{senarath_will_2019} \cite{spiekermann_inside_2019} \cite{mazeli_framework_2022} \cite{iwaya_privacy_2023} \cite{jones_profile_2020}* \\
\cite{alhirabi_security_2021}* \cite{andrade_privacy_2022}* \cite{carboni_privacy_2023}* \cite{de_chaves_privacy_2023}* \cite{hu_dark_2023}* \\
\cite{iwaya_privacy_2023-1}* \cite{wilkowska_interdisciplinary_2023}* \cite{iwaya_privacy_2024}*
\end{tabular} &
Privacy Enhancing Technologies (PETs) &
\begin{tabular}[t]{@{}l@{}}
\cite{pal_privacy_2019} \cite{iwaya_security_2020} \cite{noauthor_guest_2020} \cite{zhang_security_2020} \cite{liu_privacy_2021} \\
\cite{beg_data_2022} \cite{jordan_selecting_2022} \cite{lo_systematic_2022} \cite{marcolla_survey_2022} \cite{mazmudar_cache_2022} \\
\cite{sun_toward_2022} \cite{wu_privacy-preserved_2022} \cite{zhao_survey_2022} \cite{issa_blockchain-based_2023} \cite{khalid_enhancing_2023} \\
\cite{li_longitudinal_2023} \cite{nguyen_federated_2023} \cite{pramod_privacy-preserving_2023} \cite{witt_decentral_2023} \cite{zukaib_blockchain_2023} \\
\cite{almarshoud_security_2024} \cite{du_privategaze_2024} \cite{liang_identity_2024} \cite{melzi_overview_2024} \cite{mo_security_2024} \\
\cite{odema_privynas_2024} \cite{solis_exploring_2024} \cite{wang_pp-csa_2024} \cite{weng_faster_2024} \cite{yang_approaching_2024} \\
\cite{yang_secudb_2024} \cite{yu_dop-sql_2024} \cite{zhang_ppfed_2024} \cite{zhao_scenario-based_2024} \cite{alghuried_blockchain_2025} \\
\cite{ali_privacy-preserved_2025} \cite{chen_advancements_2025} \cite{wang_security_2025} \cite{zhao_visual_2025} \cite{hoel2019privacy}* \\
\cite{sun_data_2020}* \cite{benhamida_pyff_2021}* \cite{daoudagh_data_2021}* \cite{hammoudeh_service-oriented_2021}* \cite{saksena_rebooting_2021}* \\
\cite{semantha_conceptual_2021}* \cite{zhang_privacy-by-design_2022}* \cite{mastrolembo_ventura_enhancing_2023}* \cite{salem_comprehensive_2023}* \cite{wilkowska_interdisciplinary_2023}* \\
\cite{liu_privacy-preserving_2024}* \cite{yu_insights_2024}*
\end{tabular}
\\
\addlinespace[0.8ex]
Modeling \& Specification &
\begin{tabular}[t]{@{}l@{}}
\cite{alshammari_privacy_2018} \cite{caiza_reusable_2019} \cite{gharib2020ontology} \cite{sun_data_2020} \cite{alhirabi_security_2021} \\
\cite{alkhariji_synthesising_2021} \cite{benhamida_pyff_2021} \cite{chen_holistic_2021} \cite{hammoudeh_service-oriented_2021} \cite{semantha_conceptual_2021} \\
\cite{salem_comprehensive_2023} \cite{herwanto_leveraging_2024} \cite{herwanto_toward_2024} \cite{barhamgi_user-centric_2018}* \cite{islam_assurance_2018}* \\
\cite{hoel2019privacy}* \cite{wagner_technical_2019}* \cite{jones_profile_2020}* \cite{zhang_security_2020}* \cite{daoudagh_data_2021}* \\
\cite{andrade_privacy_2022}* \cite{jha_internet_2022}* \cite{wu_privacy-preserved_2022}* \cite{de_chaves_privacy_2023}* \cite{iwaya_privacy_2023-1}* \\
\cite{mastrolembo_ventura_enhancing_2023}* \cite{wilkowska_interdisciplinary_2023}* \cite{almarshoud_security_2024}* \cite{cejas2024compai}* \cite{elkourdi_exploring_2024}* \\
\cite{liu_privacy-preserving_2024}* \cite{pallas2024privacy}* \cite{solis_exploring_2024}* \cite{de_chaves_user-centred_2025}*
\end{tabular} &
  &
\begin{tabular}[t]{@{}l@{}}

\end{tabular}
\\
\bottomrule
\end{tabularx}
\end{table*}

\end{document}